\documentclass[]{jfm}

\usepackage{graphicx}
\usepackage{newtxtext}
\usepackage{newtxmath}
\usepackage{natbib}
\usepackage{hyperref}
\usepackage{subfigure}
\hypersetup{
    colorlinks = true,
    urlcolor   = blue,
    citecolor  = black,
}

\newcommand{\RomanNumeralCaps}[1]

\title{Jet-edge interaction: linear and non-linear frequency-selection mechanisms}
\shorttitle{Jet-edge interactions}
\shortauthor{M. N. Stavropoulos et al.}

\author{Michael N. Stavropoulos\aff{1},
  André V.G. Cavalieri\aff{2},
  Lutz Lesshafft\aff{3},
 \and Peter Jordan\aff{1}
 \corresp{\email{peter.jordan@univ-poitiers.fr}}}

\affiliation{\aff{1}Institut Pprime, CNRS-Université de Poitiers-ENSMA, 86962 Chasseneuil‑du‑Poitou, Poitiers, France
\aff{2}Instituto Tecnológico de Aeronáutica, São José dos Campos, SP 12228-900, Brazil
\aff{3}LadHyX, CNRS, École Polytechnique, Institut Polytechnique de Paris, 91120 Palaiseau, France}

\begin{document}
\maketitle

\begin{abstract}
We consider a round turbulent jet grazing a rectangular plate angled at $45^\circ$. Through sound pressure measurements, the tonal dynamics associated with jet-edge interaction are explored in a parameter space comprising jet Mach number, $M_j$, and plate radial position, $R/D$. A variety of spectral signatures are observed and classified. The classification - based on analysis of power-spectral density and bicoherence, and on the resonance model proposed by \citet{jordan2018jet} - comprises: broadband spectra; tonal spectra associated with purely linear frequency-selection mechanisms; tonal spectra associated with both linear and non-linear frequency selection. 
The classification identifies regions in the parameter space $(M_j,R/D)$; and clarifies mechanisms underpinning regime changes. The linear frequency selection (LFS) regime comprises multiple tones, with no evidence of triad interaction. A regime involving non-linear frequency selection emerges from this state, with the strong amplification of one LFS tone, which then generates multiple harmonics. Intermediate regimes are identified involving weaker, non-harmonic triadic interactions where two LFS tones interact to generate a third tone.
In addition to these mechanisms a mode-switching mechanism is identified at $M_j =0.84$ and shown to result from the cut-on of a new upstream-travelling wave at that Mach number. The mode-switch is found to be remarkably robust, occurring in a repeatable manner over a Mach-number increment of 0.01 regardless of whether the Mach number is increased or decreased (no hysteresis is observed).
\end{abstract}


\section{Introduction}
Systems involving high-speed jets are often found to exhibit resonance behaviour. An initial disturbance, generated at a position upstream within the flow, travelling downstream encounters a boundary which converts it to an upstream-travelling disturbance. This disturbance then travels back upstream until reaching the initial upstream position, where it then generates a new downstream-travelling disturbance closing the feedback loop. Resonance may arise due to the presence of a physical boundary as in edge tones \citep{richardson1931edge,powell1961edgetone}, cavities \citep{rossiter1964,smith2024history}, and impingement \citep{ho1981dynamics,henderson2005experimental,edgington2025vortices}; or without an external boundary as in potential-core resonance \citep{towne2017acoustic}, and screech \citep{edgington2019aeroacoustic}.

Motivated by the problem of installed jet noise, we consider a subsonic turbulent jet grazing the sharp edge of an adjacent metal plate inclined at $45^\circ$. In such configurations jet-edge interactions generate both broadband and tonal sound. The tonal component has been shown by \cite{jordan2018jet} to be primarily underpinned by linear frequency-selection mechanisms associated with a feedback loop in which downstream-travelling Kelvin-Helmholtz waves are scattered by the edge into a family of dispersive, upstream-travelling, guided jet modes. Coherent structures such as these are often modelled by researchers seeking to develop explanations for the emitted sound since signatures of coherence were first observed in turbulent jets \citep{Mollo1967,crow_champagne_1971,brown_roshko_1974}.
The downstream-travelling wavepacket, modelled via linearisation of the Navier-Stokes equations about the mean flow \citep{JordanColoniusWavepacket,cavalieri2019wave}, can be used to describe the broadband noise generated by installed jets. \citet{CAVALIERI20146516,Piantanida2016,NOGUEIRA2017,Nogueira2019} used such models to reproduce experimental trends such as, the exponential decay of radiated sound with plate radial displacement, sound amplitudes, and directivity.
In \citet{lawrence2015installed} the authors considered the tonal sound generated by jet-edge interaction. They postulated that the tones were due to resonance between downstream-travelling Kelvin-Helmholtz waves and upstream-travelling sound waves. \citet{jordan2018jet} showed that the tones are associated with a linear frequency-selection mechanism that can be modelled using a cylindrical vortex sheet. The resonance mechanism involves feedback between downstream-travelling Kelvin-Helmholtz wavepackets \citep{JordanColoniusWavepacket} and upstream-travelling guided \citep{tam1989three,towne2017acoustic,schmidt2017,nogueira2024guided} waves. But that study did not explore the non-linear aspects of the tonal dynamics.

We here consider the jet-edge interaction problem using a thick, rigid plate, and with the streamwise edge position and plate angle fixed at, respectively, $L/D=2$ and $\alpha = 45^\circ$. We study the spectral signatures in a parameter space defined by the radial edge position, and the jet Mach number. We identify, characterise, and classify the spectral signatures that emerge, and explore bifurcations between the different resonance regimes.

The paper is organised as follows. In \S \ref{Sec:Method} both the experimental methodology and analysis techniques are detailed. The different dynamics observed across the parameter space are defined and discussed in \S \ref{Sec:Tonal}. Transitions and competition between the different dynamics are considered in \S \ref{Sec:Change}. In \S \ref{Sec:SPL} the prior sections' results are re-contextualised with the experimental sound pressure levels, and concluding remarks are provided in \S \ref{Sec:Conclude}.

\section{Methodology}
\label{Sec:Method}
\subsection{Experimental setup} 
\label{Sec:Setup}
\begin{figure} 
\centering
\includegraphics[clip=true, trim= 0 0 0 0, width=0.8\textwidth]{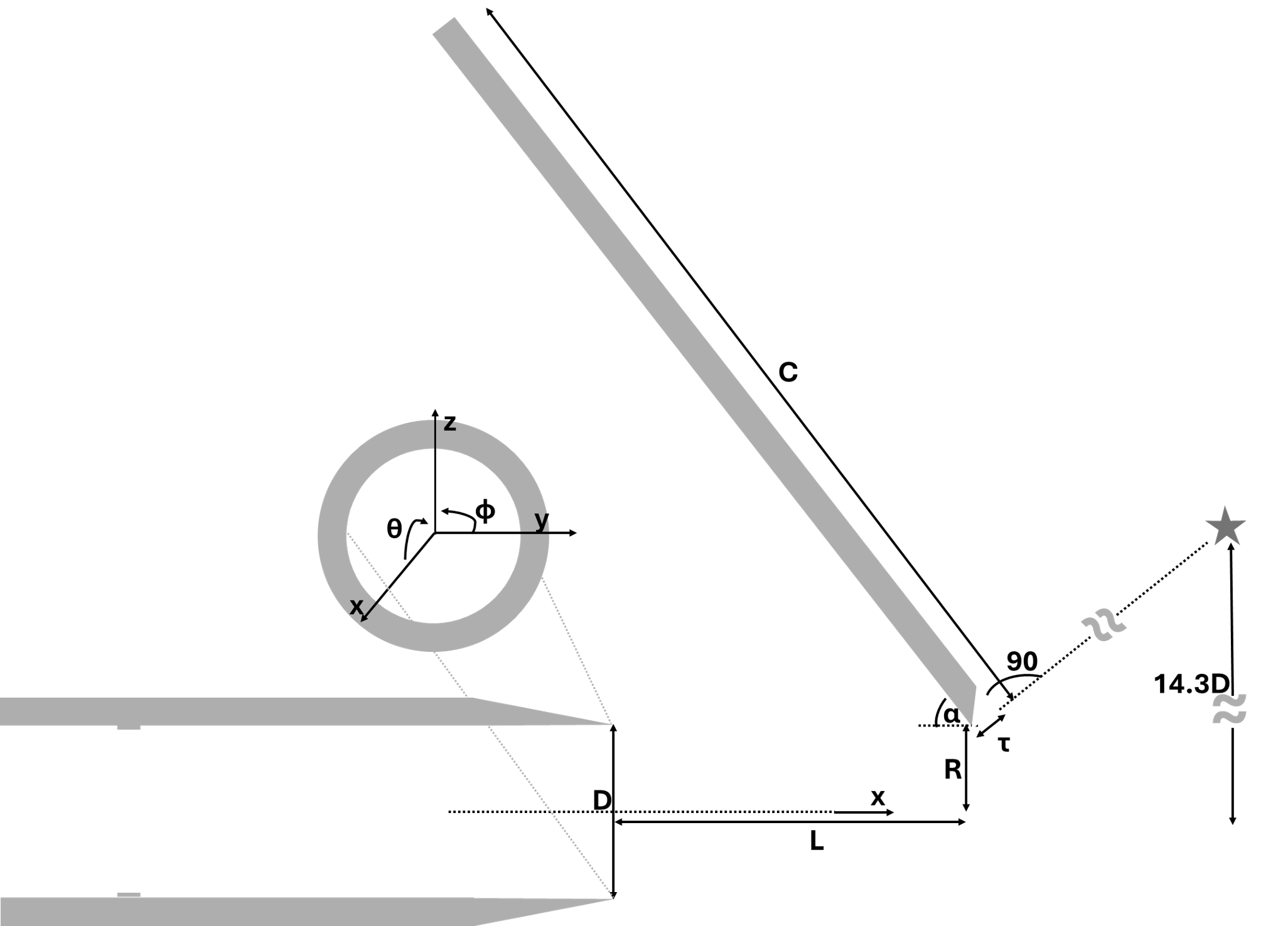}
\caption{Illustration of the installed-jet setup when viewed from above. Key variables are highlighted, and the microphone position is marked with a star.}
\label{fig:1}
\end{figure}

A schematic of the setup is shown in figure \ref{fig:1}. A jet of diameter $D$ is positioned a radial distance $R$ from a plate of span $S$ and 'chord' $C$. The plate edge is at distance $L$ from the jet exit and is inclined at an angle of $\alpha = 45^\circ$ relative to the downstream direction $x$. $R$, $L$, $S$, and $C$, are all normalised by $D$. The polar angle $\theta$ and azimuthal angle $\varphi$ are measured from the positive $x$ and $y$ axes respectively.
The experiments were performed at the \textit{Bruit \& Vent} facility of \textit{Institut Pprime} following prior jet-edge studies \citep{amaral2023,Stavropoulos2024Edge}. Isothermal conditions are imposed between the jet and ambient environment (within 0.3 K). The nozzle, the same used in previous studies \citep{cavalieri2012axisymmetric,cavalieri2013wavepackets,jordan2018jet}, comprises a convergent section, followed by a constant-diameter region before the exit. The boundary layer is tripped $2.7D$ upstream of the nozzle exit and fully turbulent at the exit plane. The nozzle has an exit diameter, $D$, of 50 mm, and the plate dimensions of $S = 15D$, $C = 5D$, and an 8 mm thickness ($\tau$). The edge of the plate has been chamfered to ensure it remains sharp. A B\&K type 4944-B 1/4-inch pressure field microphone, with frequency range of 16 Hz - 70 kHz, is placed at a radial distance of 14.3$D$ from the nozzle, positioned at $90^\circ$ to the plate edge (see figure \ref{fig:1}). Calibration of the microphone is done with a B\&K type 4231 sound calibrator. Measurements were taken with a sample rate of 200 kHz for 30 seconds, jet Mach number, $M_j$, was varied in the range 0.4:0.01:1 for $R/D$ in the range 0.5:0.01:0.71, and at $R/D = 0.8$. Streamwise edge position and plate angle where held constant at, respectively,  $L/D = 2$ and $\alpha = 45^\circ$. For each $R/D$ position data is acquired starting at $M_j = 1$, and then decreasing the jet mach number in 0.01 increments. In regions of interest, evolution of the tonal dynamics was also explored by increasing Mach number, to confirm that the direction of change of the Mach number is not important. 
\subsection{Power spectral density}
Power Spectral Density were computed from the time-series data using  the Welch method \citep{welch1967use} and a hanning window, block size of 16,384 (for 1461 total blocks), and 75\% overlap; the resultant frequency resolution is 12Hz. Following \citet{Bres2018} the PSD is presented in the non-dimensional form
\begin{equation}
    \mathit{PSD}(St) = 10\log_{10}\left(\frac{\mathit{PSD}(f)}{\mathit{p_{ref}}}\frac{U_j}{D}\frac{1}{M_j^{4}} \right),
    \label{eqn:2.1}
\end{equation}
where the reference pressure, $p_{\mathit{ref}}$, is $2\cdot10^{-5}$ Pa, and non-dimensionalisation has been applied using the jet diameter, $D$, and the jet-exit velocity, $U_j$. Equation \ref{eqn:2.1} gives units of PSD in dB/St, with $St = \frac{fD}{U_j}$ the non-dimensional frequency.
\subsection{Bicoherence}
\label{sec:Bic}
The PSD extracts information concerning fluctuation energy density magnitude only. An indicator of non-linear interaction between different frequencies is provided by the bispectrum \citep{sigl1994introduction,padois2016tonal}:
\begin{equation}
    B(f_1,f_2) = \frac{1}{N}\sum_{i=1}^{N}|X_{i}(f_1)X_{i}(f_2)X_{i}^{*}(f_1+f_2)|,
    \label{eqn:2.2}
\end{equation}
which is computed using the same block size, windowing, and overlap used for the PSD calculation. Equation \ref{eqn:2.2} reveals phase-synchronisation between $f_1$, $f_2$, and their sum $f_3$, and is an indicator of tonal triadic interaction. Equation \ref{eqn:2.2} is normalised by the real triple product ($B_{R}$)
\begin{equation}
    \mathit{B_{R}}(f_1,f_2) = \frac{1}{N}\sum_{i=1}^{N}X_{i}(f_1)X_{i}^{*}(f_1)X_{i}(f_2)X_{i}^{*}(f_2)X_{i}^{*}(f_1+f_2)X_{i}(f_1+f_2),
    \label{eqn:2.3}
\end{equation}
to provide the bicoherence
\begin{equation}
    b(f_1,f_2) = \frac{B(f_1,f_2)}{\sqrt{\mathit{B_{R}}(f_1,f_2)}}.
    \label{eqn:2.4}
\end{equation}
The bicoherence (equation \ref{eqn:2.4}) ranges from $0$ to $1$ and indicates the level of phase-synchronisation between frequencies $f_1$, $f_2$, and $f_1 + f_2$. Note that from their definitions, equations \ref{eqn:2.2}, \ref{eqn:2.3}, and \ref{eqn:2.4} are all symmetric about the line $f_1 = f_2$. This analysis technique risks false positives, in the sense that a non-zero value of equation \ref{eqn:2.4} does not immediately indicate a triadic interaction. Equation \ref{eqn:2.2} is ensemble averaged so if frequencies $f_1$ and $f_2$ are phase coupled the phase of each block will be similar, and thus combine constructively leading to a large $B$ magnitude. Conversely, uncoupled frequency pairs will produce blocks of differing phase which, when averaged, combine destructively and tend towards a magnitude of zero; in practice this value will be small but non-zero. As such, it is necessary to set a minimum $b$ value below which all values are treated as zero. Values of $b$ computed over a region where the system exhibits purely broadband noise are first considered as a baseline case, since any positive $b$ value here cannot correspond to non-linear generation of tones. For $R/D = 0.8$ across all $M_j$ (a solely broadband region) the mean maximal $b$ is 0.29 and overall maximum $b$ of 0.32. From this, we propose the minimum threshold $b > 0.35$ as an indication of a triadic interaction. Any $b$ calculated below this threshold will be treated as if it were zero. Contours of $b$ will also be overlain with markers due to the $St$ localisation of $b$ peaks (see Appendix \ref{App:A}).
\subsection{Linear modelling}
\label{sec:Model}
\begin{figure} 
    \centering
    \includegraphics[clip=true, trim= 0 0 0 0, width=0.7\textwidth]{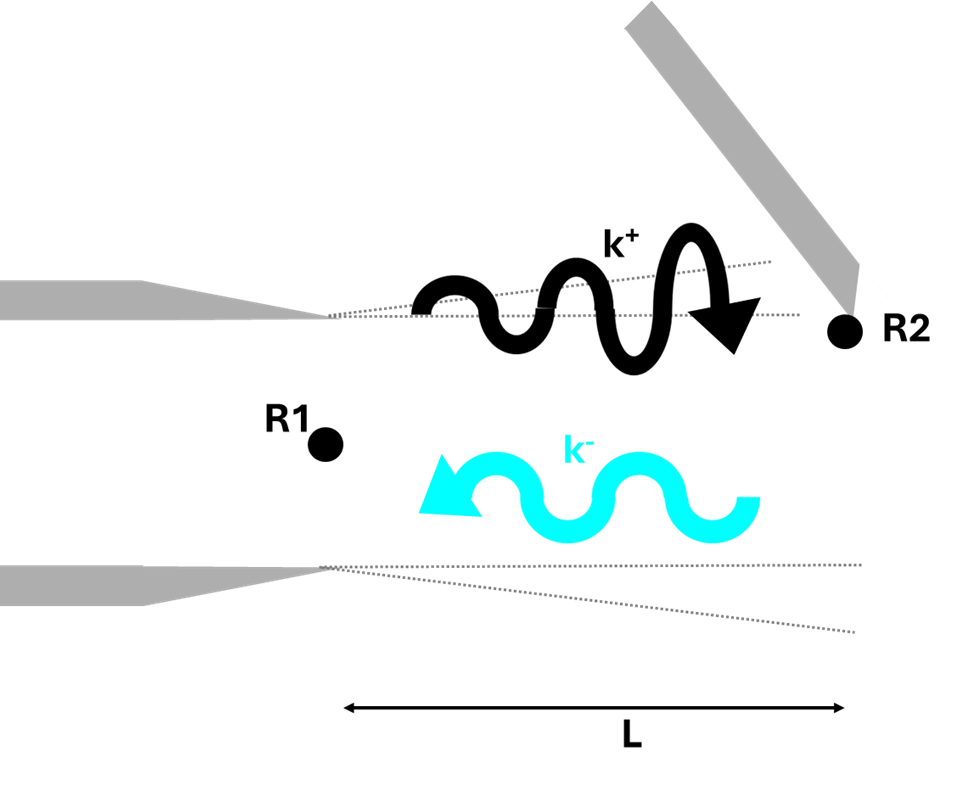}
    \caption{Feedback loop considered for the linear frequency-prediction model.}%
    \label{fig:2}%
\end{figure}

To identify tones whose frequency is determined by a linear mechanism, we use the model proposed by \citet{jordan2018jet}. The same model has been successfully used to understand other resonant fluid-mechanics phenomena, for example screech tones in supersonic flows \citep{mancinelli2021,stavropoulos2023}. The model assumes a feedback loop between downstream- and upstream-travelling waves, respectively, $k^+$ and $k^-$, that couple via two reflection points (characterised by complex-valued reflection coefficients) one upstream ($R1$), one downstream ($R2$), and separated by a distance $L$. For the installed-jet configuration these locations are the nozzle exit plane (upstream) and the plate edge (downstream), respectively. The $k^+$ wave travels from the nozzle lip to the plate edge ($R2$), where it is converted to a $k^-$ wave, which travels to the nozzle lip ($R1$) where it is scattered into a $k^+$ wave. An illustration of the feedback mechanism is provided in figure \ref{fig:2}.
From this feedback mechanism equations describing the magnitude and phase of the resonance may be found. Following prior studies \citep{jordan2018jet,mancinelli2019screech} only the phase component is considered here, expressed as
\begin{equation}
    \Delta k = \frac{(2p\pi+\phi)}{L}.
    \label{eqn:2.5}
\end{equation}
Here $\Delta k$ is the difference between the real components of the wavenumbers $k^{+}$ and $k^{-}$, $p$ is a positive integer representing the number of concurrent resonance cycles considered, $\phi$ is the phase difference between the two reflection coefficients imposed as either $0$ (in-phase) or $\pi$ (out-of-phase), and $L$ is the distance from the nozzle exit plane to the downstream edge. Linear stability theory is used to obtain values of $\Delta k$ as functions of jet Mach number and frequency. These values are compared to the right-hand-side of equation \ref{eqn:2.5} for different values of $\phi$ and $p$. Where $\frac{(2p\pi+\phi)}{L}$ intersects with $\Delta k(St,M_j)$ indicates a prediction of tonal frequency for that $St$ value.

$\Delta k$ is modelled using a cylindrical vortex sheet as in \citet{jordan2018jet}. The normal-mode ansatz, is, for the pressure,
\begin{equation}
    \tilde{P}(x,r,\theta,t) = P(r)e^{i(kx-\omega t+m\theta)},
    \label{eqn:2.6}
\end{equation}
where $m$ is the azimuthal Fourier mode number. Substituting into the Euler equations, an analytical dispersion relation is found:
\begin{equation}
\mathcal{D} = \frac{1}{\left(1-\frac{k M}{\omega}\right)^2}+\frac{1}{T}\frac{I_m\left(\frac{\gamma_i}{2}\right)\left[\frac{\gamma_0}{2}K_{m-1}\left(\frac{\gamma_0}{2}\right)+m K_{m}\left(\frac{\gamma_0}{2}\right)\right]}{K_m\left(\frac{\gamma_0}{2}\right)\left[\frac{\gamma_i}{2}I_{m-1}\left(\frac{\gamma_i}{2}\right)-m I_{m}\left(\frac{\gamma_i}{2}\right)\right]} = 0,
\label{eqn:2.7}
\end{equation}
where
\begin{eqnarray}
\gamma_{i} = \sqrt{k^2 -\frac{1}{T}(\omega-Mk)^2}, \nonumber \\ 
\gamma_{0} = \sqrt{k^2 -\omega^2}.
\label{eqn:2.8}
\end{eqnarray}
Here, $k$ is the streamwise wavenumber, $\omega$ the frequency, $I_m$ and $K_m$ are the modified Bessel functions of first and second kind, the subscripts $i$ and $0$ refer to inside and outside the jet core, $T$ the temperature ratio between jet and ambient (kept equal to 1 to match the isothermal condition of the experiment), and $M$ the acoustic Mach number (which for $T = 1$ is identical to the jet Mach number $M_j$). For a given eigenvalue pair, $(k,\omega)$, eigenfunctions are given by,
\begin{equation}
P(r)=\begin{cases}
\begin{aligned}
& A_{m}I_{m}\left(\gamma_i\frac{r}{D}\right), \qquad\quad \ \ \,\,\, 0 \leq\frac{r}{D} \leq 0.5\\ 
& B_{m}K_{m}\left(\gamma_0\frac{r}{D}\right), \qquad\qquad \ \ \ \,\,\, \frac{r}{D} \geq 0.5\\ 
\end{aligned}
\end{cases}
\label{eqn:2.9}
\end{equation}
with $A_m$ and $B_m$ constants.

\begin{figure}
\centering
\includegraphics[clip=true, trim= 0 0 0 0, width=0.7\textwidth]{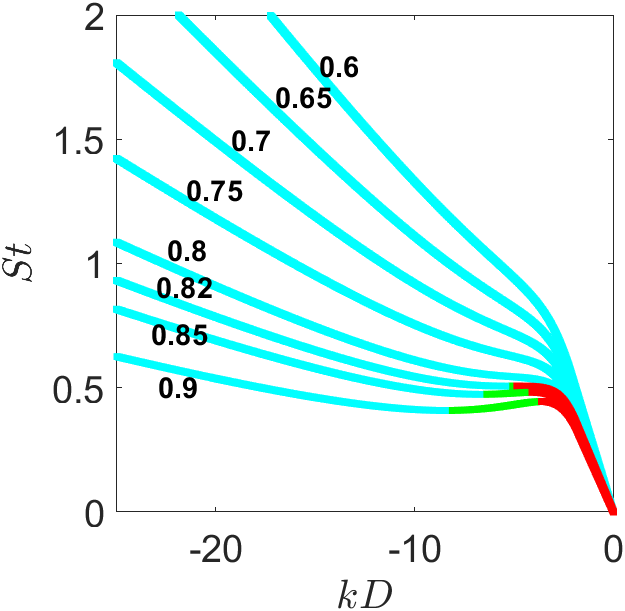}
\caption{Selected dispersion relation curves for $M_j$ 0.6, 0.65, 0.7, 0.75, 0.8, 0.82, 0.85, 0.9 where, for $M_j < 0.82$ $k^{-}_{th}$ waves (cyan) are present and for $M_j \geq 0.82$ $k^{-}_{d}$ (cyan), $k^{-}_{p}$ (red), and $k^{+}_{T}$ (green) waves are present.}%
\label{fig:3}%
\end{figure}
Selected dispersion relation curves for real frequency and wavenumber are shown in figure \ref{fig:3} to highlight the different families of waves supported by the vortex-sheet model and some of which play an important role in the tonal dynamics observed in the data. The eigenvalue loci shown correspond to the $(0,1)$ mode following the convention $(m,n)$, where $m$ is the azimuthal mode number and $n$ the radial mode number - associated with the number of anti-nodes in the pressure eigenfunction \citep{tam1989three,towne2017acoustic,gojon2018oscillation,edgington2018upstream}. For $M_j < 0.82$, figure \ref{fig:3} illustrates that there is a single upstream-travelling wave available to close the feedback loop: $k_{th}^{-}$ (cyan); whereas, for $M_j > 0.82$ both the $k_{d}^{-}$ (cyan) and $k_{p}^{-}$ (red) upstream-travelling waves exist to close resonance. Also shown in figure \ref{fig:3} are the downstream-propagating $k_{T}^{+}$ (green) waves. For a more complete discussion of the waves supported by the vortex sheet, and observed in turbulent jet data, the reader can refer to \citet{towne2017acoustic,jordan2018jet}. Recent work by \citet{nogueira2024guided} has provided a detailed discussion of the $k^{-}_{p}$ wave, which can be understood, as can all of the other guided waves of figure \ref{fig:3}, in terms of reflection and transmission of pressure fluctuations at the shear layer. The upstream-travelling, guided modes have been shown to be important in a large variety of resonant fluid-mechanics systems, including impinging jets \citep{tam_ahuja_1990,tam1992impingement,bogey2017feedback,jaunet2019dynamics}, subsonic jets \citep{towne2017acoustic,schmidt2017}, jet-edge interaction \citep{jordan2018jet}, supersonic screeching jets \citep{edgington2018upstream,gojon2018oscillation,mancinelli2019screech}, twin jets \citep{nogueira_edgington-mitchell_2021,stavropoulos2023}, among others.
\begin{figure}
\centering
\includegraphics[clip=true, trim= 0 0 0 0, width=0.7\textwidth]{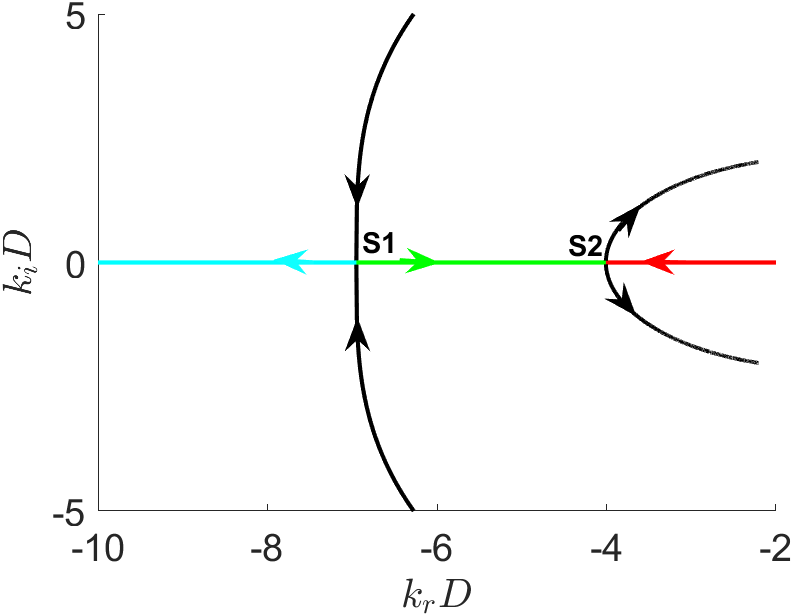}
\caption{Illustration of saddle-point formation within the complex k-plane with arrows indicating the direction of increasing $St$.}%
\label{fig:4}%
\end{figure}

In figure \ref{fig:3} the points where a curve, for $M_j \geq 0.82$, changes from $k^{-}_d$ to $k^{+}_T$, and $k^{+}_T$ to $k^{-}_p$ are saddle points in the complex $k$-plane. This is illustrated in figure \ref{fig:4}. At low $St$ the $k^{-}_d$ and $k^{+}_T$ modes are evanescent, with increasing frequency their eigenvalues approach and attain the real axis at which point they become propagative. With further increase in frequency the $k_{T}^+$ modes move to smaller negative wavenumber, whereas the $k_{d}^-$ mode moves to larger negative wavenumber. As $St$ increases further the $k^{+}_T$ and $k^{-}_p$ waves meet at a second saddle point, and then become evanescent for greater values of $St$. The frequencies of the $S1$ and $S2$ saddle points are important, as for $St < S1$ only $k^{-}_p$ waves are propagative ($k^{-}_d$ and $k^{+}_T$ evanescent), whilst for $St > S2$ only $k^{-}_d$ waves are propagative ($k^{-}_p$ and $k^{+}_T$ evanescent). This behaviour will be important in classifying the tonal mode switching observed in the experiment.

Equation \ref{eqn:2.5} will be used with $k^{-}_{th}$, $k^{-}_d$, and $k^{-}_p$ as the $k^{-}$ wave, the specific choice depending on the $M_j$ and $St$ considered, and with the Kelvin-Helmholtz mode as the $k^{+}$ wave.
\section{Tonal dynamics}
\label{Sec:Tonal}
\subsection{Spectral signatures}
\label{Sec:Spec}
\subsubsection{Broadband}
\label{Sec:BB}
\begin{figure}
\centering
\subfigure[PSD]{\includegraphics[clip=true, trim= 0 0 0 0, width=0.47\textwidth]{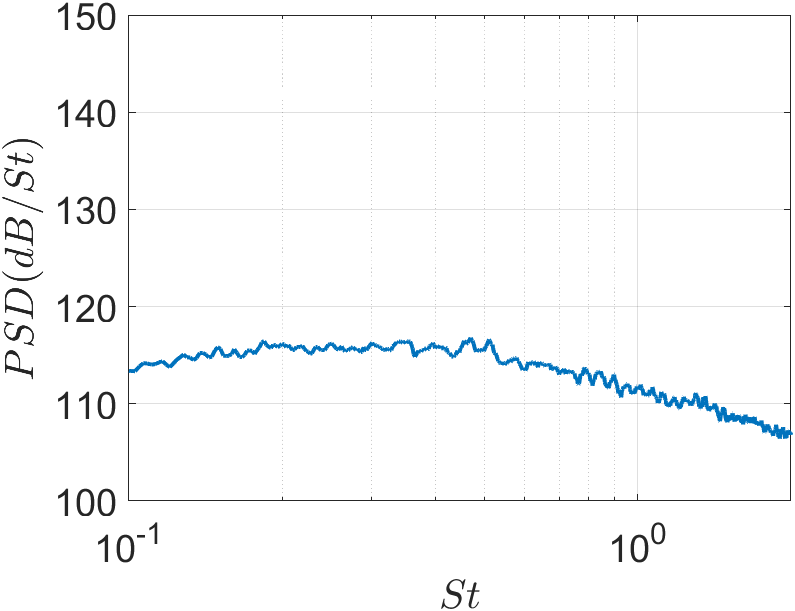}}\subfigure[$b$]{\includegraphics[clip=true, trim= 0 0 0 0, width=0.5\textwidth]{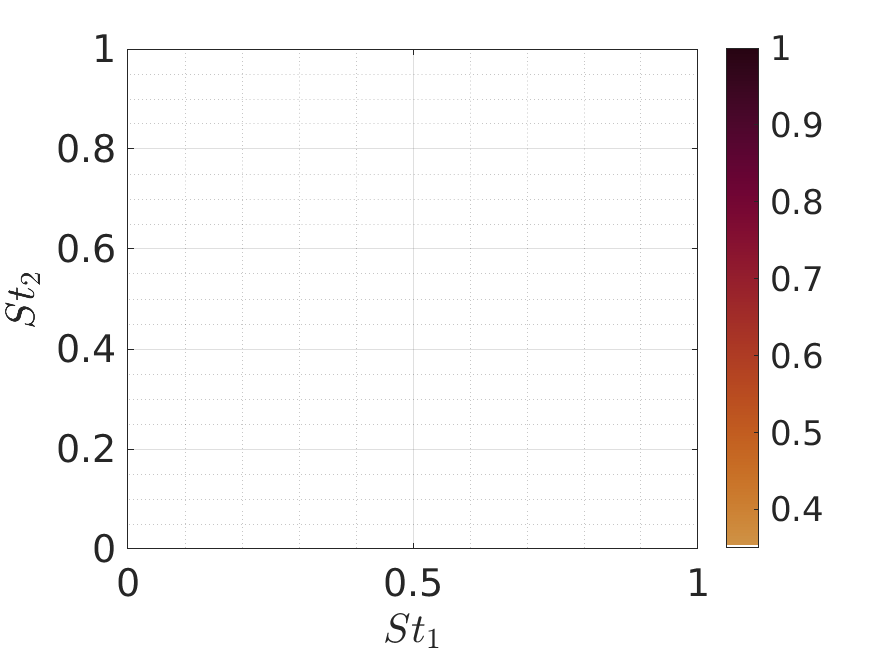}}
\caption{Broadband spectra for $(M_j,R/D) = (0.9,0.8)$.}%
\label{fig:5}%
\end{figure}
Broadband spectra are observed at $M_j$ values near 1. A typical spectrum is shown in figure \ref{fig:5}(a). In the absence of tones, the bicoherence is zero for all frequencies considered.
\subsubsection{Linear frequency selection}
\label{Sec:LFS}
\begin{figure}
\centering
\subfigure[PSD]{\includegraphics[clip=true, trim= 0 0 0 0, width=0.47\textwidth]{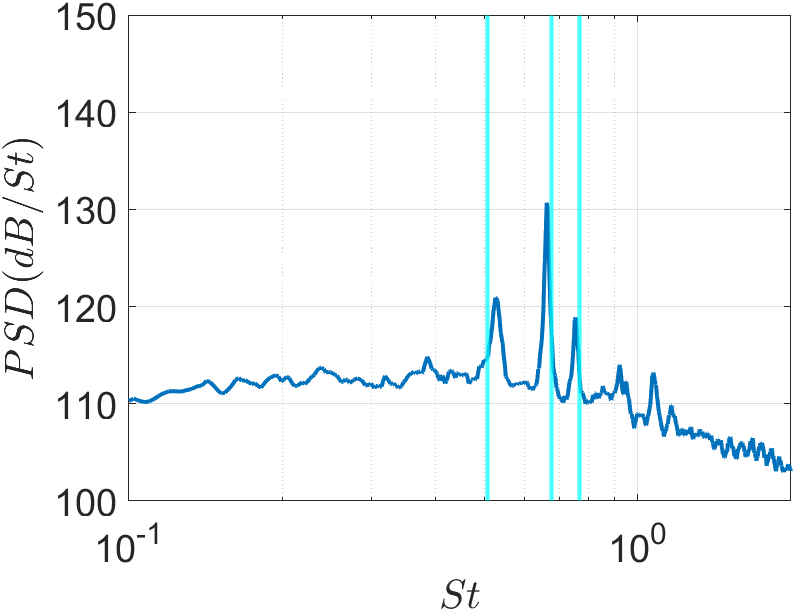}}\subfigure[$b$]{\includegraphics[clip=true, trim= 0 0 0 0, width=0.5\textwidth]{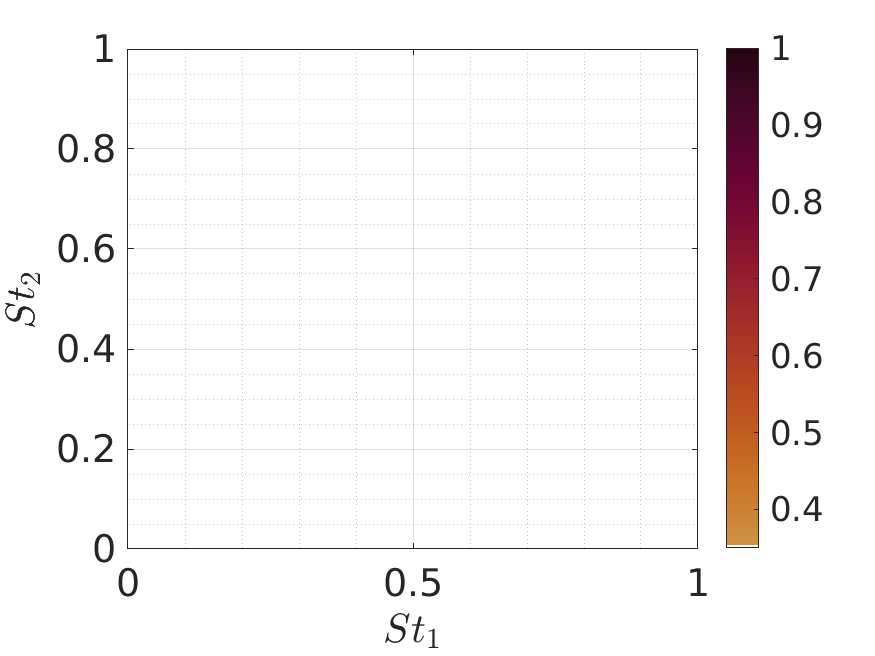}}
\caption{Linear frequency selection, LFS, shown for $(M_j,R/D) = (0.70,0.56)$. Predictions in cyan use $p$ = 2, 3, 4 and $\phi = 0$.}%
\label{fig:6}%
\end{figure}
The linear frequency-selection mechanism (LFS) regime is characterised by multiple spectral peaks at non-harmonic frequencies and with zero bicoherence. Figure \ref{fig:6} shows an example for $(M_j,R/D) = (0.7,0.56)$, where three tones dominate the spectrum. In figure \ref{fig:6}(a) predictions from linear stability theory using equation \ref{eqn:2.5} with $k^{-}_{th}$ as the upstream wave (as $M_j < 0.82$), are indicated by the vertical cyan lines. The agreement between linear model and data, the inharmonic peak distribution (which the linear model shows to be due to the dispersive character of the upstream-travelling wave), and the absence of bicoherence, supports the LFS hypothesis.
\subsubsection{Non-linear frequency selection}
\label{Sec:NLFS}
\begin{figure}
\centering
\subfigure[PSD]{\includegraphics[clip=true, trim= 0 0 0 0, width=0.47\textwidth]{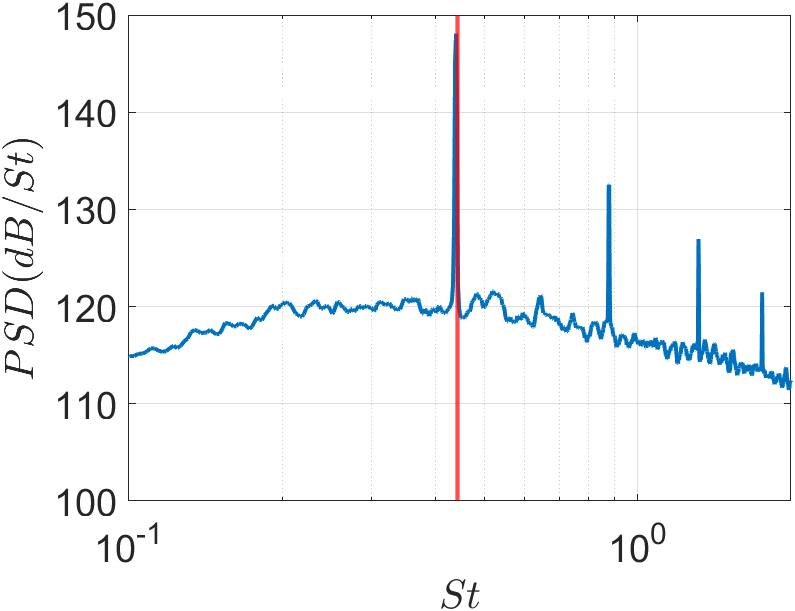}}\subfigure[$b$]{\includegraphics[clip=true, trim= 0 0 0 0, width=0.5\textwidth]{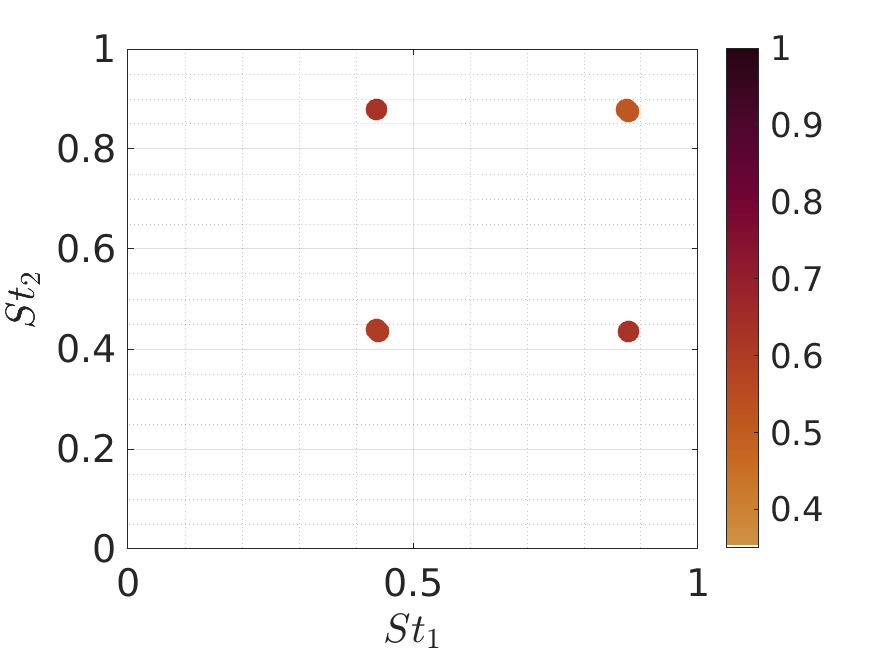}}
\caption{Non-linear frequency selection, NLFS, shown for $(M_j,R/D) = (0.87,0.5)$. Prediction in red uses $p$ = 2 and $\phi = 0$.}%
\label{fig:7}%
\end{figure}
Tones associated with non-linear frequency selection (NLFS) are characterised by a strong fundamental and its harmonics. An example is shown for $(M_j,R/D) = (0.87,0.5)$ in figure \ref{fig:7}(a). Overlaid on the fundamental tone is a prediction using equation \ref{eqn:2.5} for $k_p^-$ waves with $p = 2$, and $\phi = 0$. These values of $p$ and $\phi$ are the same which provide alignment with the LFS peak at $M_j = 0.88$, indicating a continuity when moving from $M_j = 0.88$ to 0.87.
The alignment between experiment and model demonstrates that in this NLFS regime, the fundamental tone corresponds to an LFS tone that has become strongly amplified. A strong-oscillator dynamic results, in which other LFS tones are suppressed and harmonics emerge (through interactions of the same frequency) alongside tones generated via non-harmonic triadic interactions (of two different frequencies). The magnitude of the fundamental tone far exceeds ($>15dB$) those of the LFS tones (\S \ref{Sec:LFS}). Non-linear interaction associated with the fundamental and its harmonics is clearly visible in the bicoherence, figure \ref{fig:7}(b). A strong coupling is observed at $St_1 = St_2 = 0.44$ which produces the first harmonic at $St = 0.88$. An interaction may then be seen between $St_1 = 0.44$ and $St_2 = 0.88$ which produces the second harmonic at $St = 1.32$, and so forth.
\begin{figure}
\centering
\subfigure[PSD]{\includegraphics[clip=true, trim= 0 0 0 0, width=0.47\textwidth]{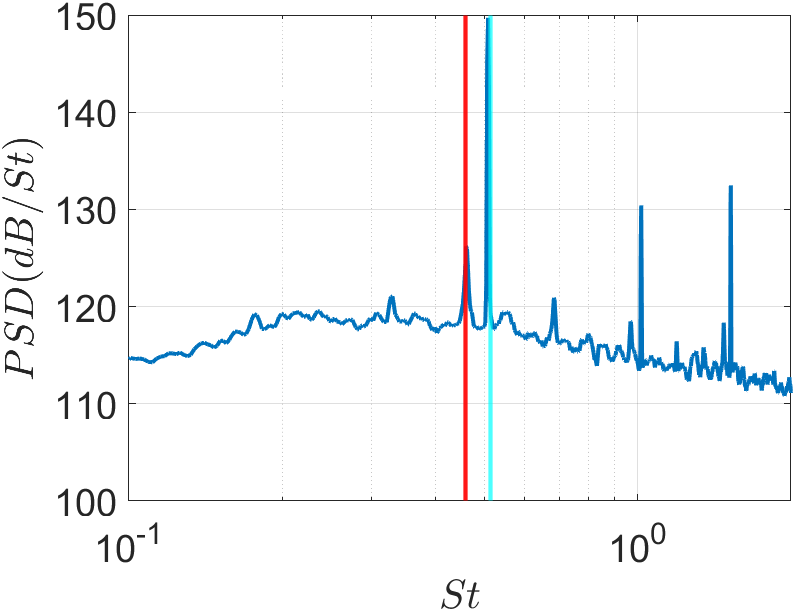}}\subfigure[$b$]{\includegraphics[clip=true, trim= 0 0 0 0, width=0.5\textwidth]{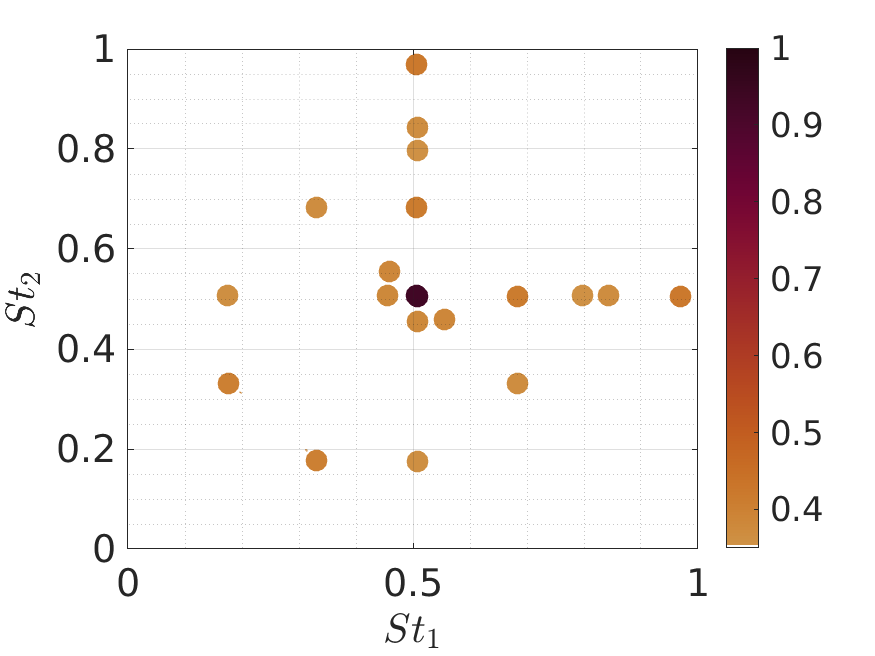}}\\
\subfigure[Triad 1]{\includegraphics[clip=true, trim= 0 0 0 0, width=0.47\textwidth]{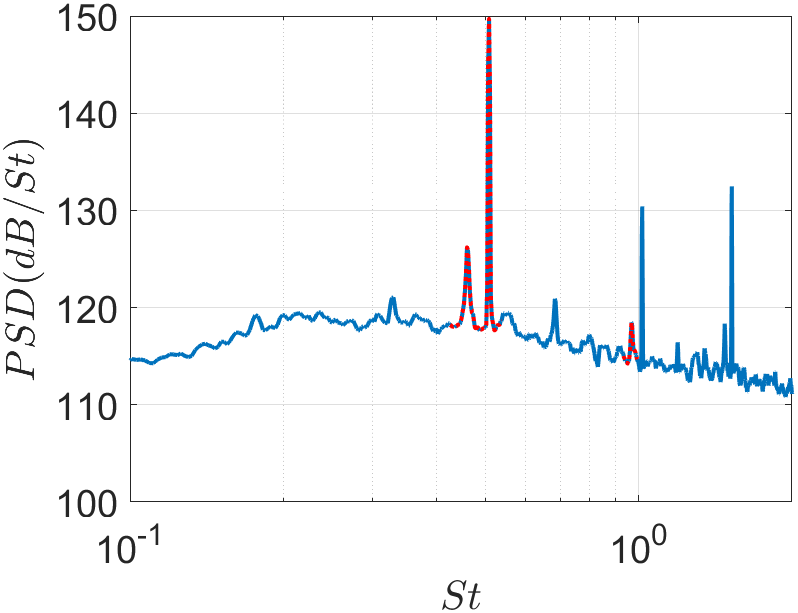}}\subfigure[Triad 2]{\includegraphics[clip=true, trim= 0 0 0 0, width=0.47\textwidth]{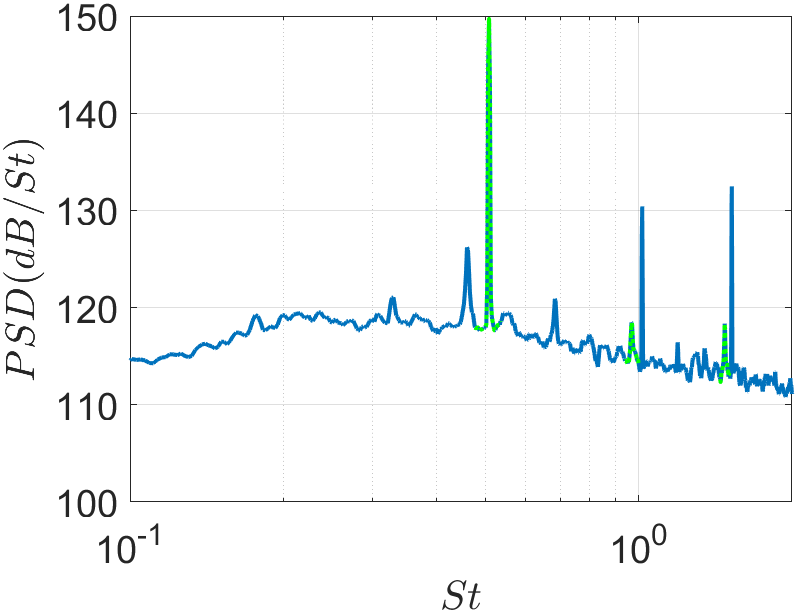}}
\caption{Non-linear frequency selection, NLFS, with multiple triadic interactions shown for $(M_j,R/D) = (0.83,0.5)$. Predictions in cyan use $p$ = 4 and $\phi = 0$, and in red use $p$ = 2 and $\phi = 0$.}%
\label{fig:8}%
\end{figure}

An example of NLFS dynamics with, in addition to the harmonic triadic interactions, multiple non-harmonic triadic interactions is illustrated in figure \ref{fig:8}, for $(M_j,R/D) = (0.83,0.5)$. In figure \ref{fig:8}(a) LFS predictions from equation \ref{eqn:2.5} are overlain with, $k^{-}_{d}$ (cyan), and $k^{-}_p$ (red) as the upstream waves considered. As before, the values of $p$ and $\phi$ used for the linear model predictions continuously align with the peaks across $M_j$ into regions of LFS dynamics. This agreement again demonstrates an LFS mechanism has led to a strong-oscillator dynamic - which dominates the flow. In figure \ref{fig:8}(b) multiple distinct triadic interactions are manifest, with both harmonic and non-harmonic interactions. The non-harmonic triads here have amplitudes multiple orders of magnitude lower than the main oscillator dynamics, and for this reason we refer to them as passive. Two of these non-harmonic triads are highlighted in figure \ref{fig:8}(c) and (d). The first (figure \ref{fig:8}(c)) involves an interaction between the fundamental (associated with $k_d^-$ feedback), a tone generated through $k_p^-$ feedback, and an interaction tone at $St \sim 0.95$. The bicoherence does not indicate the direction of energy transfer for a given triadic interaction; however, as two of the tones involved are shown to arise through LFS mechanisms, the third tone ($St \sim 0.95$) can be understood to arise through this interaction. A second non-harmonic example is provided in figure \ref{fig:8}(d). Here, the tone generated from the triad just discussed interacts with the fundamental to generate a new tone at their sum frequency.
\subsubsection{LFS with weak non-linear interactions}
\label{Sec:LFS+}
\begin{figure}
\centering
\subfigure[PSD]{\includegraphics[clip=true, trim= 0 0 0 0, width=0.47\textwidth]{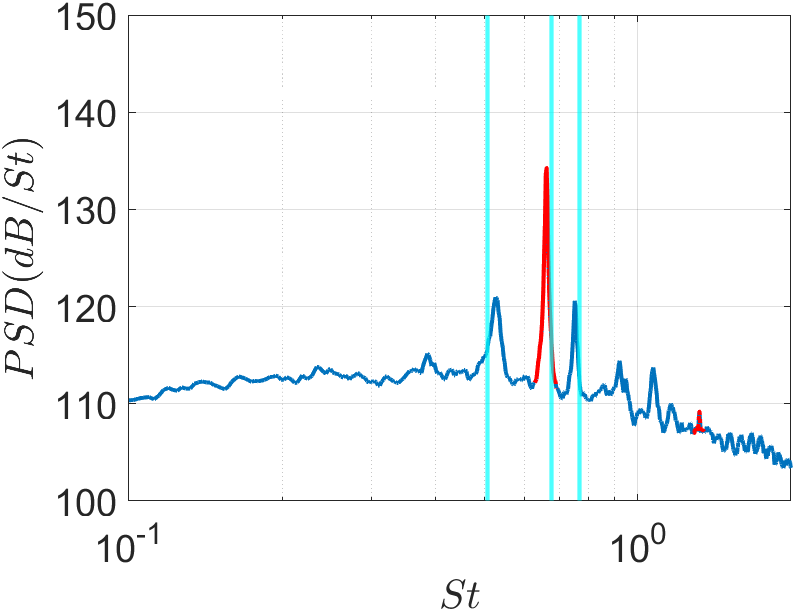}}\subfigure[$b$]{\includegraphics[clip=true, trim= 0 0 0 0, width=0.5\textwidth]{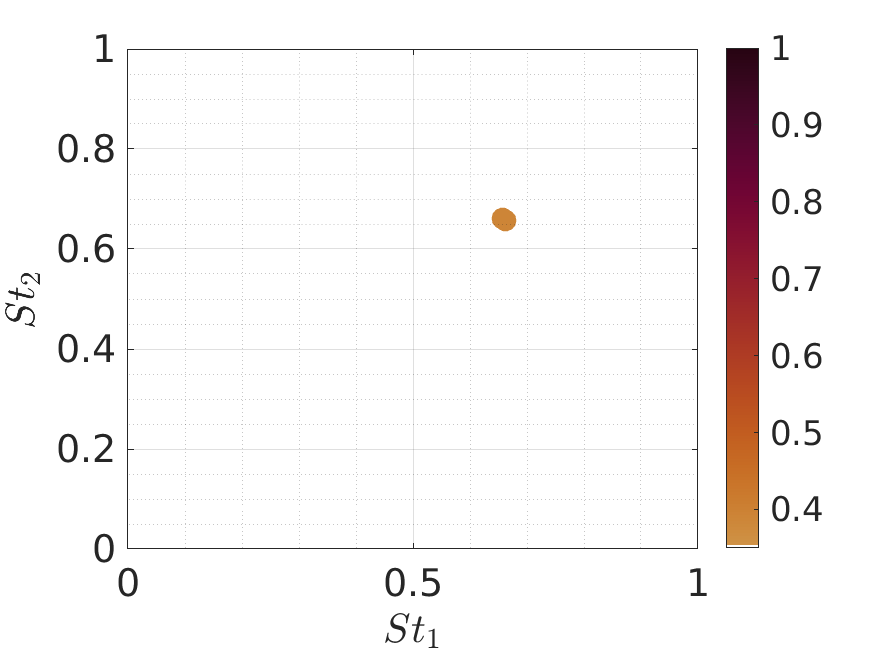}}
\caption{Multiple LFS tones with a singular harmonic (red) shown for $(M_j,R/D) = (0.7,0.55)$. Predictions in cyan use $p$ = 2, 3, 4 and $\phi = 0$.}%
\label{fig:9}%
\end{figure}
We observe a regime in which multiple LFS tones are present, but only one is accompanied by its harmonic, or is involved in a non-harmonic interaction. One such example, for $(M_j,R/D) = (0.7,0.55)$, is shown in figure \ref{fig:9}. Three tones are here associated with LFS (as shown by the linear tone prediction in cyan), but the high-amplitude tone, at $St=0.66$ is shown, by the bicoherence, to generate a harmonic at $St=1.32$ (both highlighted in red). The LFS is underpinned by the $k^{-}_{th}$ upstream wave (as $M_j < 0.82$)
\begin{figure}
\centering
\subfigure[PSD]{\includegraphics[clip=true, trim= 0 0 0 0, width=0.47\textwidth]{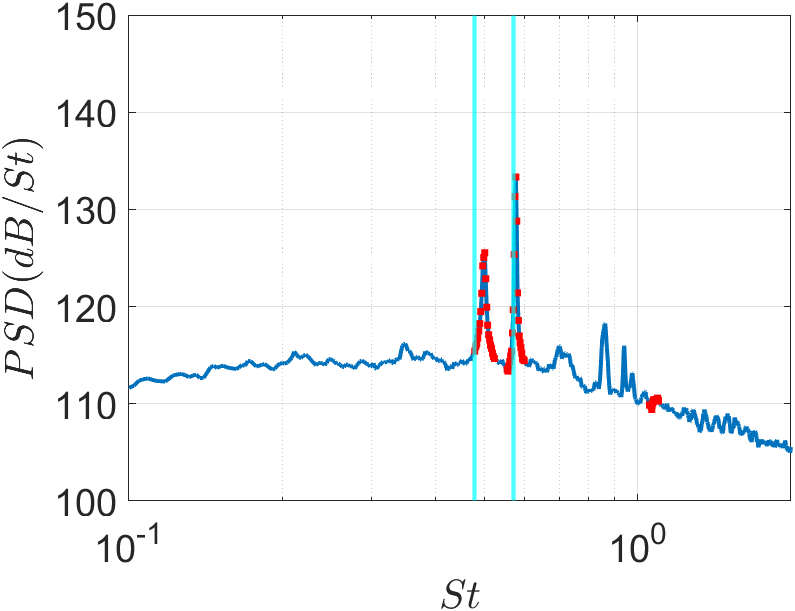}}\subfigure[$b$]{\includegraphics[clip=true, trim= 0 0 0 0, width=0.5\textwidth]{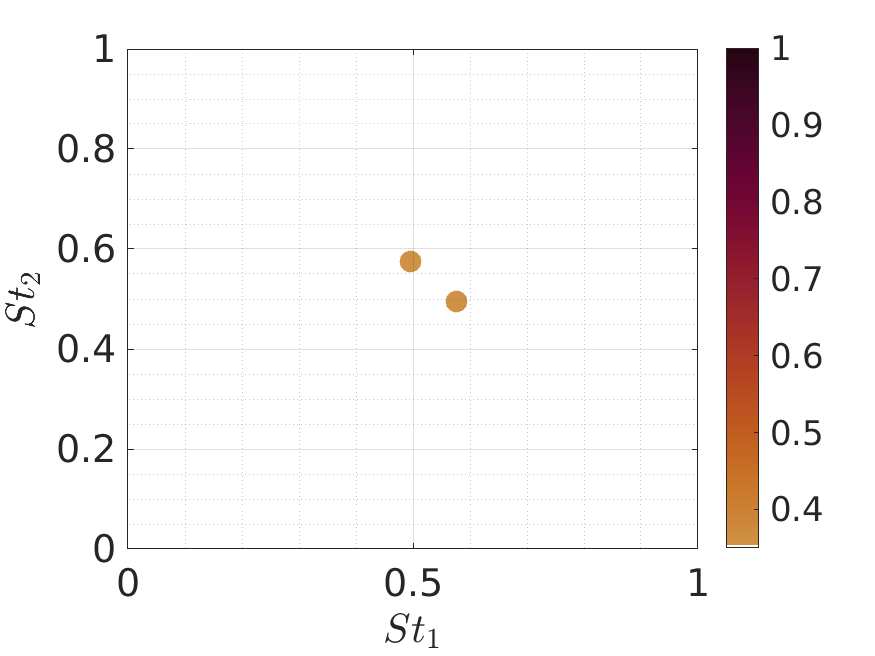}}
\caption{Multiple LFS tones with a singular triad (red) shown for $(M_j,R/D) = (0.78,0.57)$. Predictions use $p$ = 2, 3 and $\phi = 0.2\pi$.}%
\label{fig:10}%
\end{figure}

A further example, where the system exhibits LFS behaviour with the addition of a triadic interaction, is illustrated in figure \ref{fig:10} for $(M_j,R/D) = (0.78,0.57)$. In the PSD, figure \ref{fig:10}(a), predictions from equation \ref{eqn:2.5} align well with the tones of the spectrum. As in \S \ref{Sec:LFS}, this implies tones who arise through a linear frequency-selection mechanism. Predictions here used $k^{-}_{th}$ (cyan) as the upstream wave. The bicoherence in figure \ref{fig:10}(b) suggests a single non-linear interaction. This is indicated in red in figure \ref{fig:10}(a) to highlight the tones involved, in particular as the peak generated via the triadic interaction is of very low amplitude relative to the broadband noise. The interaction is not producing a harmonic ($St_1 \neq St_2$). In contrast, the fundamental tones of the NLFS (figures \ref{fig:7} and \ref{fig:8}) dynamics, were observed to be of greater magnitude in the PSD.
\subsection{Regime classification}
\label{Sec:Class}
\begin{figure} 
    \centering
    \includegraphics[clip=true, trim= 0 0 0 0, width=0.7\textwidth]{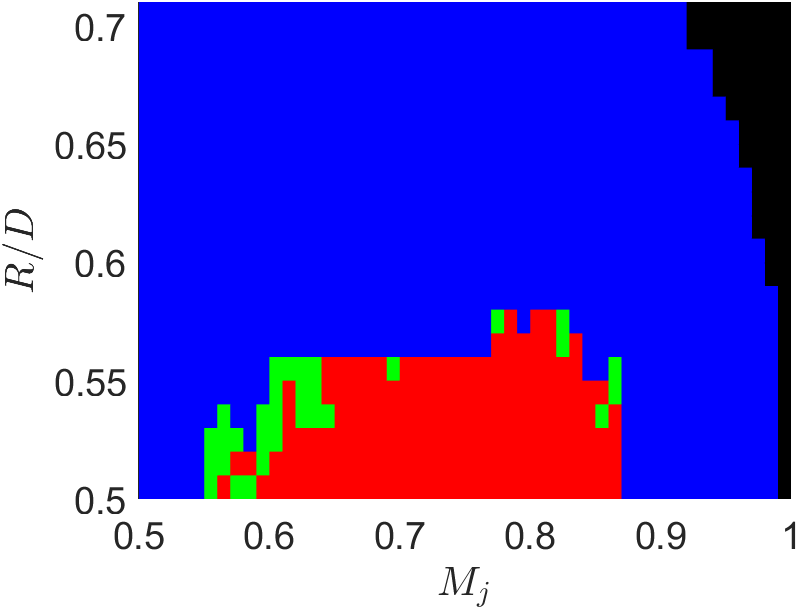}
    \caption{Classification of the jet-edge interaction dynamics across the parameter space considered. These are divided as, broadband (black), LFS (blue), LFS with weak non-linear interactions (green), and NLFS (red).}%
    \label{fig:11}%
\end{figure}
\begin{figure} 
    \centering
    \includegraphics[clip=true, trim= 0 0 0 0, width=0.7\textwidth]{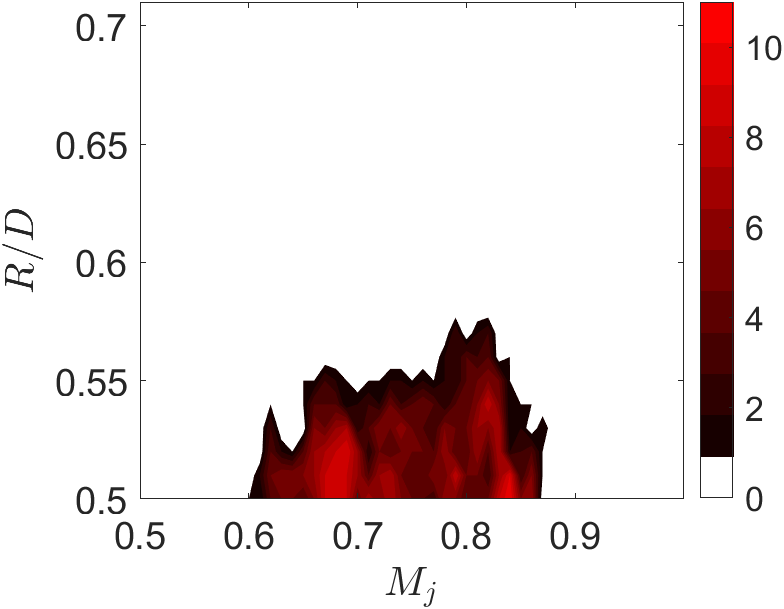}
    \caption{Classification of the number of non-harmonic triadic interactions occurring across the parameter space considered. These are divided as no triads (white), a single triad (black), through to the most triads observed (11,red).}%
    \label{fig:12}%
\end{figure}
The above classification is used to identify regions of the parameter space associated with the different dynamics. This is shown in figure \ref{fig:11}. Several trends are observed. At $M_j > 0.99$ for $R/D = 0.5$ the dynamics are broadband, with the size of the broadband region increasing with increasing $R/D$. As $M_j$ is decreased, LFS is observed. For $R/D\geq0.58$ once the flow transitions from broadband to LFS, tonal dynamics remain driven by LFS dynamics as $M_j$ is decreased further. And for $M_j<0.55$ resonance is again LFS regardless of the radial position of the edge. Non-linear dynamics involving LFS and harmonics and/or non-harmonic triads are observed in the Mach number range $0.56 \leq M_j \leq 0.87$, and for $R/D \leq 0.57$. Moving from high to low $M_j$ the system passes directly from LFS to NLFS dynamics. From \S \ref{Sec:LFS} and \ref{Sec:NLFS}, the fundamental amplitude may increase by $>15dB$ when comparing regions of LFS to NLFS dynamics. As such, figure \ref{fig:11} indicates a sharp change in the fundamental tone amplitude across the LFS-NLFS boundary as $M_j$ is decreased. Conversely, at the low $M_j$ limit of the NLFS region, the system does not transition back directly from NLFS to LFS so abruptly. Instead there is an extended region of LFS with secondary, weak, triadic interactions (green in figure \ref{fig:11}). Within the NLFS region the number of non-harmonic triads present varies and is presented in figure \ref{fig:12}. The largest number of non-harmonic triadic interactions are found localised near $M_j = 0.84$ and $M_j = 0.69$ for $R/D = 0.5$, and is observed to be repeatable (see Appendix \ref{App:B}).
\section{Regime switching}
\label{Sec:Change}
\begin{figure}
\centering
\subfigure[PSD]{\includegraphics[clip=true, trim= 0 0 0 0, width=0.47\textwidth]{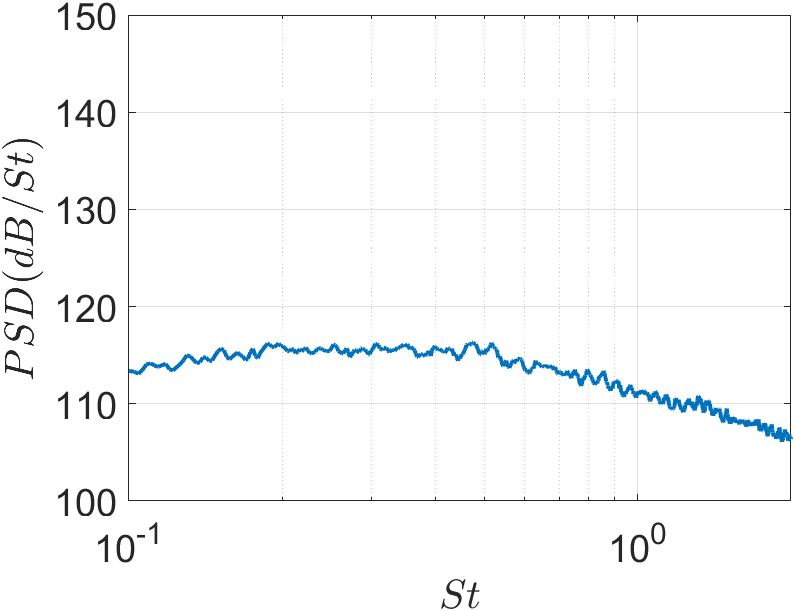}}\subfigure[$b$]{\includegraphics[clip=true, trim= 0 0 0 0, width=0.5\textwidth]{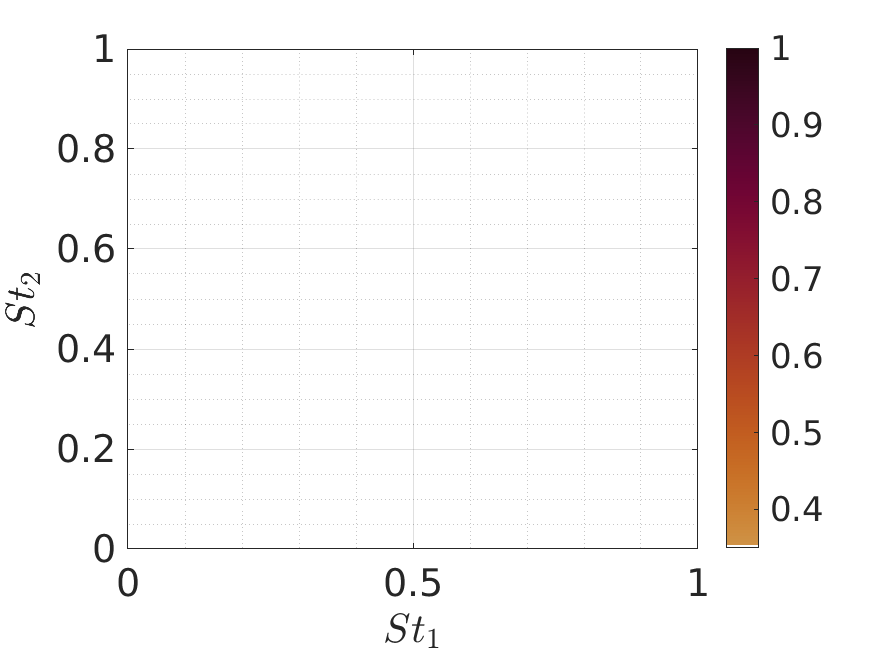}}\\
\subfigure[PSD]{\includegraphics[clip=true, trim= 0 0 0 0, width=0.47\textwidth]{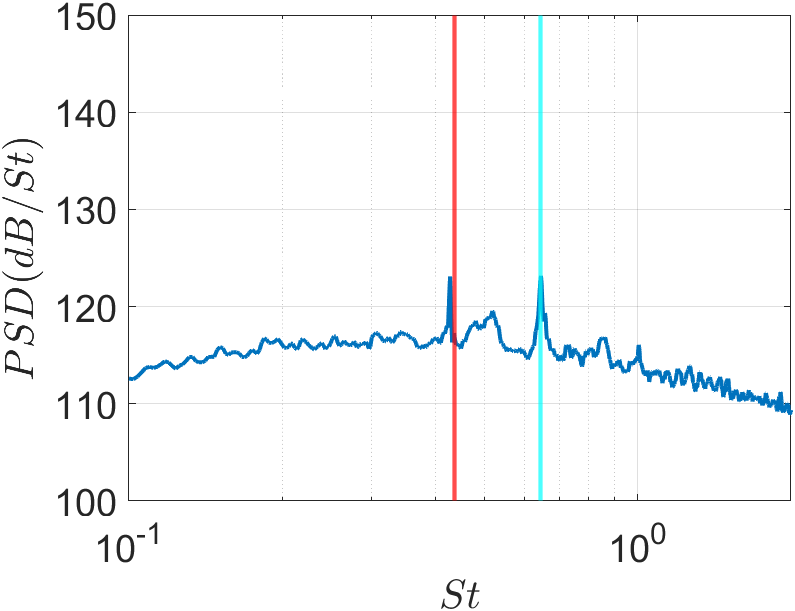}}\subfigure[$b$]{\includegraphics[clip=true, trim= 0 0 0 0, width=0.5\textwidth]{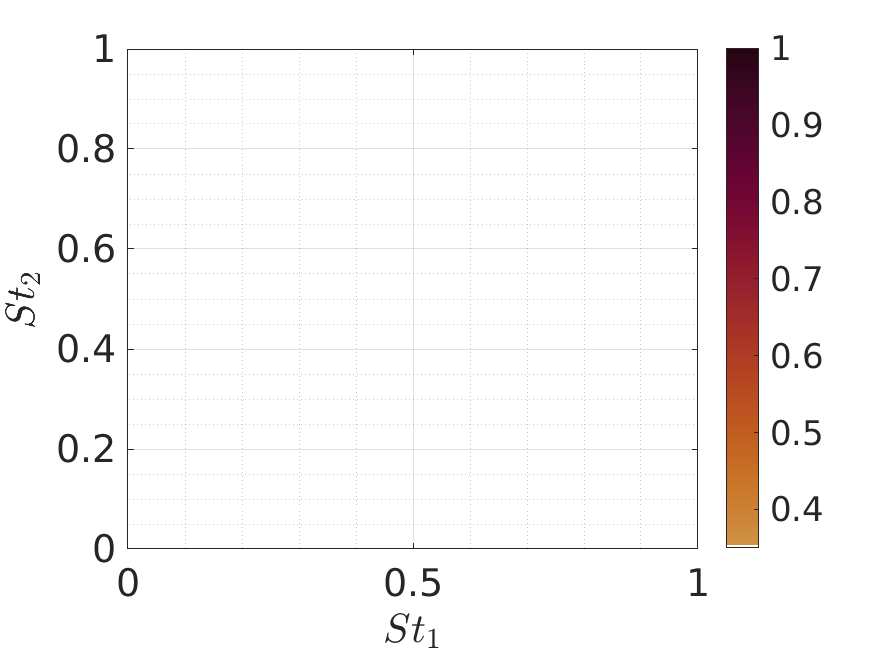}}\\
\caption{System transition between broadband noise for $(M_j,R/D) = (0.89,0.8)$ (a),(b) and LFS for $(M_j,R/D) = (0.89,0.5)$ (c),(d). Frequency prediction overlaid in (c) uses $\phi = 0$, and $p = 2$ (red, $k_p^-$) 9 (cyan, $k_{d}^-$).}%
\label{fig:13}%
\end{figure}
We here look more closely at the different regime-change characteristics. Figure \ref{fig:13} illustrates the switch from broadband to LFS, via comparison of spectral signatures at $(M_j,R/D) = (0.89,0.8)$, and $(M_j,R/D) = (0.89,0.5)$. From the broadband spectrum two LFS tones emerge. The lower-frequency tone corresponds to feedback closure by the $k^{-}_p$ (red) upstream wave, and the higher-frequency tone by the $k_d^-$ (cyan) upstream wave. Bicoherence plots, figure \ref{fig:13}(b) and (d), are zero for both broadband and LFS dynamics. Recall from \S \ref{sec:Bic} that the computation of $b$ uses the full $30s$ acquisition length divided into 16384 point blocks, and so we are confident when stating that no values of $b$ above the threshold to be considered non-zero ($b>0.35$) were found.

\begin{figure}
\centering
\subfigure[PSD]{\includegraphics[clip=true, trim= 0 0 0 0, width=0.47\textwidth]{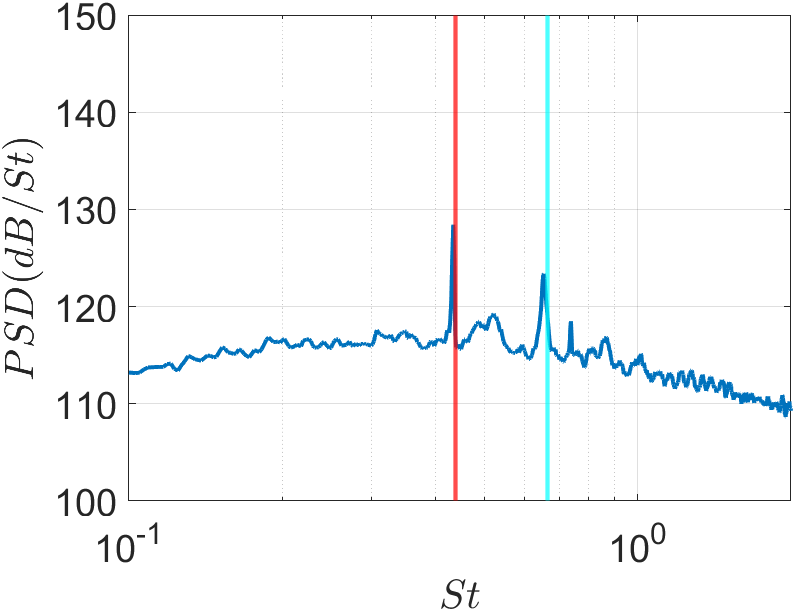}}\subfigure[$b$]{\includegraphics[clip=true, trim= 0 0 0 0, width=0.5\textwidth]{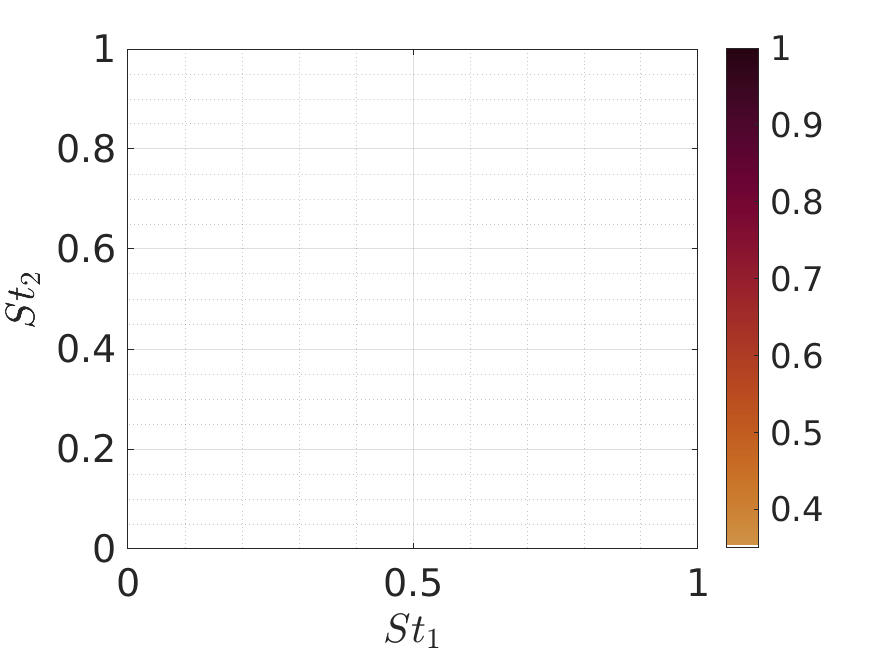}}\\
\subfigure[PSD]{\includegraphics[clip=true, trim= 0 0 0 0, width=0.47\textwidth]{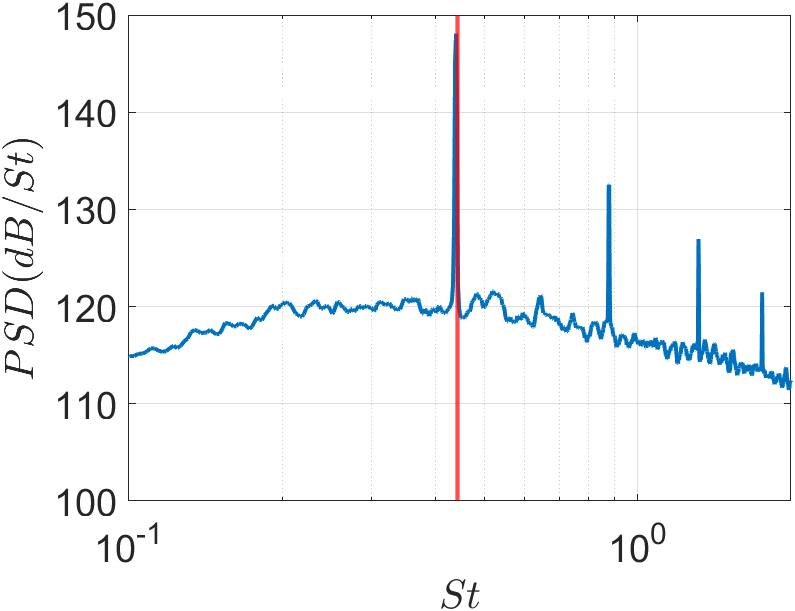}}\subfigure[$b$]{\includegraphics[clip=true, trim= 0 0 0 0, width=0.5\textwidth]{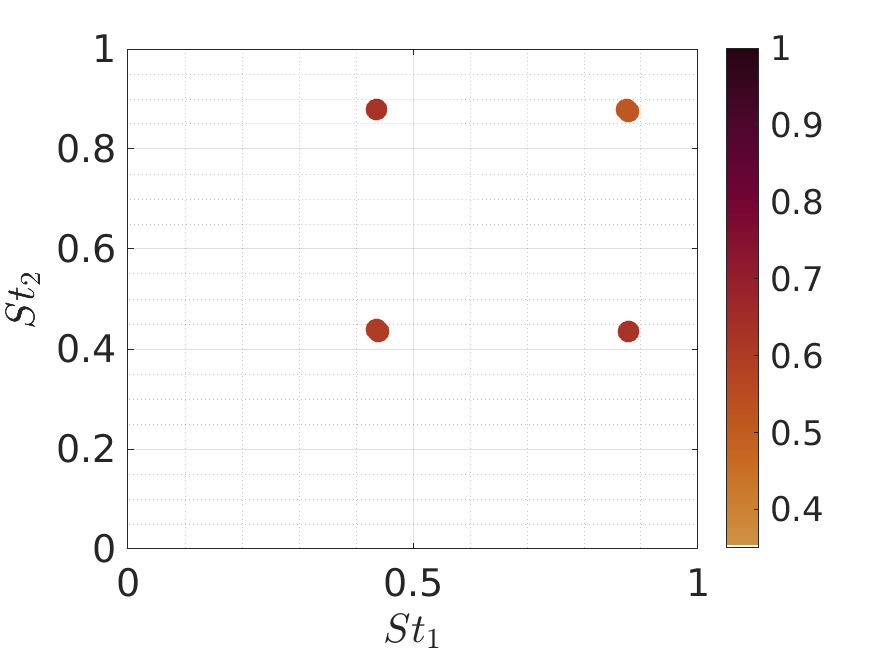}}\\
\caption{System transition between LFS for $(M_j,R/D) = (0.88,0.5)$ (a),(b) and NLFS for $(M_j,R/D) = (0.87,0.5)$ (c),(d). Frequency prediction overlaid in (a),(c) use $p = 2$ and $\phi = 0$ (red, $k_p^-$), and $p = 9$ and $\phi = 0$ (cyan, $k_d^-$).}%
\label{fig:14}%
\end{figure}

The sharp transition from LFS to NLFS is shown in figure \ref{fig:14} (from LFS at $(M_j,R/D) = (0.88,0.5)$, to NLFS at $(M_j,R/D) = (0.87,0.5)$). The bicoherence changes from zero across all $St$, figure \ref{fig:14}(b), to harmonic interactions ($St_1 = St_2$), figure \ref{fig:14}(d), corresponding to the harmonic series observed in the PSD, figure \ref{fig:14}(c). Both PSD plots, figure \ref{fig:14}(a) and (c), contain a prediction line using equation \ref{eqn:2.5}, with feedback closure by the $k^{-}_p$ (red) upstream wave; note that the LFS tone closed by the $k_d^-$ wave at $M_j = 0.88$ has here been suppressed, the tone near $St \sim 0.73$ relates to $m = 1$ resonance with non-axisymmetric resonance the focus of a future work. At $M_j = 0.88$ the model agreement and zero bicoherence confirms the LFS classification. For $M_j = 0.87$ the amplitude of the LFS fundamental has increased by two orders of magnitude, dominating the dynamics with a strong-oscillator behaviour and generating a series of harmonics. This dramatic change, occurring with a change of only 0.01 in $M_j$, is repeatable, occurs whether data is acquired with increasing or decreasing Mach number, and does not exhibit hysteresis (see Appendix \ref{App:B}).

\begin{figure}
\centering
\subfigure[PSD]{\includegraphics[clip=true, trim= 0 0 0 0, width=0.47\textwidth]{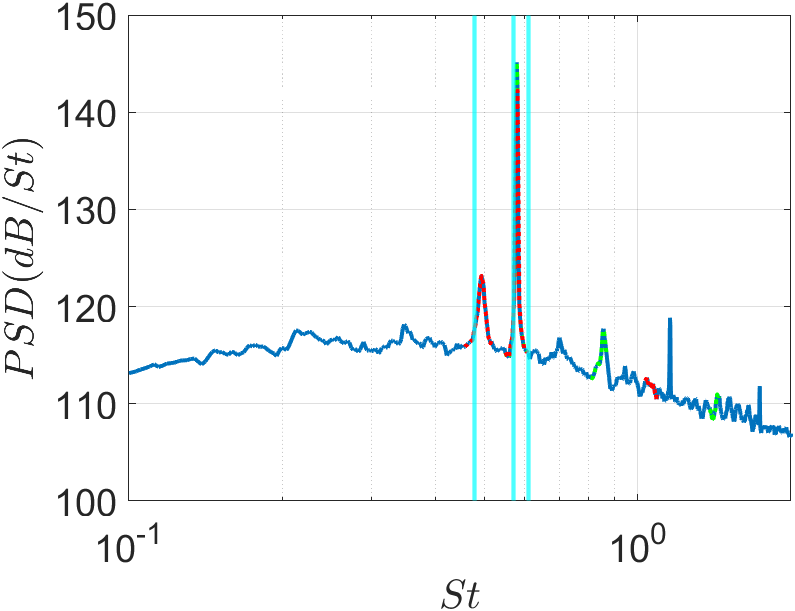}}\subfigure[$b$]{\includegraphics[clip=true, trim= 0 0 0 0, width=0.5\textwidth]{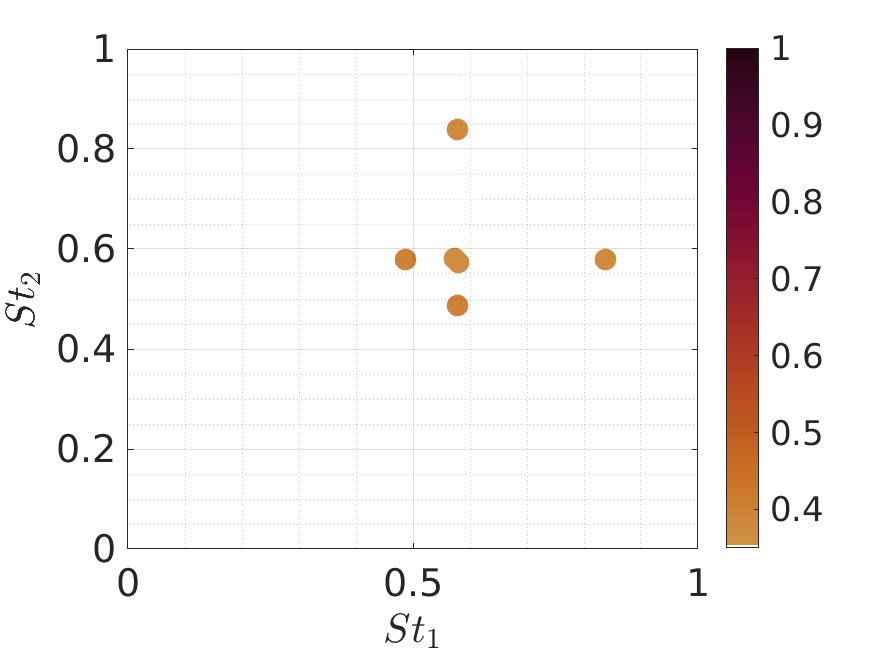}}\\
\subfigure[PSD]{\includegraphics[clip=true, trim= 0 0 0 0, width=0.47\textwidth]{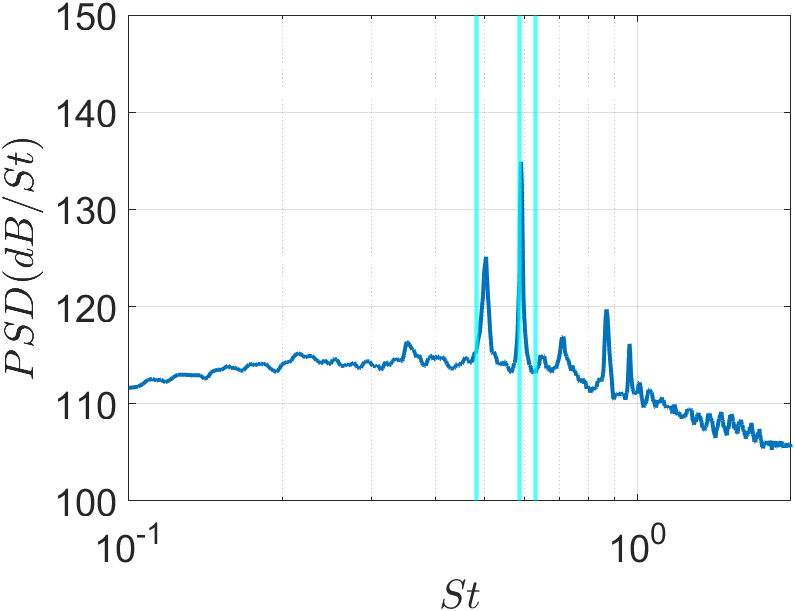}}\subfigure[$b$]{\includegraphics[clip=true, trim= 0 0 0 0, width=0.5\textwidth]{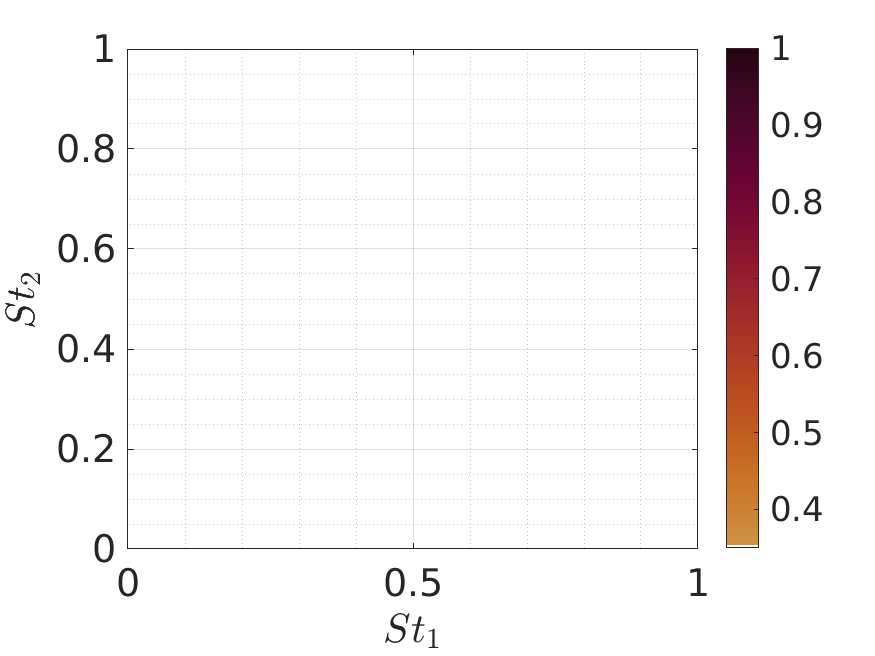}}\\
\caption{System transition between NLFS for $(M_j,R/D) = (0.78,0.56)$ (a),(b) and LFS for $(M_j,R/D) = (0.77,0.56)$ (c),(d). Frequency prediction overlaid in (a),(c) use $p = 2,3,4$ and $\phi = 0$.}%
\label{fig:15}%
\end{figure}

A final transition example is from NLFS with non-harmonic triads, to LFS. This is shown in figure \ref{fig:15} for $(M_j,R/D) = (0.78,0.56)$, NLFS, and $(M_j,R/D) = (0.77,0.56)$, LFS. The PSD plots, figure \ref{fig:15}(a) and (c), are overlain with predictions using equation \ref{eqn:2.5} for $k^{-}_{th}$ (cyan) as the upstream wave. These predictions highlight again that the fundamental tone in NLFS dynamics arises originally through an LFS mechanism, with an identical prediction line linking the fundamental tone at $M_j = 0.78$ with the second linear mode at $M_j = 0.77$. For the NLFS case the bicoherence, figure \ref{fig:15}(b), reveals three non-linear interactions (one harmonic and two non-harmonic), with the triadic interactions coloured on figure \ref{fig:15}(a). The triad highlighted in red is an interaction between two tones arising through LFS mechanisms (with one now being the fundamental), imparting energy to a third tone at their sum frequency. The green triad involves the fundamental tone and an $m = 1$ tone at $St \sim 0.8$ (this tone will be demonstrated to be $m = 1$ in a future work) imparting energy to an $m = 1$ mode at their sum frequency. This transition, from NLFS to LFS, can be compared with the more dramatic transition, across the frontier at $M_j = 0.87$, from LFS to NLFS. At the lower Mach number limit the change is less drastic. The amplitude reduction of the fundamental tone is one order of magnitude (as compared with two orders of magnitude for the high-Mach-number frontier). As discussed earlier, regarding figure \ref{fig:11} in \S \ref{Sec:Class}, the regime switch at the low Mach-number frontier is less abrupt than the high-Mach-number transition. The transitions discussed here occur whether $M_j$ is increased or decreased, and has been observed to be both robust and repeatable (see Appendix \ref{App:B}).
\subsection{Competing feedback loops}
\label{Sec:Comp}
\begin{figure}
\centering
\subfigure[PSD]{\includegraphics[clip=true, trim= 0 0 0 0, width=0.3\textwidth]{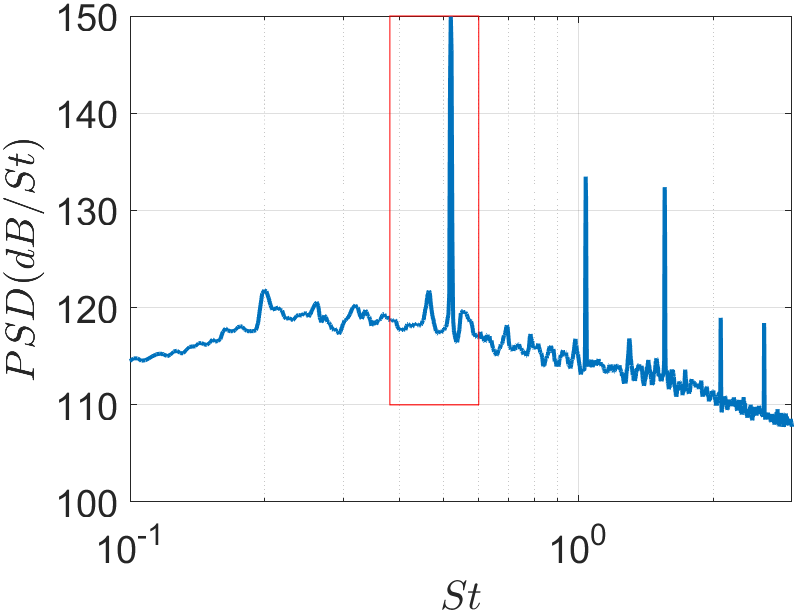}}\subfigure[Zoomed region]{\includegraphics[clip=true, trim= 0 0 0 0, width=0.3\textwidth]{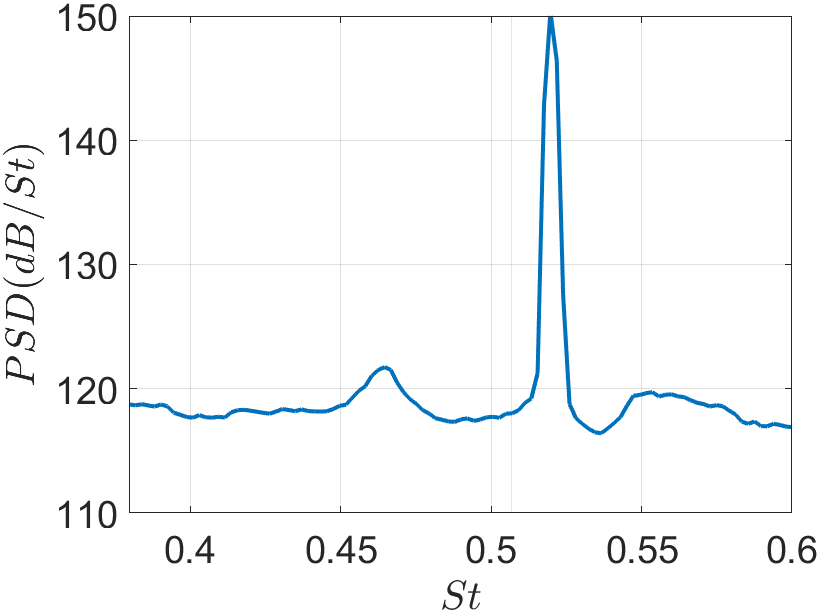}}\subfigure[Dispersion relation curves]{\includegraphics[clip=true, trim= 0 0 0 0, width=0.3\textwidth]{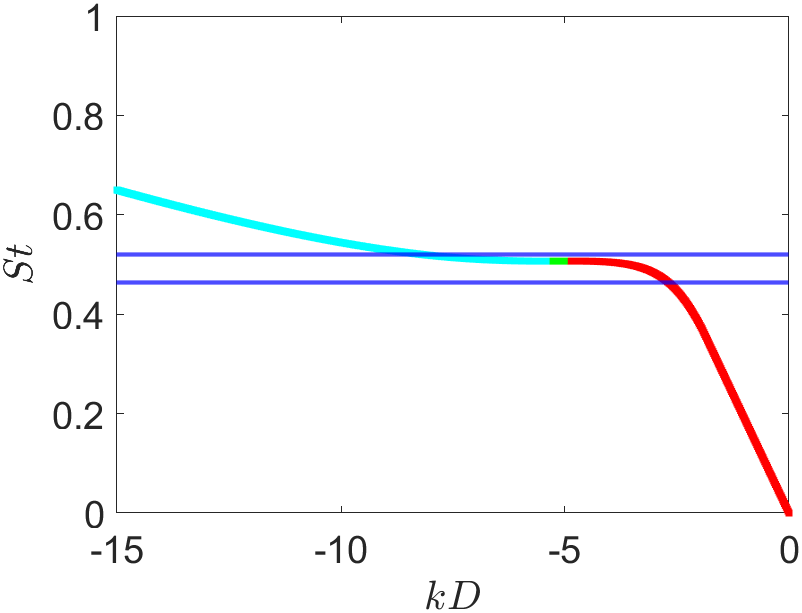}}\\
\subfigure[]{\includegraphics[clip=true, trim= 0 0 0 0, width=0.3\textwidth]{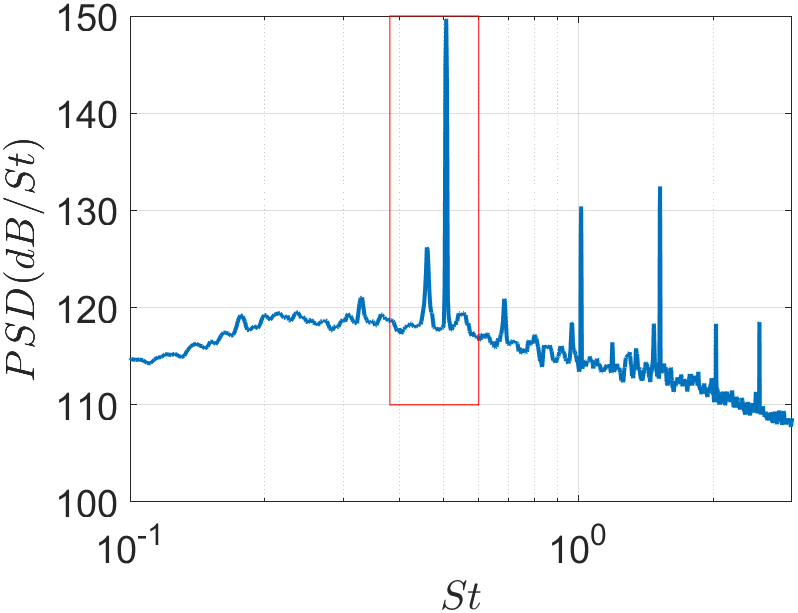}}\subfigure[]{\includegraphics[clip=true, trim= 0 0 0 0, width=0.3\textwidth]{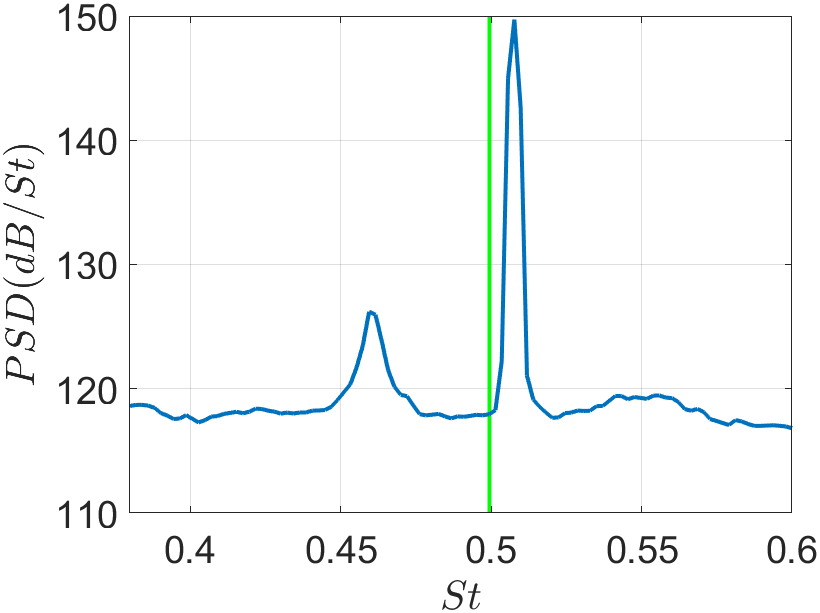}}\subfigure[]{\includegraphics[clip=true, trim= 0 0 0 0, width=0.3\textwidth]{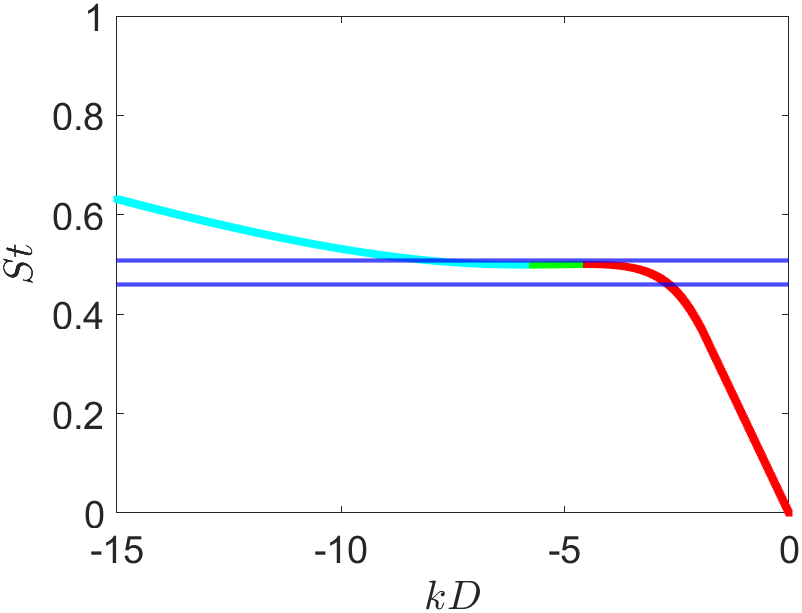}}\\
\subfigure[]{\includegraphics[clip=true, trim= 0 0 0 0, width=0.3\textwidth]{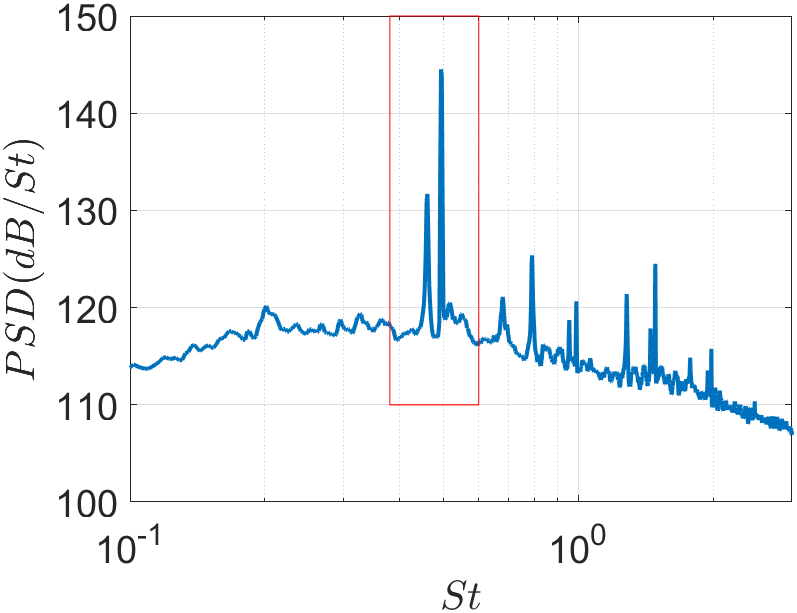}}\subfigure[]{\includegraphics[clip=true, trim= 0 0 0 0, width=0.3\textwidth]{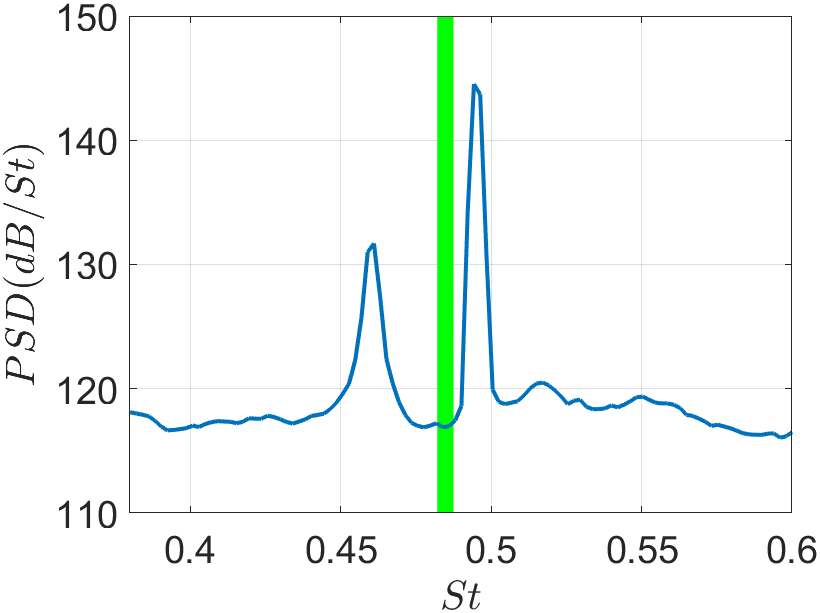}}\subfigure[]{\includegraphics[clip=true, trim= 0 0 0 0, width=0.3\textwidth]{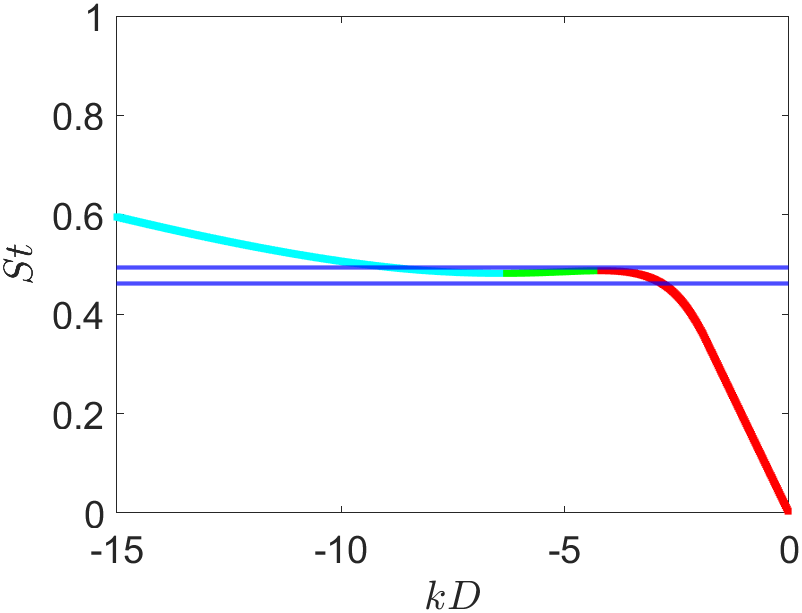}}\\
\subfigure[]{\includegraphics[clip=true, trim= 0 0 0 0, width=0.3\textwidth]{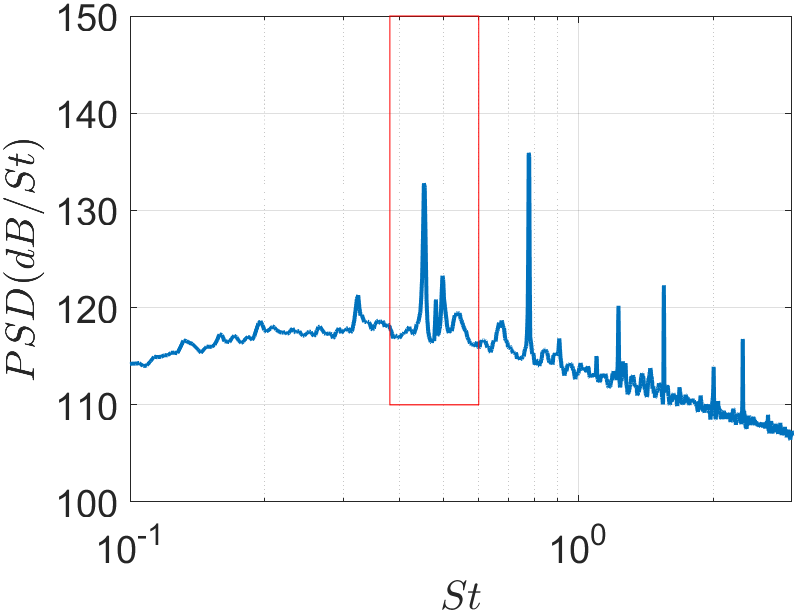}}\subfigure[]{\includegraphics[clip=true, trim= 0 0 0 0, width=0.3\textwidth]{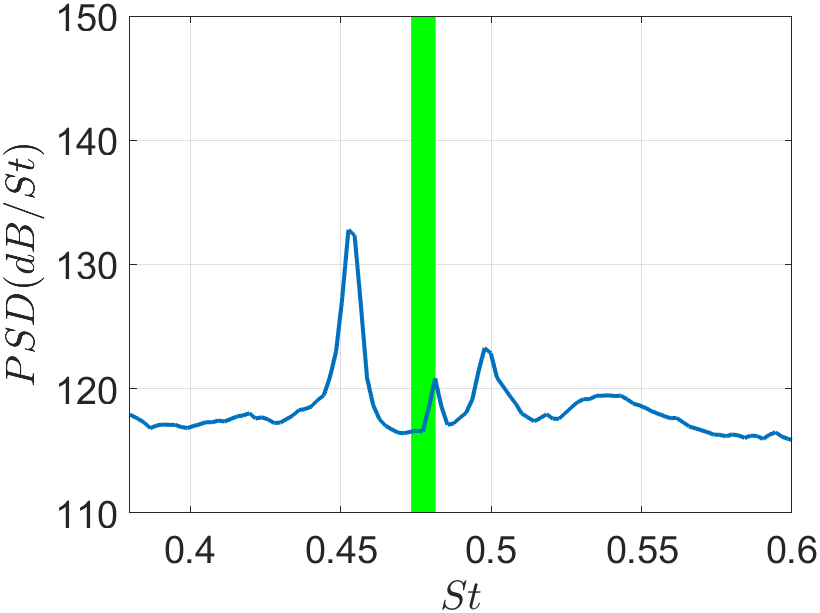}}\subfigure[]{\includegraphics[clip=true, trim= 0 0 0 0, width=0.3\textwidth]{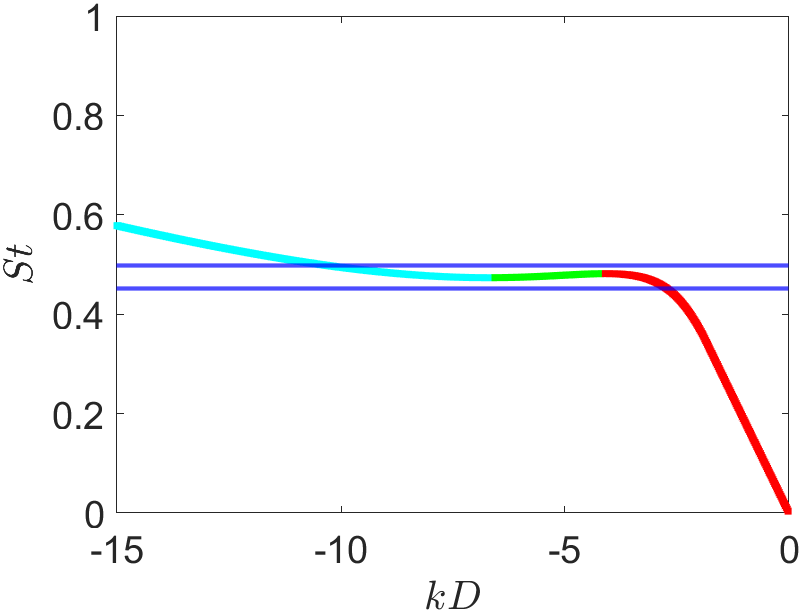}}\\
\subfigure[]{\includegraphics[clip=true, trim= 0 0 0 0, width=0.3\textwidth]{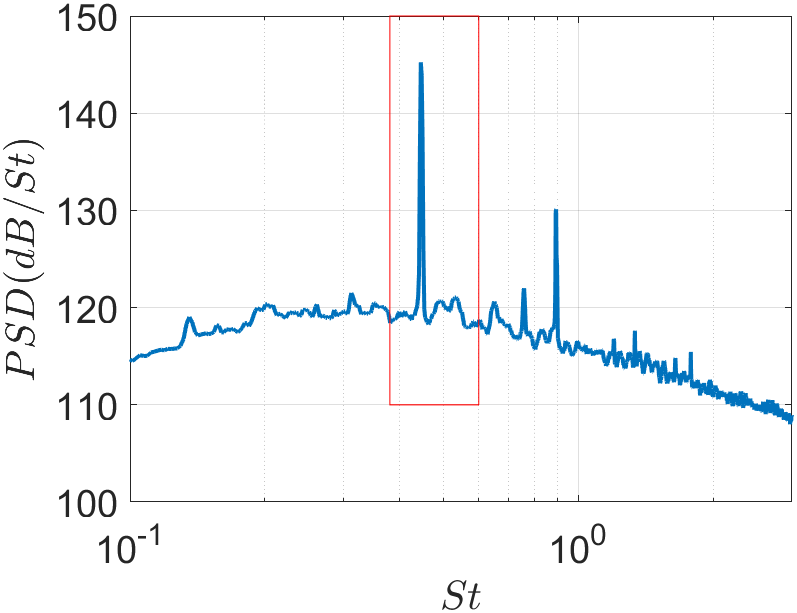}}\subfigure[]{\includegraphics[clip=true, trim= 0 0 0 0, width=0.3\textwidth]{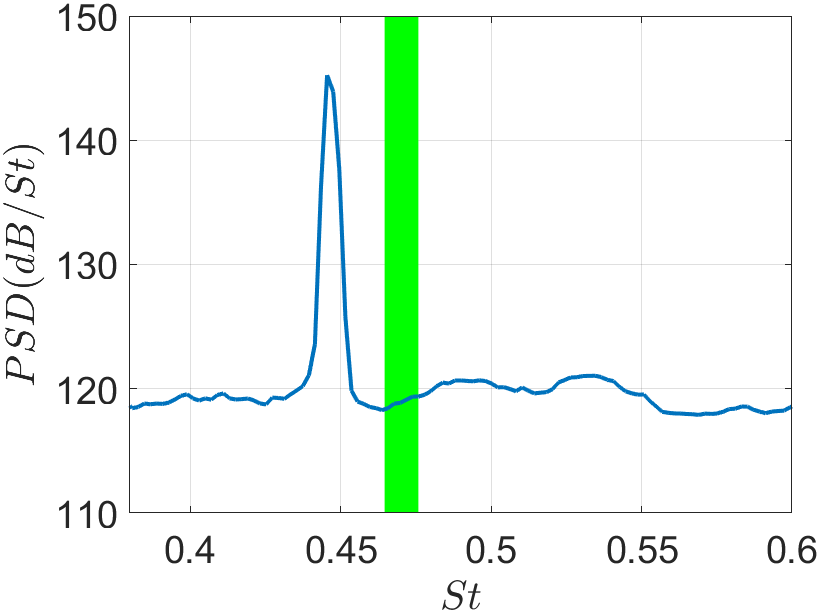}}\subfigure[]{\includegraphics[clip=true, trim= 0 0 0 0, width=0.3\textwidth]{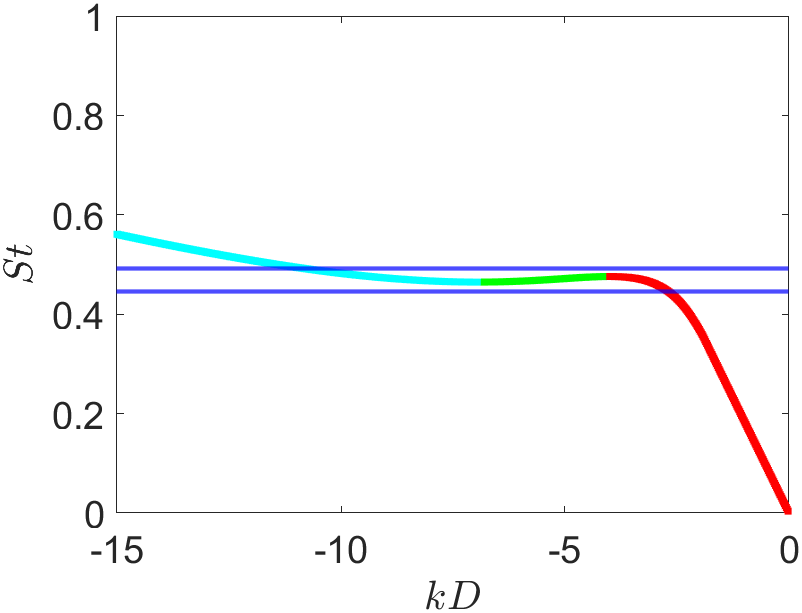}}
\caption{The PSD spectra (a), (d), (g), (j), (m), zoomed view of the red region in the PSD (b), (e), (h), (k), (n) with the region between the S1 and S2 saddle points shaded green, and the dispersion relation highlighting the $k^{-}_{d}$ (cyan), $k^{-}_{p}$ (red), and $k^{+}_{T}$ (green) waves with the frequencies of the two tones in the zoomed region overlain (c), (f), (i), (l), (o). These are plotted for $M_j = 0.82$ (a)-(c), $M_j = 0.83$ (d)-(f), $M_j = 0.84$ (g)-(i), $M_j = 0.85$ (j)-(l), and $M_j = 0.86$ (m)-(o).}%
\label{fig:16}%
\end{figure}
As seen above, for NLFS cases, the fundamental tone emerges via an LFS mechanism. There are cases where competition between two LFS tones is observed, and indeed in one particular case, a strong oscillator dynamics associated with a saturated LFS and its harmonics can transition to a new limit cycle due to the emergence of a new LFS tone. This occurs in the Mach number range, $0.82 < M_j <0.86$ and is illustrated in  figure \ref{fig:16}. Three plots are shown for each $M_j$. The first shows the PSD and has the region containing the two competing tones highlighted in red. This highlighted region is shown in the second plot with a region shaded green indicating the frequency range bounded by the $S1$ and $S2$ saddle points (cf. figure \ref{fig:4}). The third plot shows the vortex-sheet dispersion relation for that $M_j$ highlighting the $k^{-}_{d}$ (cyan), $k^{-}_{p}$ (red), and $k^{+}_{T}$ (green) waves, with the frequencies of the two competing tones overlaid (blue). For $M_j = 0.82$, figure \ref{fig:16}(b), the higher-frequency tone is dominant. As the Mach number is increased from 0.82 to 0.86 a switch is observed: the low-frequency tone comes to dominate the dynamics and the high frequency tone is suppressed. Over a range of only 0.04 in $M_j$, a strong-oscillator behaviour with NLFS harmonics changes to a new oscillator dynamics, and this switch is underpinned by a linear frequency-selection mechanism: this switch is robust and free from hysteresis (it is repeatable for increasing and decreasing Mach-number acquisitions - see Appendix \ref{App:B}).

Note that the two tones lies on either side of the saddle-point-bounded frequency range, indicated by the green shaded region. As discussed in \S \ref{sec:Model}, at frequencies below $S1$ only the $k_p^-$ wave is propagative, while above the $S2$ saddle point only the $k_d^-$ wave is propagative. Different feedback closure mechanisms are thus available for frequencies above and below the $S1-S2$ frequency range. The lower-frequency tone is due to resonance underpinned by the $k^{-}_{p}$ wave. The higher-frequency tone is associated with a feedback loop underpinned by the $k^{-}_{d}$ wave. This may be further illustrated in figure \ref{fig:16}(c), (f), (i), (l), and (o), where the tonal frequencies are overlaid on the dispersion relation. Each tonal frequency intersects the curve only once. The higher-frequency tone intersects the $k^{-}_{d}$ portion of the curve (cyan), whilst the lower-frequency tone intersects the $k^{-}_{p}$ portion (red), again suggesting a competition between two different feedback loops. 
\begin{figure}
\centering
\subfigure[$M_j$ 0.8]{\includegraphics[clip=true, trim= 0 0 0 0, width=0.33\textwidth]{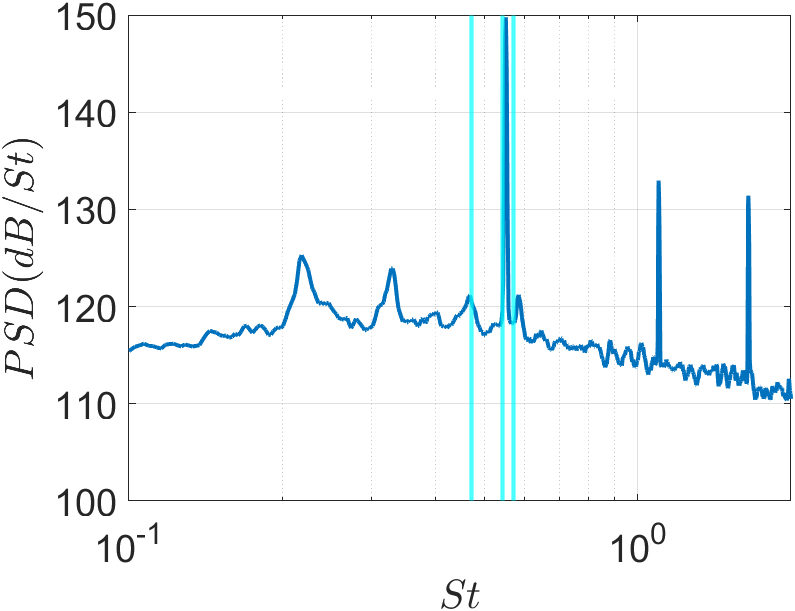}}\subfigure[$M_j$ 0.81]{\includegraphics[clip=true, trim= 0 0 0 0, width=0.33\textwidth]{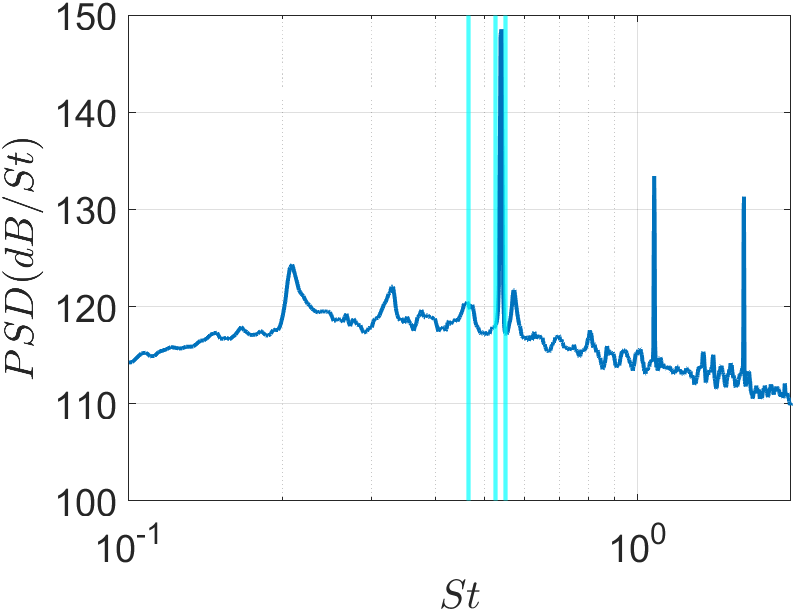}}\subfigure[$M_j$ 0.82]{\includegraphics[clip=true, trim= 0 0 0 0, width=0.33\textwidth]{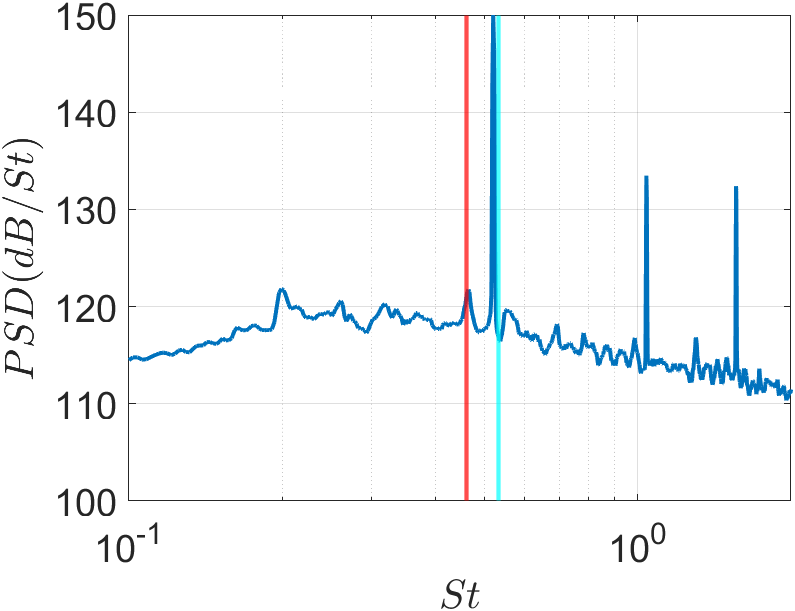}}
\caption{PSD spectra with predictions from linear stability theory overlain, (a) $M_j = 0.8$, (b) $M_j = 0.81$, and (c) $M_j = 0.82$. Lines in cyan and red utilise $k^{-}_{d}$ and $k^{-}_{p}$ waves respectively. Frequency predictions overlaid in (a),(b) uses $p = 2,3,4$ and $\phi = 0$, and (c) uses $p = 4$ and $\phi = 0$ (cyan) and $p = 2$ and $\phi = 0$ (red).}%
\label{fig:17}%
\end{figure}
\begin{figure}
\centering
\subfigure[]{\includegraphics[clip=true, trim= 0 0 0 0, width=0.25\textwidth]{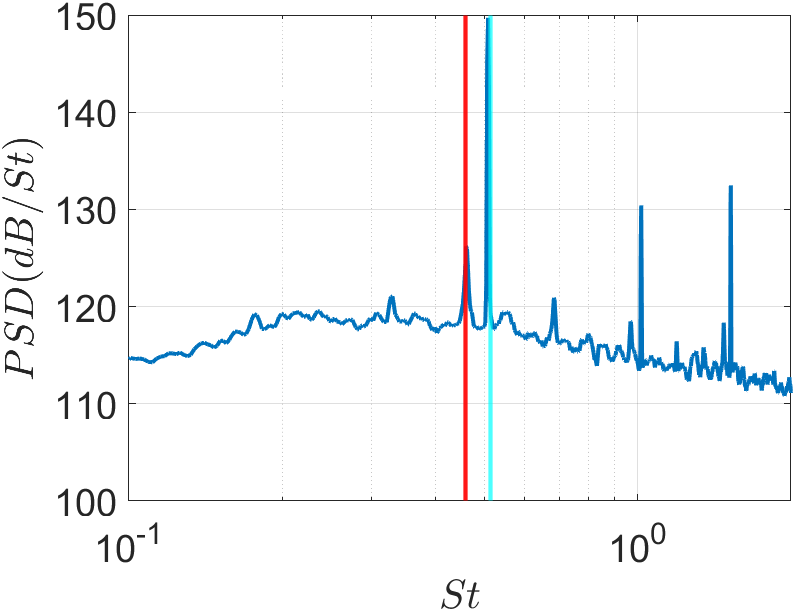}}\subfigure[]{\includegraphics[clip=true, trim= 0 0 0 0, width=0.25\textwidth]{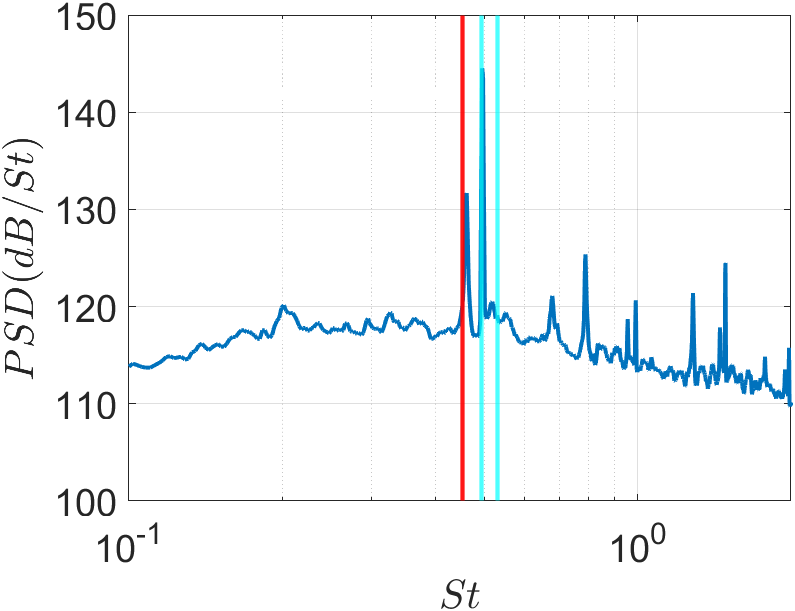}}\subfigure[]{\includegraphics[clip=true, trim= 0 0 0 0, width=0.25\textwidth]{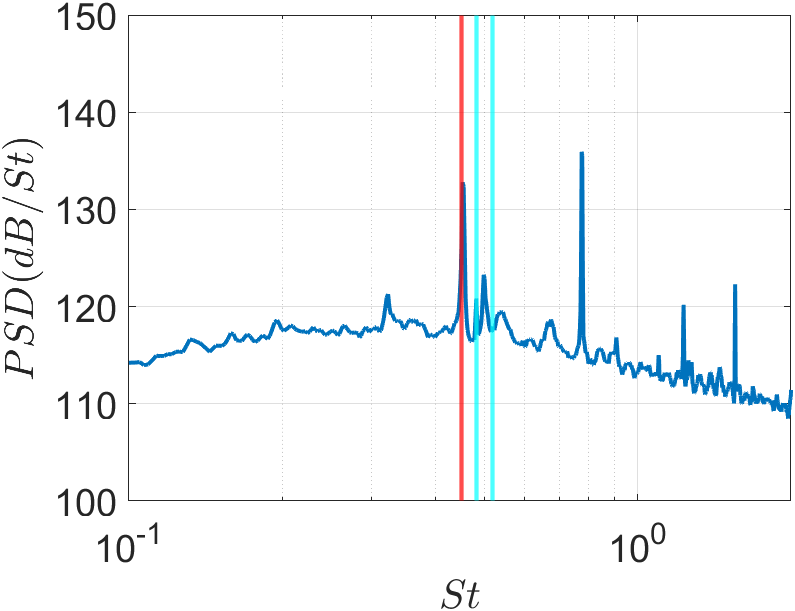}}\subfigure[]{\includegraphics[clip=true, trim= 0 0 0 0, width=0.25\textwidth]{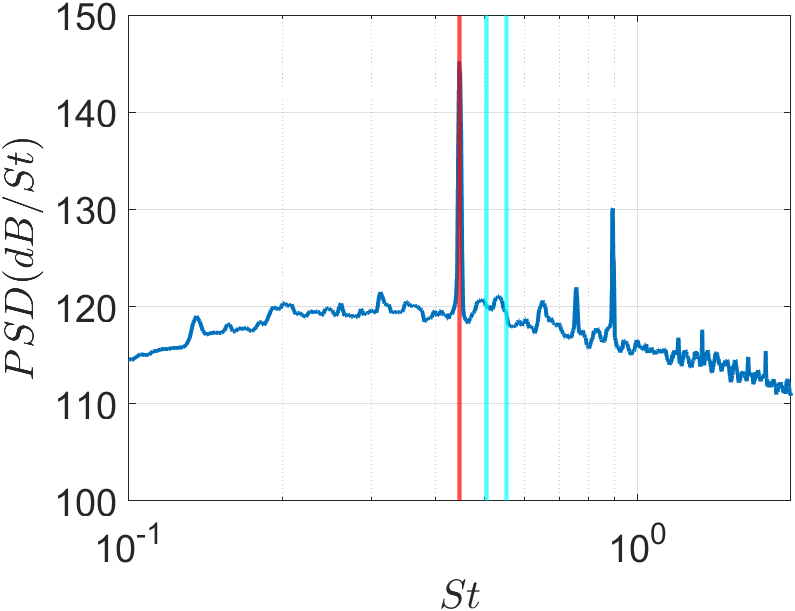}}\\
\subfigure[]{\includegraphics[clip=true, trim= 0 0 0 0, width=0.25\textwidth]{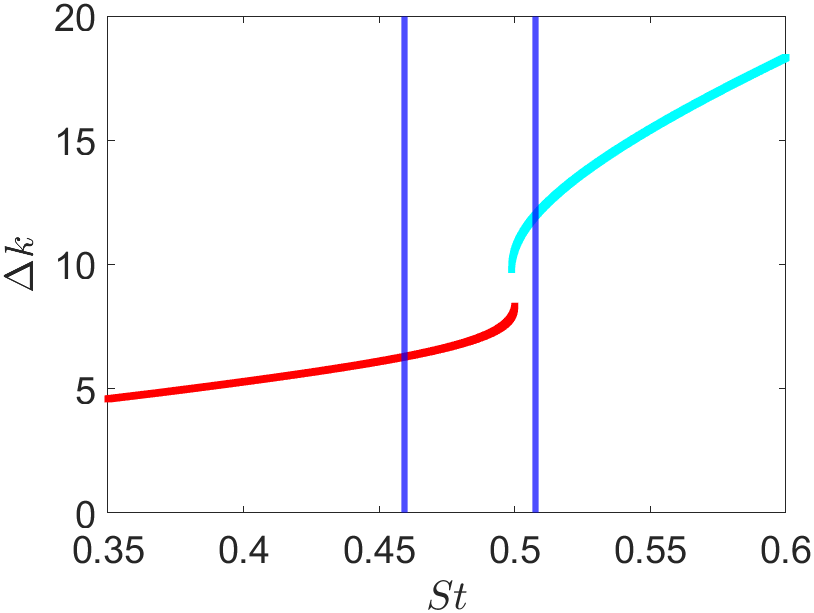}}\subfigure[]{\includegraphics[clip=true, trim= 0 0 0 0, width=0.25\textwidth]{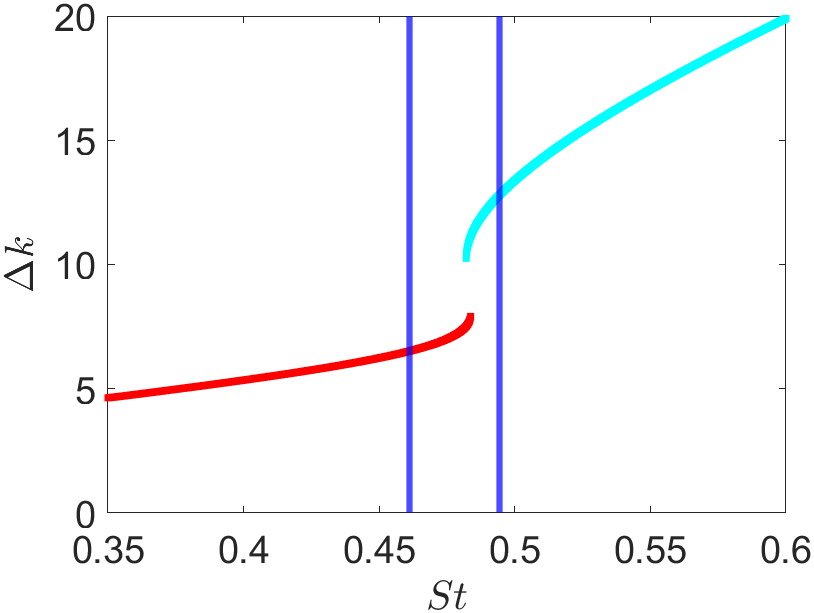}}\subfigure[]{\includegraphics[clip=true, trim= 0 0 0 0, width=0.25\textwidth]{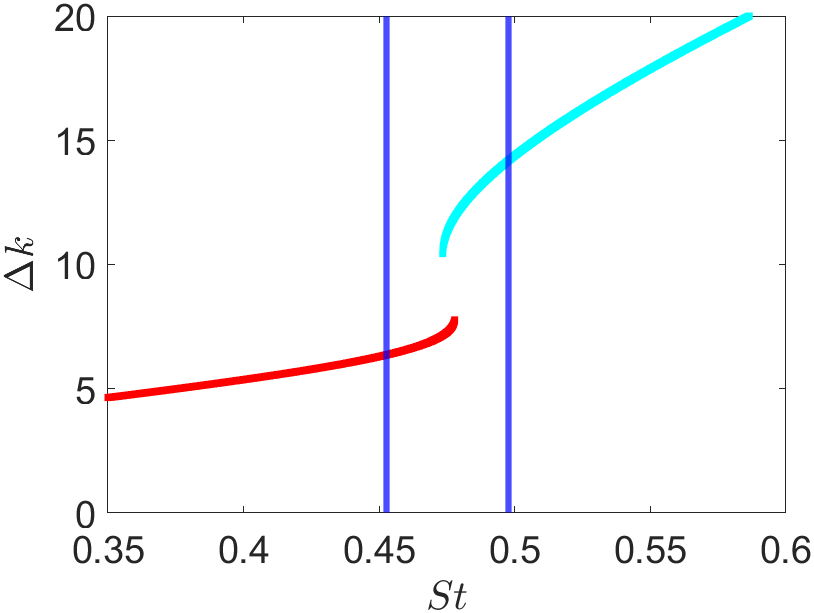}}\subfigure[]{\includegraphics[clip=true, trim= 0 0 0 0, width=0.25\textwidth]{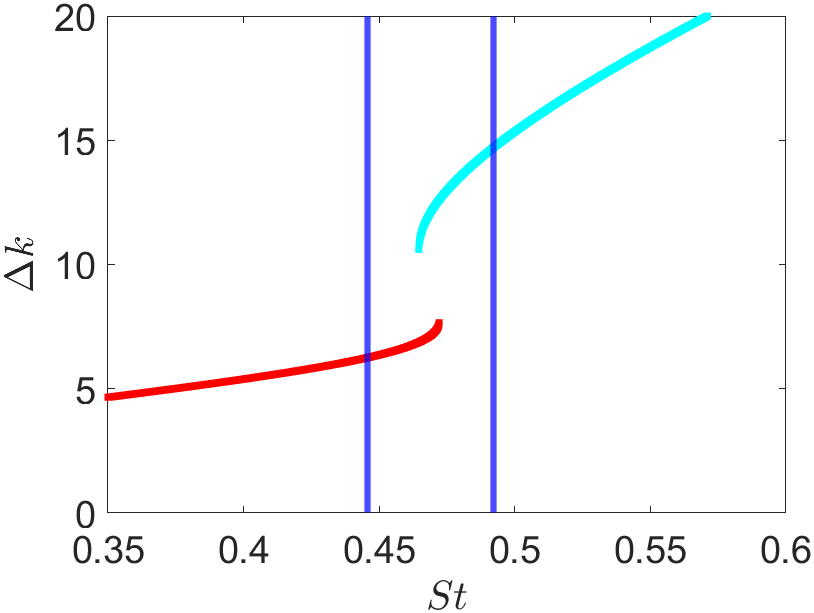}}
\caption{PSD spectra with predictions from linear stability theory overlain (a)-(d), and the value $\Delta k$ as a function of $St$ with the two tonal frequencies of interest overlain (e)-(h). Computed for $M_j = 0.83$ (a) and (e), $M_j = 0.84$ (b) and (f), $M_j = 0.85$ (c) and (g), $M_j = 0.86$ (d) and (h). Lines in cyan and red utilise $k^{-}_{d}$ and $k^{-}_{p}$ waves respectively. Frequency predictions overlaid use $p = 2$ and $\phi = 0$ (red) (a)-(d), 
$p = 4$ and $\phi = 0$ (cyan) (a), $p = 4,5$ and $\phi = 0$ (cyan) (b),(c), and $p = 5,6$ and $\phi = 0$ (cyan) (d).}%
\label{fig:18}%
\end{figure}

That these tones arise due to a linear frequency-selection mechanism is supported by their prediction by equation \ref{eqn:2.5}, as shown in figure \ref{fig:17}, where predictions are made using $k^{-}_{th}$ (cyan) as the upstream wave in figure \ref{fig:17}(a) and (b), and using $k^{-}_d$ (cyan), and $k^{-}_p$ (red) in figure \ref{fig:17}(c). In the first two plots, $M_j = 0.8,0.81$, a single upstream-travelling wave, $k^{-}_{th}$, exists to close resonance. At $M_j = 0.82$, figure \ref{fig:17}(c), both $k^{-}_{d}$ and $k^{-}_{p}$ waves are available to close the feedback loop - but the $k^{-}_{d}$ predictions aligns best with the fundamental tone. In figure \ref{fig:18} predictions using both feedback loops are shown in the range for which mode competition is observed ($M_j$ 0.83-0.86). The tones align well with the predictions. In particular the lower-frequency tone is matched with predictions using $k^{-}_{p}$, whilst the higher-frequency tone is matched with predictions using $k^{-}_{d}$. This supports the discussion of figure \ref{fig:16} where the tone switching was argued to be due to the existence of two competing feedback mechanisms; as the system moves past $M_j = 0.84$ the fundamental tone is no longer closed by resonance involving a $k^{-}_{d}$ or $k^{-}_{th}$, but by a $k^{-}_{p}$ wave. In addition figure \ref{fig:18}(e)-(h) shows $\Delta k$ (the left-hand-side of equation \ref{eqn:2.5}) as a function of $St$ for both $k^{-}_{d}$ (cyan), and $k^{-}_{p}$ (red). Overlain are the frequencies of the tones from the experiment in blue, which intersect respectively with the cyan (higher-frequency tone), and the red (lower-frequency tone) curves, indicating the closure mechanism associated with each and illustrating again that the competing LFS tones are associated with distinct feedback loops, each closed by a different $k^-$ wave. A simplified argument as to why $k_p^-$ resonance might be preferred by the system to $k_d^-$ resonance is put forth in Appendix \ref{App:C}.
 
\subsection{Mode switching for other $R/D$}
\label{Sec:Robust}
\begin{figure}
\centering
\subfigure[$M_j$ 0.82]{\includegraphics[clip=true, trim= 0 0 0 0, width=0.33\textwidth]{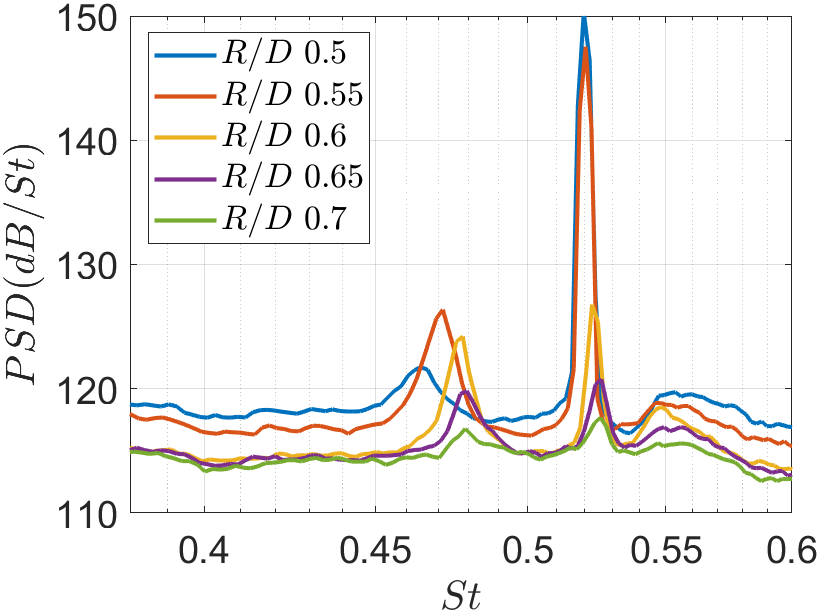}}\subfigure[$M_j$ 0.83]{\includegraphics[clip=true, trim= 0 0 0 0, width=0.33\textwidth]{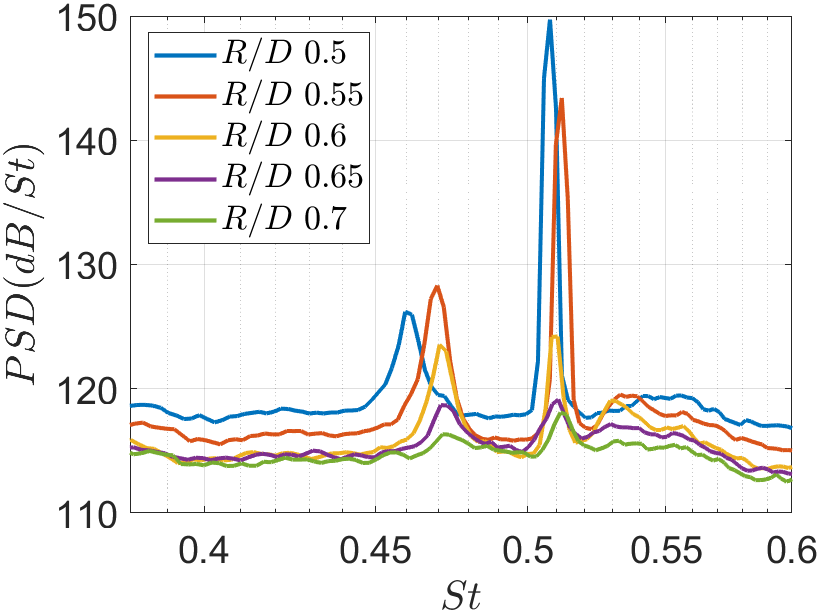}}\\
\subfigure[$M_j$ 0.84]{\includegraphics[clip=true, trim= 0 0 0 0, width=0.33\textwidth]{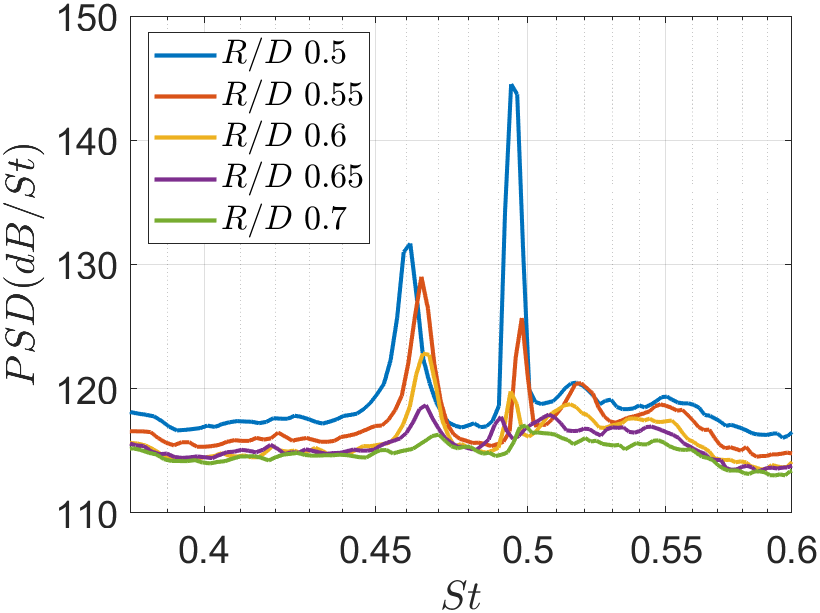}}\subfigure[$M_j$ 0.85]{\includegraphics[clip=true, trim= 0 0 0 0, width=0.33\textwidth]{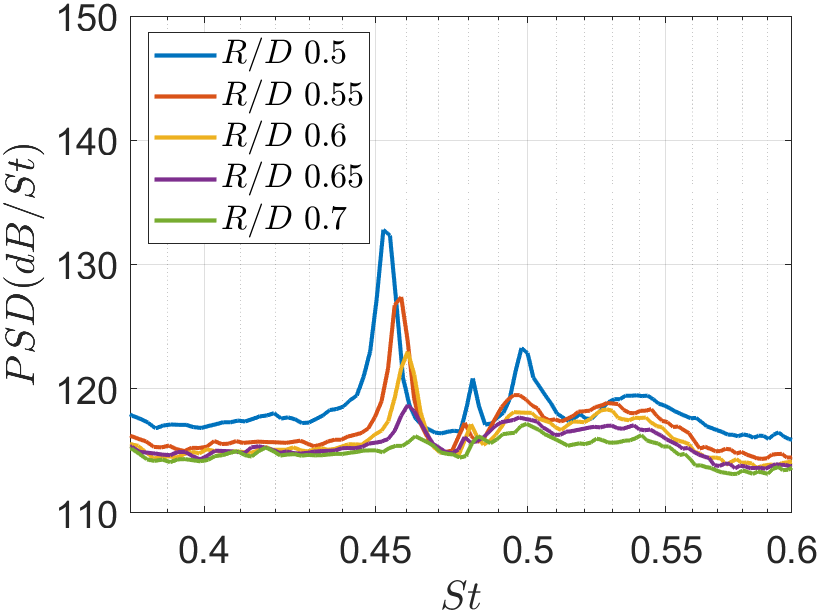}}\subfigure[$M_j$ 0.86]{\includegraphics[clip=true, trim= 0 0 0 0, width=0.33\textwidth]{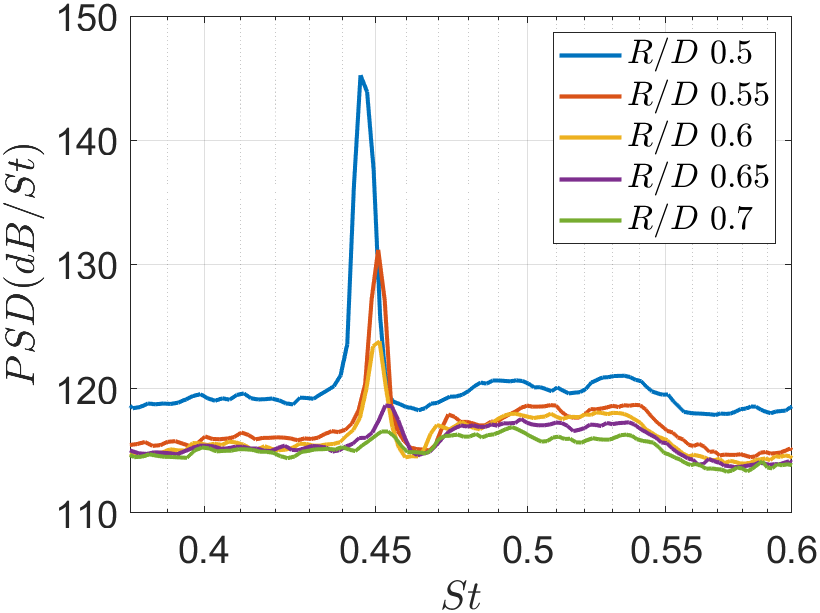}}
\caption{PSD spectra for $M_j 0.83 - 0.86$ across $R/D$. The competing tones may still be observed out to $R/D$ of 0.65.}%
\label{fig:19}%
\end{figure}
The analysis above focused on $R/D = 0.5$. In figure \ref{fig:19} PSD for $M_j$ 0.82-0.86 are shown for a range of $R/D$ values. The plots, which correspond to the zoomed region of figure \ref{fig:16}, show the same tone competition for $R/D$ 0.5-0.65. Some differences are the slightly lower value of $M_j$ at which the tone switch happens, and the lower tone amplitudes. Beyond $R/D = 0.7$ the tone competition is no longer observed.

\section{Tone amplitudes}
\label{Sec:SPL}
\begin{figure}
    \centering
    \includegraphics[clip=true, trim= 0 0 0 0, width=0.7\textwidth]{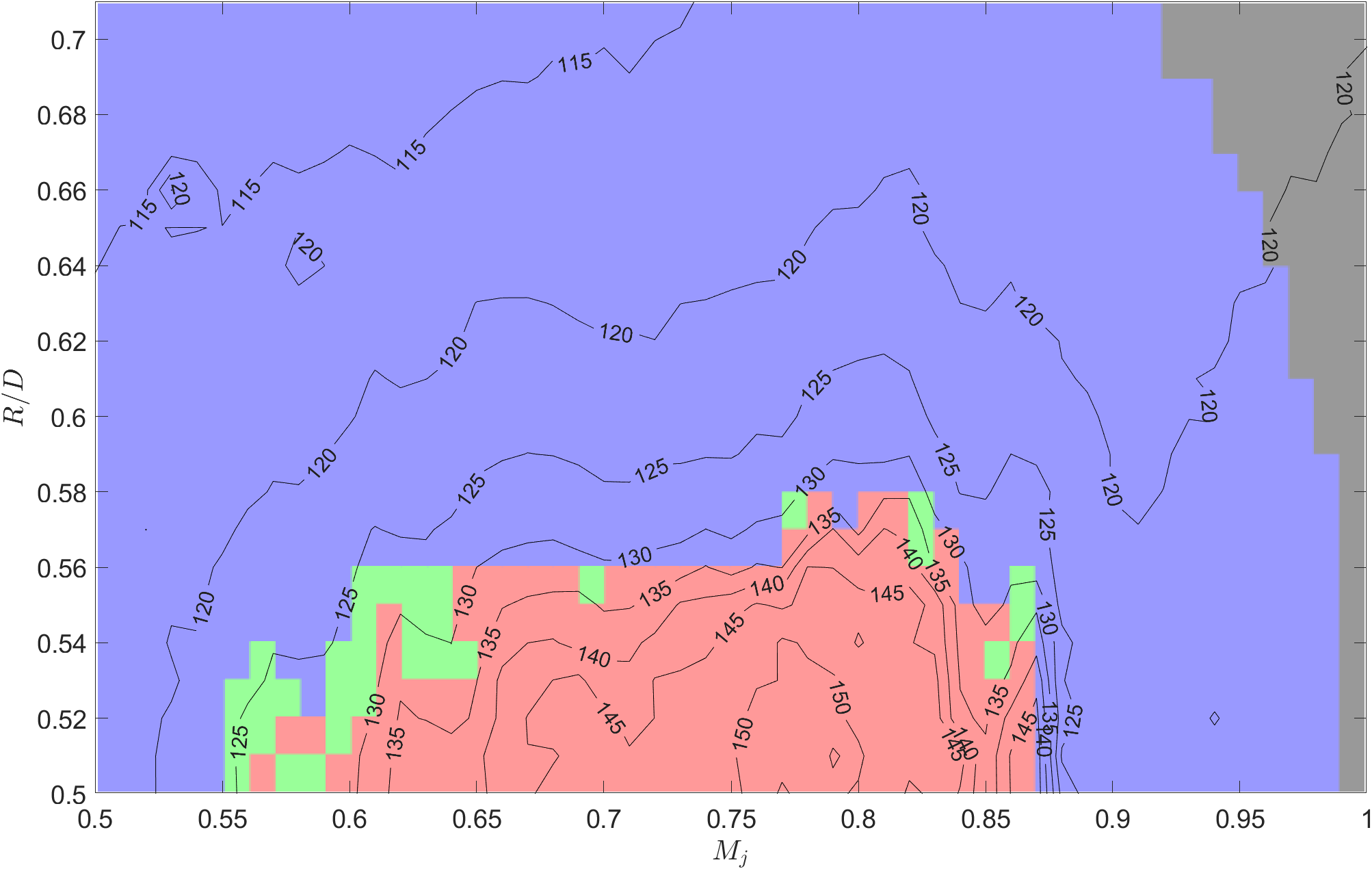}
    \caption{The spectral signatures classification from figure \ref{fig:11} overlain with a contour of maximum SPL at each position.}%
    \label{fig:20}%
    \end{figure}
\begin{figure} 
    \centering
    \includegraphics[clip=true, trim= 0 0 0 0, width=0.7\textwidth]{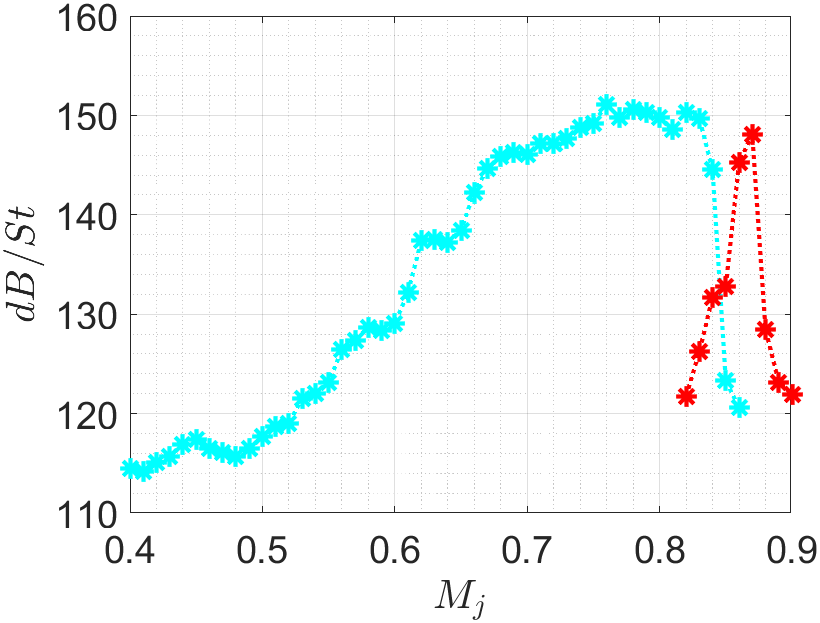}
    \caption{Comparison of the fundamental tone SPL amplitude across $M_j$ for $R/D = 0.5$ separated by feedback loop. For $M_j < 0.82$ (cyan) the feedback loop is closed via $k^{-}_{th}$ waves, for $M_j \geq 0.82$ (cyan) by $k^{-}_{d}$ waves, and $M_j \geq 0.82$ (red) by $k^{-}_{p}$ waves.}%
    \label{fig:21}%
\end{figure}
The classification in the parameter space $(M_j, R/D)$ can be complemented by considering the maximum tone amplitudes achieved by resonance. This is done in figure \ref{fig:20} which shows the maximal amplitudes superposed on the mode-classification map. Figure \ref{fig:20} shows that the SPL is greatest in regions corresponding to NLFS system dynamics (red region); here, the measured amplitudes reach $150dB$. An exception is at $M_j = 0.85$ and $R/D \leq 0.55$, where there is a noticeable decrease in amplitude compared to the SPL at $M_j = 0.84$. This is associated with $m = 1$ resonance, to be discussed in a future work. When moving away from the NLFS region the levels decrease quickly, in particular when passing over the boundaries between the different dynamics.
In figure \ref{fig:21} the amplitude of the fundamental tone is plotted across $M_j$ for $R/D = 0.5$. For $M_j \geq 0.82$, the amplitudes of the two competing linear tones, closed by $k^{-}_{d}$ (cyan) and $k^{-}_{p}$ (red) waves, are shown. The tone amplitudes are seen to undergo exponential variations with Mach number, of up to three orders of magnitude. And in the case of the $k_p^-$ closure, this growth and decay of the amplitude over a short Mach-number range is quite spectacular. We note also the correspondence between amplitude growth of the $k_p^-$ resonance and the amplitude decay of the $k_d^-$ mechanism, during the tone-switching Mach-number range.

\newpage
\section{Conclusions}
\label{Sec:Conclude}
We have considered interactions from a turbulent round jet grazing upon the edge of an angled metal plate as a function of jet Mach number, $M_j$, and radial plate position, $R/D$. The far-field sound was recorded throughout the $(M_j,R/D)$ parameter space with a combination of power spectral density (PSD), bicoherence, and linear stability analyses used to characterise the different spectral signatures observed. These comprise: broadband spectra; tones arising through a linear frequency-selection mechanism (LFS); LFS accompanied by harmonic or non-harmonic triadic interactions of linear tones; and non-linear frequency-selection mechanism (NLFS). Non-linear tone interactions are observed only in a region close to the jet $R/D \leq 0.57$, and in the Mach-number range $0.56 \leq M_j \leq 0.87$. Transition between LFS and NLFS regimes can involve changes in SPL of many orders of magnitude, across Mach-number increments of 0.01. Such changes are robust: repeatable and free from hysteresis. Using a linear frequency-prediction model, the fundamental tone of strong non-linear oscillator dynamics is shown to arise through an LFS mechanism. In the Mach-number range $0.82 \leq M_j \leq 0.86$ there is competition between two such dynamics. Using the linear model we show that this is underpinned by a switch in feedback mechanisms; where cut-on of a new upstream-travelling wave leads to the establishment of a new feedback loop. It is argued that the dominant feedback loop is due to the stronger reflection mechanisms associated with the new upstream-travelling wave. The study raises the question as to why the dynamics switch from saturated LFS tones, with no observable triadic interaction, to strong NLFS oscillator dynamics. This robust (repeatable, no hysteresis) regime switching, which occurs at two Mach numbers ($M_j$ = 0.57 and 0.87) is the subject of ongoing research.  

\appendix
\section{Bicoherence localisation}
\label{App:A}
\begin{figure}
\centering
\subfigure[]{\includegraphics[clip=true, trim= 0 0 0 0, width=0.5\textwidth]{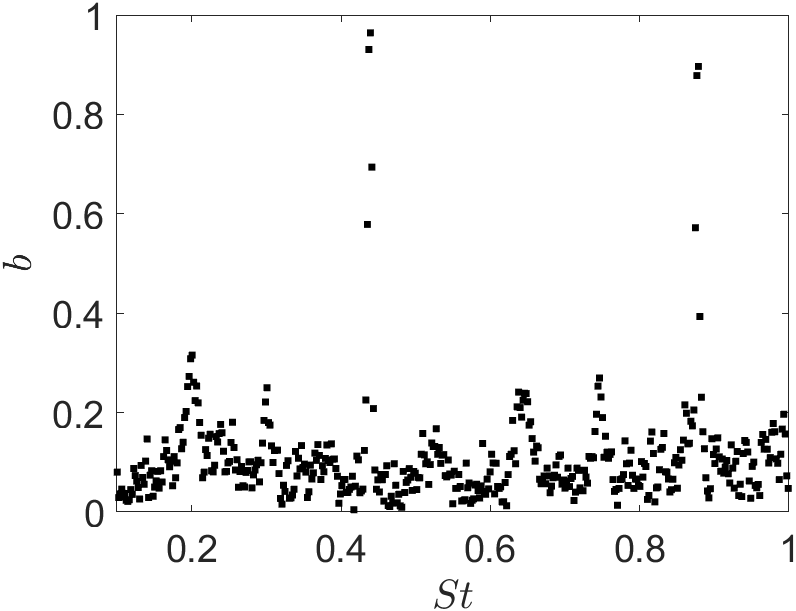}}\subfigure[]{\includegraphics[clip=true, trim= 0 0 0 0, width=0.5\textwidth]{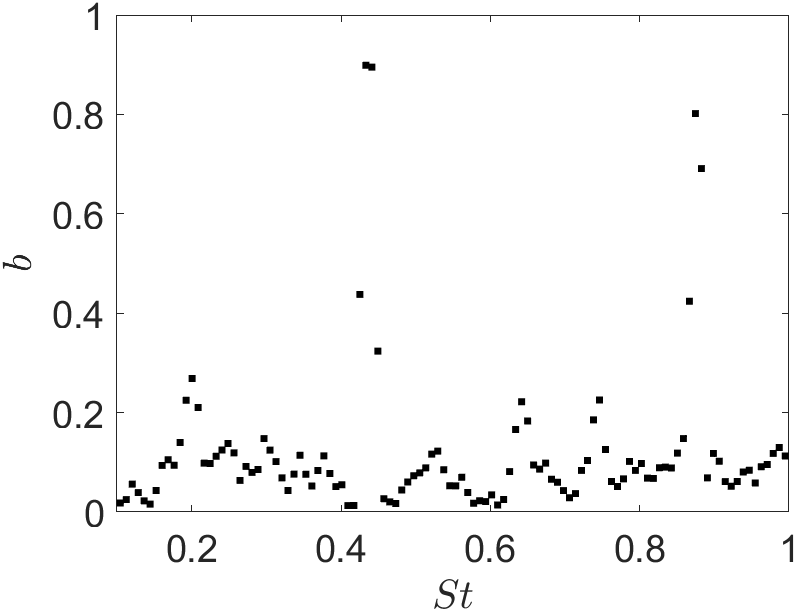}}
\caption{Value of $b$ at the fundamental frequency across $St$ using 16384 (a) and 4096 (b) points respectively in the fast Fourier transform.}%
\label{fig:22}%
\end{figure}
Here the localisation of $b$ peaks in $St$ is shown for $(M_j,R/D) = (0.87,0.5)$ in figure \ref{fig:22}. It is seen that triadic interactions detected at a given frequency only leak into 2-3 frequencies on either side of the peak, with $b$ values at all other $St$ being below the significance threshold ($b<0.35)$. As such, this presents a difficulty in displaying contours of $b$ that can be easily visualised. To mitigate this, the $b$ contours presented in this paper are overlain with markers at each $b$ peak to allow the contour plots to be read more clearly.

\section{Repeatability and hysteresis}
\label{App:B}
\begin{figure}
\centering
\subfigure[$M_j$ 0.83]{\includegraphics[clip=true, trim= 0 0 0 0, width=0.5\textwidth]{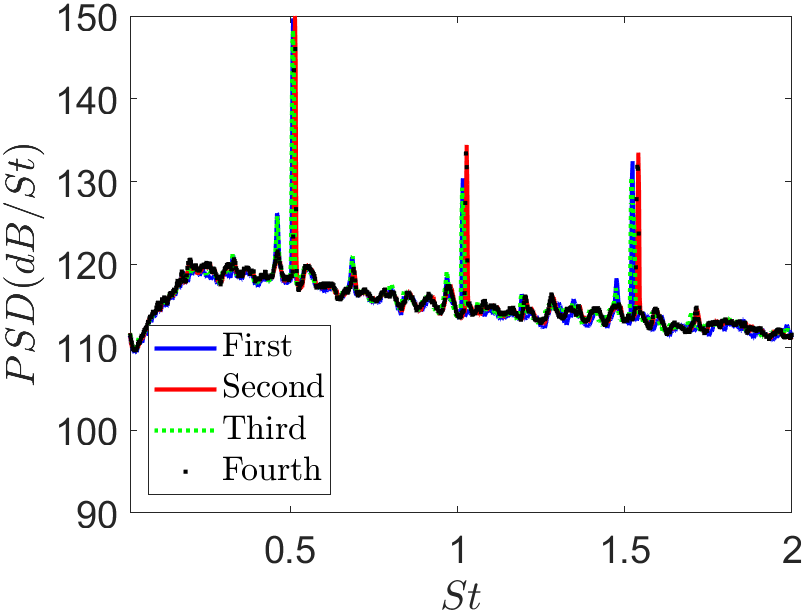}}\subfigure[$M_j$ 0.84]{\includegraphics[clip=true, trim= 0 0 0 0, width=0.5\textwidth]{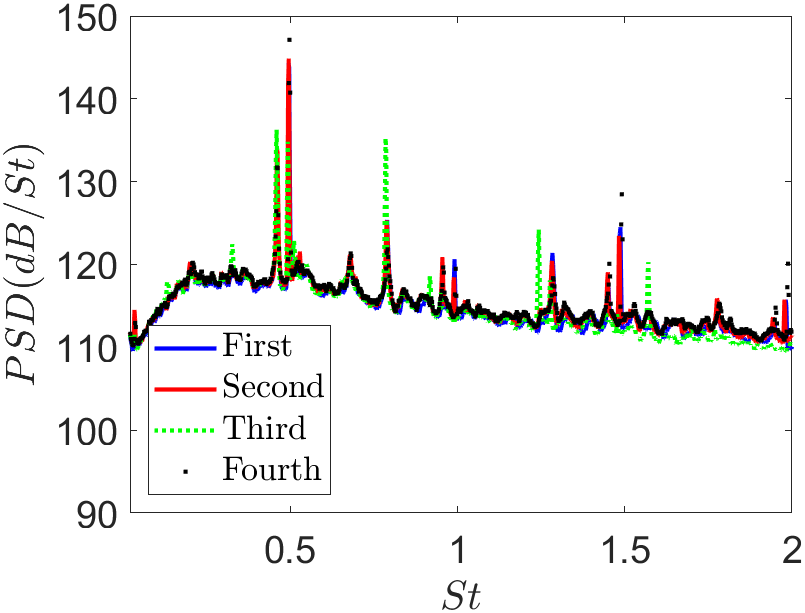}}\\
\subfigure[$M_j$ 0.85]{\includegraphics[clip=true, trim= 0 0 0 0, width=0.5\textwidth]{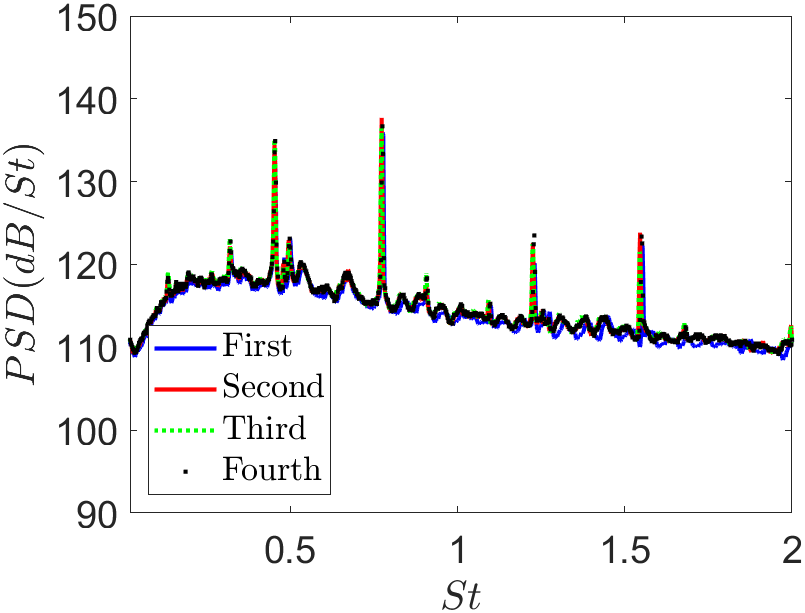}}\subfigure[$M_j$ 0.86]{\includegraphics[clip=true, trim= 0 0 0 0, width=0.5\textwidth]{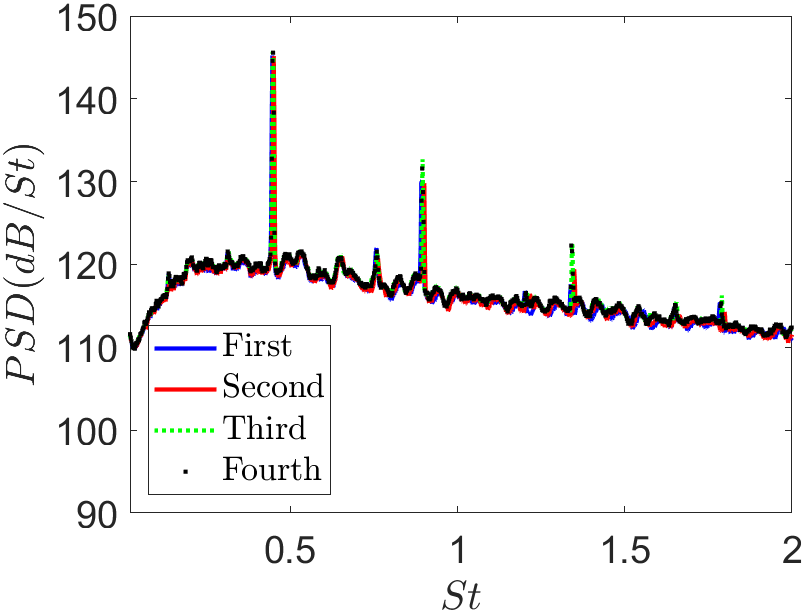}}\\
\subfigure[$M_j$ 0.87]{\includegraphics[clip=true, trim= 0 0 0 0, width=0.5\textwidth]{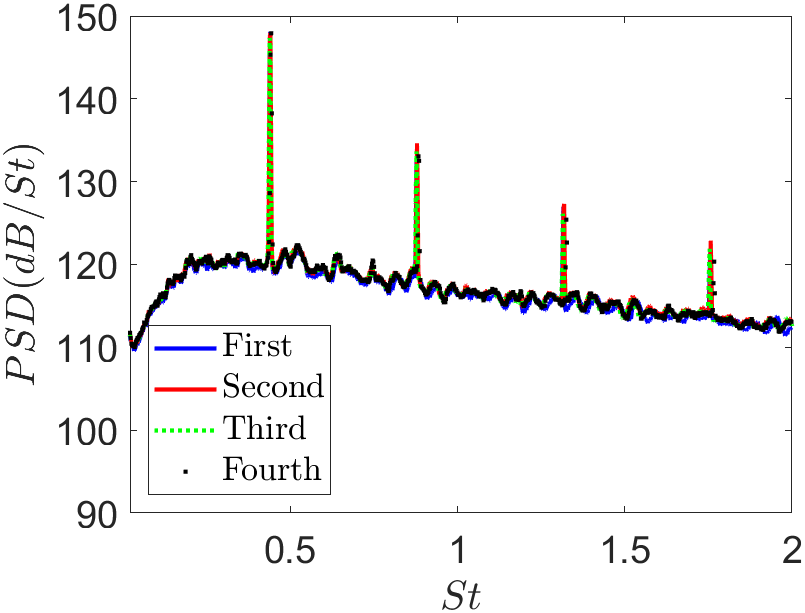}}\subfigure[$M_j$ 0.88]{\includegraphics[clip=true, trim= 0 0 0 0, width=0.5\textwidth]{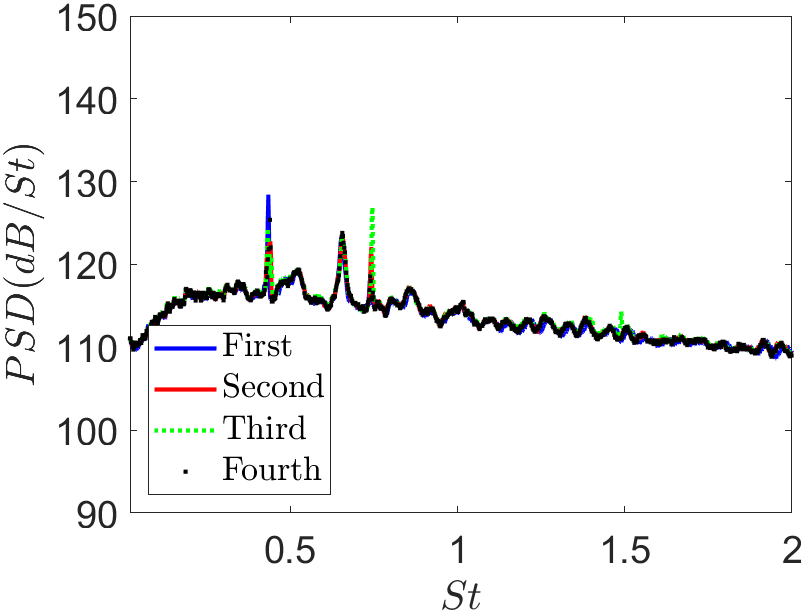}}
\caption{PSD spectra $(M_j,R/D) = (0.83 - 0.88,0.5)$ computed for four separate acquisitions.}%
\label{fig:23}%
\end{figure}
\begin{figure} 
    \centering
    \includegraphics[clip=true, trim= 0 0 0 0, width=0.7\textwidth]{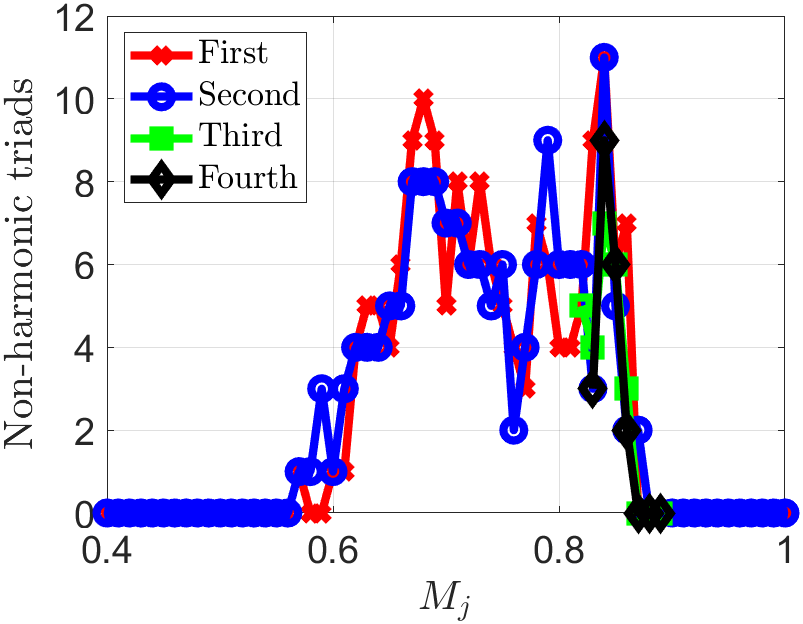}
    \caption{Comparison of the number of non-harmonic triads computed for $R/D = 0.5$ over four separate acquisitions.}%
    \label{fig:24}%
\end{figure}

In figure \ref{fig:23} the PSD spectra for $(M_j,R/D)$ = (0.83-0.88,0.5), are shown for four different acquisitions. All of these acquisitions were obtained using the same facility and microphone as detailed in \S \ref{Sec:Setup}. The acquisition labelled \textit{First} was acquired at the beginning of the experimental campaign and the remaining acquisitions were acquired during a single day three months later; there were no changes in the experimental facility during this time frame. Acquisitions \textit{First} and \textit{Second} both began at $M_j = 1$ and decreased in 0.01 increments to $M_j = 0.4$. Acquisition \textit{Third} starts at $M_j = 0.89$ and moves in 0.01 increments to $M_j = 0.82$, without shutting off the jet $M_j$ is then increased back to $M_j = 0.89$ in 0.01 increments leading to acquisition \textit{Fourth}. For all four acquisitions in figure \ref{fig:23} the competition between the two linear modes is observed, with the lower-frequency tone surpassing the higher-frequency tone. In addition, the different spectra for $M_j$ 0.83, 0.84 and 0.85, figure \ref{fig:23}(a), (c) and (d), are essentially identical. However there is a difference observed for acquisition \textit{Third} at $M_j = 0.84$, figure \ref{fig:23}(b). This spectrum shows the lower-frequency tone having a slightly greater amplitude than the higher-frequency tone at an $M_j$ of 0.84, it has overtaken the higher-frequency tone at a lower $M_j$ than the other acquisitions. It is not clear what caused that transition to take place at an earlier $M_j$. Perhaps multiple states are possible for the system at this operating condition, and if a sufficiently long measurement is taken switching between the states may even be observed. However, it is clear that this transition leads to a 10 dB difference between acquisition \textit{Third} and the others. Triggering an earlier transition could lead to significant noise reduction for $M_j = 0.84$, and perhaps could also for $M_j$ 0.82 and 0.83 if earlier transitions can be triggered there also. The transition between LFS and NLFS dynamics across the $M_j = 0.87-0.88$ boundary is seen in figures \ref{fig:23}(e) and (f), demonstrating the robustness of the transition. Importantly, there is no hysteresis found once the system reaches an NLFS regime - the strong-oscillator behaviour may be disrupted simply by changing the $M_j$ of the system by 0.01. A sole difference is found within the \textit{Third} acquisition for $M_j = 0.88$, where the amplitude of the $m=1$ tone has increased and it produces a small harmonic visible within the spectrum. In this case the system would not be classified as LFS, but LFS with weak non-linear interactions. It is unclear what caused the $m=1$ tone to be more energised during that run, and not the others. In figure \ref{fig:24} the number of non-harmonic triads present during the four acquisitions, for $R/D = 0.5$, are plotted across $M_j$. The exact number of triads present differs across the different acquisitions; however, this is to be expected as the majority are \textit{weak} triads and thus their emergence may be easily disrupted. The general trend across $M_j$ remains consistent, with peaks observed in the spectrum near $M_j = 0.84, 0.69$. Thus, while we would not expect a future acquisition to exactly match the number of non-harmonic triads found in this study - we would expect the trend in $M_j$ to be consistent with what we observe in figure \ref{fig:24}.
\section{The preferred linear tone}
\label{App:C}
\begin{figure} 
    \centering
    \includegraphics[clip=true, trim= 0 0 0 0, width=0.7\textwidth]{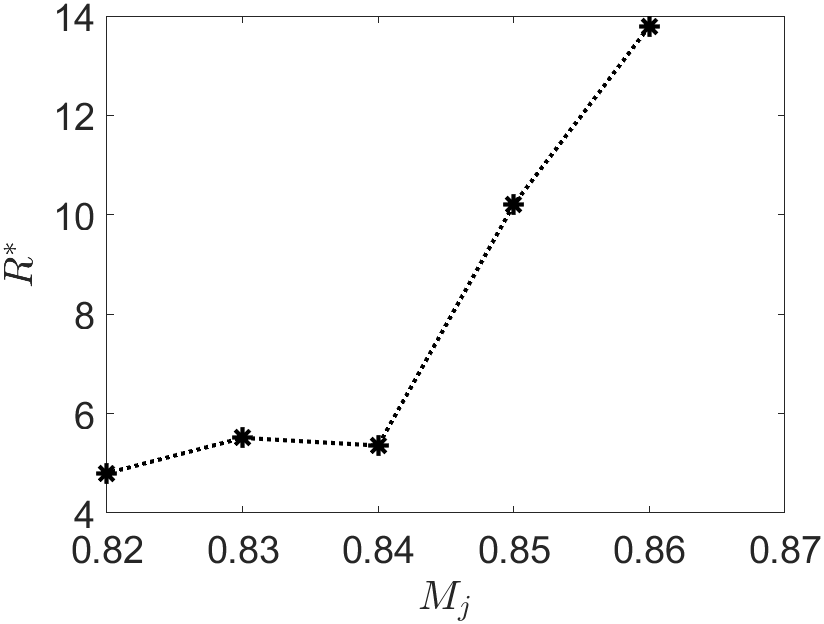}
    \caption{Comparison of reflected to incident amplitude between $k^{-}_{p}$ and $k^{-}_{d}$ resonance. $R^{*} > 1$ indicates that $k^{-}_{p}$ resonance will produce a greater amplitude increase than $k^{-}_{d}$ resonance.}%
    \label{fig:25}%
\end{figure}

To explore why one feedback loop is preferred over another in the competition detailed in \S \ref{Sec:Comp}, we consider two potential reflection mechanisms: impedance mismatch in the nozzle plane, and scattering via the nozzle lip. For the impedance mismatch the wave reflection coefficients, which determine the gain of the resonance loop, are considered over the Mach-number range 0.82-0.86 - where the mode switching occurs. The simplified model of \citet{jordan2018jet} (detailed in Appendix C of that paper), used to determine the cut-off frequency for $k^{-}_{th}$ waves, is applied. The formulation models the nozzle and jet as, respectively, hard- and soft-walled cylindrical ducts. An upstream-travelling wave, of amplitude $I$, incident on the nozzle produces a transmitted wave, of amplitude $T$, and a reflected wave, of amplitude $R$ (within this formulation reflection is considered only at the nozzle plane). The ratio of reflected to incident amplitudes can be written
\begin{equation}
    \frac{R}{I} = \left(\frac{1-\frac{k_{j}^{+}M}{\omega}}{1-\frac{k_{j}^{-}M}{\omega}}\right)\left(\frac{k_{n}^{-}-k_{j}^{-}}{k_{j}^{+}-k_{n}^{-}}\right),
    \label{eqn:C1}
\end{equation}
with $k^{-}_{n}$ the transmitted wave, $k^{-}_{j}$ the incident wave, and $k^{+}_{j}$ the reflected wave. Following \citet{jordan2018jet} the $k^{-}_{n}$ waves in equation \ref{eqn:C1} satisfy
\begin{equation}
    k_{n}^{-} = \frac{-\omega M -\sqrt{\omega^2}}{1-M^2} = -2\pi M St \left(\frac{M+1}{1-M^2}\right).
    \label{eqn:C2}
\end{equation}
Equation \ref{eqn:C1} is used to compute reflection-coefficient ratios with $k_p^-$ and $k_d^-$ considered as the incident waves, and the following ratio computed
\begin{equation}
    R^{*} = \frac{\left(\frac{R}{I}\right)_{k_{p}^{-}}}{\left(\frac{R}{I}\right)_{k_{d}^{-}}},
    \label{eqn:C3}
\end{equation}
giving, after substituting equation \ref{eqn:C2}
\begin{equation}
    R^{*} = \left(\frac{1-\frac{k_{d}^{-}}{2\pi St_d}}{1-\frac{k_{p}^{-}}{2\pi St_p}}\right)\left(\frac{1-\frac{KH_{p}}{2\pi St_d}}{1-\frac{KH_{d}}{2\pi St_p}}\right)\left(\frac{-2\pi M St_p \left(\frac{M+1}{1-M^2}\right)-k_{p}^{-}}{-2\pi M St_d \left(\frac{M+1}{1-M^2}\right)-k_{d}^{-}}\right)\left(\frac{KH_d+2\pi M St_p \left(\frac{M+1}{1-M^2}\right)}{KH_p+2\pi M St_d \left(\frac{M+1}{1-M^2}\right)}\right),
    \label{eqn:C4}
\end{equation}
where $St_d$ and $St_p$ are the frequencies of the competing tones, and $KH_p$ and $KH_d$ are the real wavenumber components of the Kelvin-Helmholtz waves that form resonance with the $k_p^-$ and $k_d^-$ waves respectively. This modelling approach carries with it important assumptions about the nature of the waves involved in resonance. In particular, this formulation treats all of the $k_d^-$, $k_p^-$, and KH waves as duct modes. For the $k_d^-$ wave, as its radial structure does not have support beyond the shear layer, the approximation as a duct mode would be suitable. Conversely, both the $k_p^-$ and KH modes maintain support beyond the shear layer, and thus their approximation as duct modes would be less valid. Whilst the derivation of equation \ref{eqn:C4} involves a duct acoustics model, the wavenumbers and frequencies for the KH, $k_p^{-}$, and $k_d^{-}$ waves are solved via the vortex-sheet dispersion relation (equation \ref{eqn:2.7}). $R^{*}$ is shown in figure \ref{fig:25} for $M_j$ values where competition occurs. Across all $M_j$ values $R^{*} > 1$, indicating that the $k^{-}_{p}$ wave generates a downstream-travelling wave with greater amplitude than that generated by the $k^{-}_{d}$ wave, and this predominance of the $k_p^-$ wave increases with $M_j$. The increase of $R^{*}$ with $M_j$ is stronger for $M_j > 0.84$, and this aligns with the point at which $k^{-}_{p}$ resonance overtakes $k^{-}_{d}$ resonance (see figure \ref{fig:16}).
\begin{figure} 
    \centering
    \includegraphics[clip=true, trim= 0 0 0 0, width=0.7\textwidth]{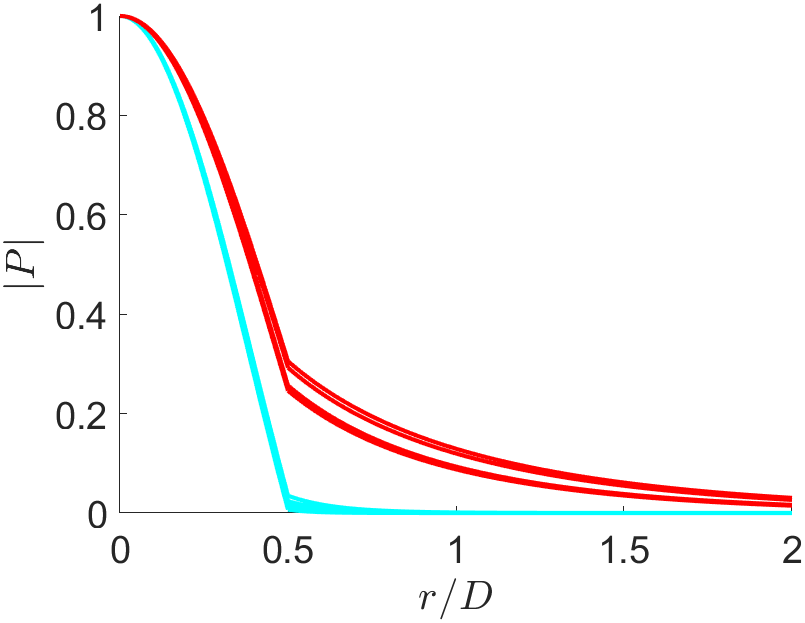}
    \caption{Normalised pressure eigenfunctions for $k^{-}_{d}$ (cyan) and $k^{-}_{p}$ (red) across $M_j = 0.82 - 0.86$ computed at the frequencies of the two tones.}%
    \label{fig:26}%
\end{figure}

For reflection in both the nozzle plane and at the plate edge, the radial support of the waves plays an important role. For the incident $k^-$ wave in the nozzle exit plane, it determines the amplitude seen by the nozzle lip (where KH receptivity is maximal). At the plate edge, a greater radial support of the $k^-$ wave will increase its receptivity to the scattering  $k^+$ wave. The pressure eigenfunctions are shown in figure \ref{fig:26} for $M_j$ 0.82-0.86 with $k^{-}_{p}$ plotted in red and $k^{-}_{d}$ in cyan. The larger radial support of the $k_p^-$ wave is clear, giving it an additional\footnote{Note that this feature is not included in the reflection coefficient ratio calculated above, which is based on solid- and soft-walled ducts, and gives an indication of the reflection mechanisms associated only with impedance mismatch in the nozzle exit plane.} advantage over the $k_d^-$ wave where energy exchange with the $k^+$ wave is concerned. For both waves the radial support decreases with increasing $M_j$. The simplified analysis presented within this section has made multiple assumptions about the nature of the waves involved in jet-edge resonance, with a more rigorous analysis requiring a non-parallel model which includes both the nozzle and plate geometries. However, for the purpose of this paper, the modelling undertaken provides some insights into the preference of $k_p^-$ resonance; namely, it is favoured over $k_d^-$ resonance both when considering the impedance mismatch in the nozzle plane (figure \ref{fig:25}), and scattering via the nozzle lip (figure \ref{fig:26}).

\section{Time-frequency analysis}
\label{App:D}
\begin{figure}
\centering
\subfigure[NLFS]{\includegraphics[clip=true, trim= 0 0 0 0, width=0.5\textwidth]{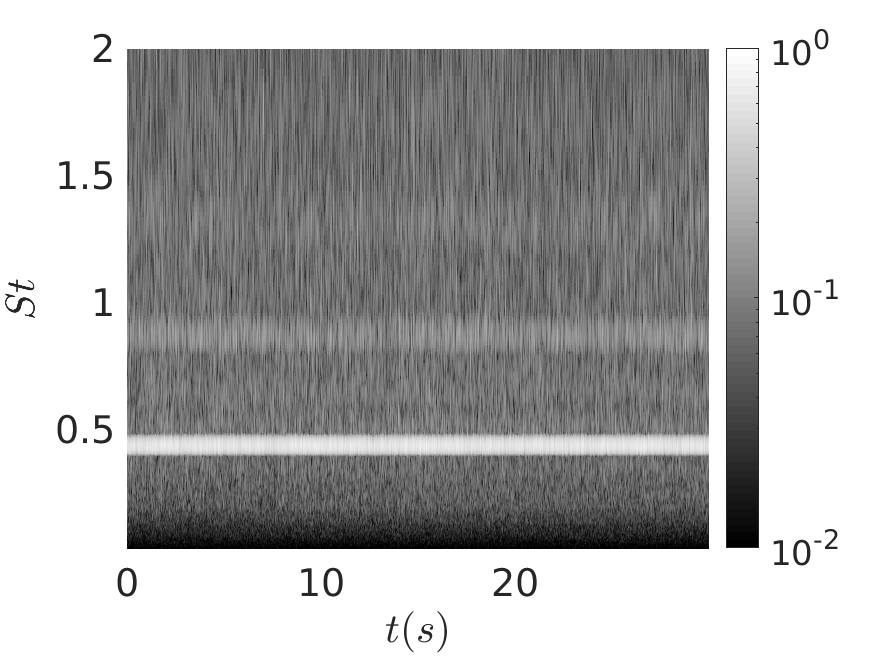}}\subfigure[LFS]{\includegraphics[clip=true, trim= 0 0 0 0, width=0.5\textwidth]{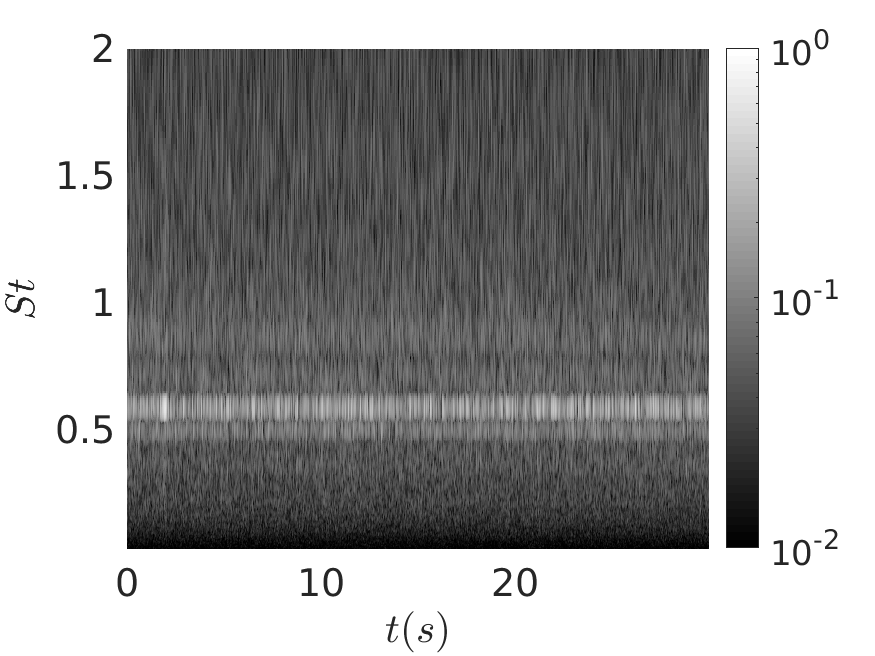}}\\
\subfigure[Axisymmetric]{\includegraphics[clip=true, trim= 0 0 0 0, width=0.5\textwidth]{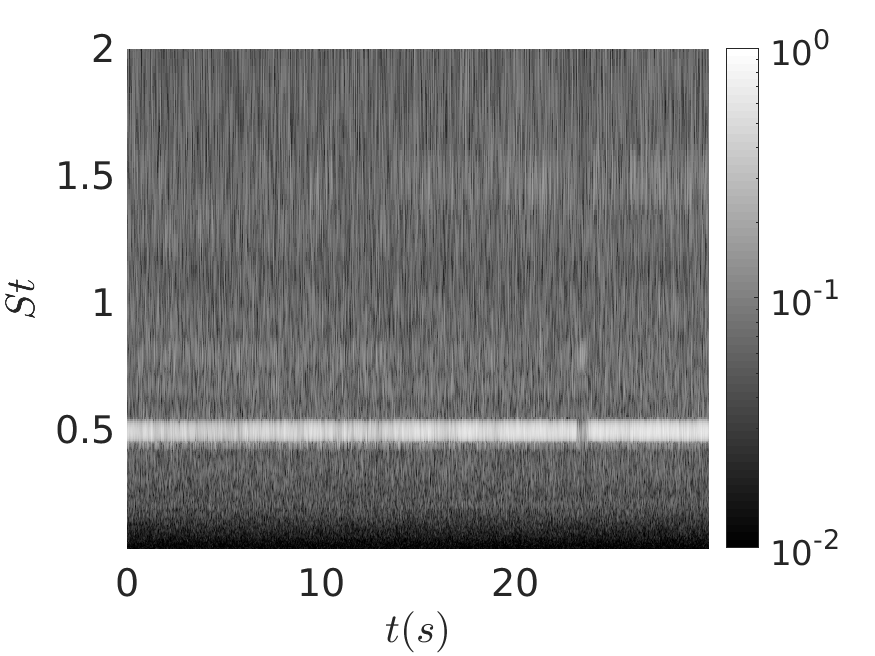}}\subfigure[Non-axisymmetric]{\includegraphics[clip=true, trim= 0 0 0 0, width=0.5\textwidth]{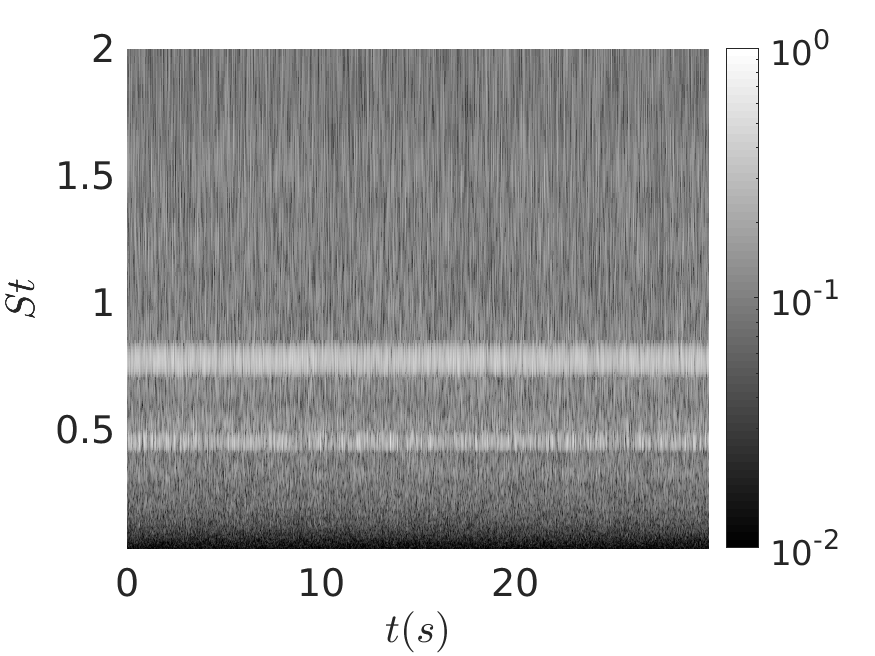}}
\caption{Wavelet spectra for $(M_j,R/D) = (0.87,0.5)$ (a), $(M_j,R/D) = (0.77,0.63)$ (b), $(M_j,R/D) = (0.84,0.5)$ (c), and $(M_j,R/D) = (0.85,0.5)$ (d). Intermittency between axisymmetric and non-axisymmetric resonance may be observed briefly in (c).}%
\label{fig:27}%
\end{figure}
Wavelet analysis \citep{Farge1992} has been used in recent studies on high-speed jets to identify intermittency between frequencies \citep{jordan2018jet,mancinelli2019screech,wong2023}. The methodology involves a convolution between $x(t)$ and an analysing wavelet that produces wavelet coefficients as a function of scale (inversely proportional to frequency) and time. Following previous studies \citep{jordan2018jet,mancinelli2019screech} a bump wavelet is chosen herein as the analysing wavelet which has a Fourier transform given by
\begin{equation}
    \hat{\Psi}(\omega) = \left(e^{1- \frac{1}{1-\sigma^2(\omega-\mu)^2}} \right)F(\omega),
    \label{eqn:D1}
\end{equation}
 where
 \begin{equation}
    F(\omega) =\begin{cases}
    \begin{aligned}
    & 1, & \mu-\frac{1}{\sigma}\leq \omega \leq \mu+\frac{1}{\sigma} \\
    & 0, & \textrm{else}
    \end{aligned}
    \end{cases}
     \label{eqn:D2}
 \end{equation}
 with $\omega = 2\pi f$. This work uses $\sigma = \frac{1}{0.6}$ and $\mu = 5$ for equations \ref{eqn:D1} and \ref{eqn:D2}, which were found to provide sufficient resolution in both time and frequency \citep{jordan2018jet}. Wavelet coefficients computed are all normalised by their maximum value. Figure \ref{fig:27} provides example wavelet spectra from across the parameter space. The first spectrum, figure \ref{fig:27}(a), is for $(M_j,R/D) = (0.87,0.5)$ (NLFS) where the flow is dominated by a strong fundamental tone and subsequent harmonics that are clearly seen in the spectra. In figure \ref{fig:27}(b) the spectrum is for $(M_j,R/D) = (0.77,0.63)$ (LFS). Here the system exhibits multiple linear tones, but without a single tone dominating the dynamics. This is highlighted in the wavelet spectrum as the tonal signatures are present, but not as clear as for the NLFS case. The final two spectrums demonstrate the transition between axisymmetric and non-axisymmetric dynamics. Figure \ref{fig:27}(c), is for $(M_j,R/D) = (0.84,0.5)$ where axisymmetric tones dominate. Here the signature of the axisymmetric tone appears strongly, and is continuous across almost the entire acquisition length. The exception occurs in a region around $t\sim22s$ where the axisymmetric signature is not present. In this segment in time there is a signature present at a $St$ value corresponding to an $m=1$ mode (the nature of this $m = 1$ resonance will be explored in a subsequent work). This suggests that the system undergoing axisymmetric resonance has briefly transitioned to an $m = 1$ resonance state, before returning to a state of axisymmetric resonance. It is likely that such transitions may also be possible at other locations, but simply did not occur within the time frame of the acquisitions taken. In figure \ref{fig:27}(d), $(M_j,R/D) = (0.85,0.5)$, the $m=1$ tone is dominant within the spectra; however, the signature of the axisymmetric tone may still be observed at a lesser magnitude. There is little intermittency observed within the parameter space when compared with the prior dataset analysed in \citet{amaral2023,Stavropoulos2024Edge}. As a key difference between the two setups is the flexibility of the plate, this would suggest that plate flexibility plays an important role in facilitating tonal intermittency and warrants further study.

\backsection[Acknowledgements]{We are grateful to Airbus for their support throughout this work. The authors would like to thank Anton Lebedev for his invaluable guidance and expertise with the experimental facility. M.N.S. wishes to thank Matteo Mancinelli for the warm support provided during the project. The bicoherence colour maps make use of the CMasher package \citep{vanderveldenCMasher2020}.}

\backsection[Funding]{This work was supported by the French government program "France 2030" (LABEX INTERACTIFS, reference ANR-11-LABX-0017-01), by the DGAC (\textit{Direction Générale de l’Aviation Civile}), by the PNRR (\textit{Plan National de Relance et de Résilience Français}) and by NextGeneration EU via the project MAMBO (\textit{Méthodes Avancées pour la Modélisation du Bruit moteur et aviOn}).}

\backsection[Declaration of interests]{The authors report no conflict of interest.}

\backsection[Author ORCIDs]{\\
Michael N. Stavropoulos https://orcid.org/0000-0003-3802-238X; \\
André V.G. Cavalieri https://orcid.org/0000-0003-4283-0232; \\
Lutz Lesshafft https://orcid.org/0000-0002-2513-4553; \\
Peter Jordan https://orcid.org/0000-0001-8576-5587.}

\bibliographystyle{jfm}
\bibliography{jfm}

@article{Bres2018, title={Importance of the nozzle-exit boundary-layer state in subsonic turbulent jets}, volume={851}, journal={Journal of Fluid Mechanics}, author={Brès, G.A. and Jordan, P. and Jaunet, V. and Le Rallic, M. and Cavalieri, A.V.G. and Towne, A. and Lele, S.K. and Colonius, T. and Schmidt, O.T.}, year={2018}, pages={83–124}}

@article{jordan2018jet,
  title={Jet--flap interaction tones},
  author={Jordan, P. and Jaunet, V. and Towne, A. and Cavalieri, A.V.G. and Colonius, T. and Schmidt, O. and Agarwal, A.},
  journal={Journal of Fluid Mechanics},
  volume={853},
  pages={333--358},
  year={2018},
  publisher={Cambridge University Press}
}

@article{towne2017acoustic,
  title={Acoustic resonance in the potential core of subsonic jets},
  author={Towne, A. and Cavalieri, A.V.G. and Jordan, P. and Colonius, T. and Schmidt, O. and Jaunet, V. and Br{\`e}s, G.A.},
  journal={Journal of Fluid Mechanics},
  volume={825},
  pages={1113--1152},
  year={2017},
  publisher={Cambridge University Press}
}

@inproceedings{amaral2023,
author = {F.R. Amaral and A. Lebedev and P. Jordan},
title = {Experiments on installed jet noise},
booktitle = {AIAA AVIATION 2023 Forum},
chapter = {},
pages = {3830},
year = {2023},
}

@article{welch1967use,
  title={The use of fast Fourier transform for the estimation of power spectra: a method based on time averaging over short, modified periodograms},
  author={Welch, P.},
  journal={IEEE Transactions on audio and electroacoustics},
  volume={15},
  number={2},
  pages={70--73},
  year={1967},
  publisher={IEEE}
}

@article{JordanColoniusWavepacket,
author = {Jordan, P. and Colonius, T.},
title = {Wave Packets and Turbulent Jet Noise},
journal = {Annual Review of Fluid Mechanics},
volume = {45},
number = {1},
pages = {173-195},
year = {2013}
}

@article{NOGUEIRA2017,
title = {A model problem for sound radiation by an installed jet},
journal = {Journal of Sound and Vibration},
volume = {391},
pages = {95-115},
year = {2017},
author = {P.A.S. Nogueira and A.V.G. Cavalieri and P. Jordan}
}

@article{Nogueira2019,
    author = {Nogueira, P.A.S. and Sirotto, J.R.L.N. and Miotto, R.F. and Cavalieri, A.V.G. and Cordioli, J.A. and Wolf, W.R.},
    title = "{Acoustic radiation of subsonic jets in the vicinity of an inclined flat plate}",
    journal = {The Journal of the Acoustical Society of America},
    volume = {146},
    number = {1},
    pages = {50-59},
    year = {2019}
}

@article{Piantanida2016,
    author = {Piantanida, S. and Jaunet, V. and Huber, J. and Wolf, W.R. and Jordan, P. and Cavalieri, A.V.G.},
    title = "{Scattering of turbulent-jet wavepackets by a swept trailing edge}",
    journal = {The Journal of the Acoustical Society of America},
    volume = {140},
    number = {6},
    pages = {4350-4359},
    year = {2016}
}

@article{CAVALIERI20146516,
title = {Scattering of wavepackets by a flat plate in the vicinity of a turbulent jet},
journal = {Journal of Sound and Vibration},
volume = {333},
number = {24},
pages = {6516-6531},
year = {2014},
author = {A.V.G. Cavalieri and P. Jordan and W.R. Wolf and Y. Gervais}
}

@article{sigl1994introduction,
  title={An introduction to bispectral analysis for the electroencephalogram},
  author={Sigl, J.C. and Chamoun, N. G.},
  journal={Journal of clinical monitoring},
  volume={10},
  pages={392--404},
  year={1994},
  publisher={Springer}
}

@article{mancinelli2019screech,
  title={Screech-tone prediction using upstream-travelling jet modes},
  author={Mancinelli, M. and Jaunet, V. and Jordan, P. and Towne, A.},
  journal={Experiments in Fluids},
  volume={60},
  number={1},
  pages={22},
  year={2019},
  publisher={Springer}
}

@article{wong2023, title={Steady and unsteady coupling in twin weakly underexpanded round jets}, volume={964}, journal={Journal of Fluid Mechanics}, publisher={Cambridge University Press}, author={Wong, T.Y.M. and Stavropoulos, M.N. and Beekman, J.R. and Towne, A. and Nogueira, P.A.S. and Weightman, J. and Edgington-Mitchell, D.}, year={2023}, pages={A2}}

@article{schmidt2017, title={Wavepackets and trapped acoustic modes in a turbulent jet: coherent structure eduction and global stability}, volume={825}, journal={Journal of Fluid Mechanics}, publisher={Cambridge University Press}, author={Schmidt, O.T. and Towne, A. and Colonius, T. and Cavalieri, A.V.G. and Jordan, P. and Brès, G.A.}, year={2017}, pages={1153–1181}}

@article{Farge1992,
   author = "Farge, M.",
   title = "Wavelet transforms and their applications to turbulence", 
   journal= "Annual Review of Fluid Mechanics",
   year = "1992",
   volume = "24",
   number = "Volume 24, 1992",
   pages = "395-458",
   publisher = "Annual Reviews",
   issn = "1545-4479",
   type = "Journal Article",
  }

@article{mancinelli2021, title={A complex-valued resonance model for axisymmetric screech tones in supersonic jets}, volume={928}, journal={Journal of Fluid Mechanics}, publisher={Cambridge University Press}, author={Mancinelli, M. and Jaunet, V. and Jordan, P. and Towne, A.}, year={2021}, pages={A32}}

@article{stavropoulos2023, title={The axisymmetric screech tones of round twin jets examined via linear stability theory}, volume={965}, journal={Journal of Fluid Mechanics}, publisher={Cambridge University Press}, author={Stavropoulos, M.N. and Mancinelli, M. and Jordan, P. and Jaunet, V. and Weightman, J. and Edgington-Mitchell, D.M. and Nogueira, P.A.S.}, year={2023}, pages={A11}}

@article{edgington2019aeroacoustic,
  title={Aeroacoustic resonance and self-excitation in screeching and impinging supersonic jets--a review},
  author={Edgington-Mitchell, D.},
  journal={International Journal of Aeroacoustics},
  volume={18},
  number={2-3},
  pages={118--188},
  year={2019},
  publisher={SAGE Publications Sage UK: London, England}
}

@article{tam1989three,
  title={On the three families of instability waves of high-speed jets},
  author={Tam, C.K.W. and Hu, F.Q.},
  journal={Journal of Fluid Mechanics},
  volume={201},
  pages={447--483},
  year={1989},
  publisher={Cambridge University Press}
}

@article{gojon2018oscillation,
  title={Oscillation modes in screeching jets},
  author={Gojon, R. and Bogey, C. and Mihaescu, M.},
  journal={AIAA Journal},
  volume={56},
  number={7},
  pages={2918--2924},
  year={2018},
  publisher={American Institute of Aeronautics and Astronautics}
}

@article{edgington2018upstream,
  title={Upstream-travelling acoustic jet modes as a closure mechanism for screech},
  author={Edgington-Mitchell, D. and Jaunet, V. and Jordan, P. and Towne, A. and Soria, J. and Honnery, D.},
  journal={Journal of Fluid Mechanics},
  volume={855},
  year={2018},
  publisher={Cambridge University Press}
}

@inproceedings{Stavropoulos2024Edge,
author = {M.N. Stavropoulos and F.R. Amaral and A.V.G. Cavalieri and L. Lesshafft and P. Jordan},
title = {Jet-Edge Interaction Tones: Linear and Non-Linear Mechanisms},
booktitle = {30th AIAA/CEAS Aeroacoustics Conference (2024)},
chapter = {},
pages = {3311},
year = {2024}
}

@article{nogueira2024guided,
  title={Guided-jet waves},
  author={Nogueira, P.A.S. and Cavalieri, A.V.G. and Martini, E. and Towne, A. and Jordan, P. and Edgington-Mitchell, D.},
  journal={Journal of Fluid Mechanics},
  volume={999},
  pages={A47},
  year={2024},
  publisher={Cambridge University Press}
}

@article{tam_ahuja_1990, title={Theoretical model of discrete tone generation by impinging jets}, volume={214}, journal={Journal of Fluid Mechanics}, publisher={Cambridge University Press}, author={Tam, C. K. W. and Ahuja, K. K.}, year={1990}, pages={67–87}}

@article{tam1992impingement,
  title={Impingement tones of large aspect ratio supersonic rectangular jets},
  author={Tam, C. K. W. and Norum, T. D.},
  journal={AIAA journal},
  volume={30},
  number={2},
  pages={304--311},
  year={1992}
}

@article{nogueira_edgington-mitchell_2021, title={Investigation of supersonic twin-jet coupling using spatial linear stability analysis}, volume={918}, journal={Journal of Fluid Mechanics}, publisher={Cambridge University Press}, author={Nogueira, P. A. S. and Edgington-Mitchell, D. M.}, year={2021}, pages={A38}}

@inproceedings{jaunet2019dynamics,
  title={Dynamics of round jet impingement},
  author={Jaunet, V. and Mancinelli, M. and Jordan, P. and Towne, A. and Edgington-Mitchell, D.M. and Lehnasch, G. and Girard, S.},
  booktitle={25th AIAA/CEAS Aeroacoustics Conference (2019)},
  pages={2769},
  year={2019}
}

@article{bogey2017feedback,
  title={Feedback loop and upwind-propagating waves in ideally expanded supersonic impinging round jets},
  author={Bogey, C. and Gojon, R.},
  journal={Journal of Fluid Mechanics},
  volume={823},
  pages={562--591},
  year={2017},
  publisher={Cambridge University Press}
}

@inproceedings{lawrence2015installed,
  title={Installed jet-flap impingement tonal noise},
  author={Lawrence, J.L.T. and Self, R.H.},
  booktitle={21st AIAA/CEAS aeroacoustics conference (2015)},
  pages={3118},
  year={2015}
}

@article{cavalieri2019wave,
  title={Wave-packet models for jet dynamics and sound radiation},
  author={Cavalieri, A.V.G. and Jordan, P. and Lesshafft, L.},
  journal={Applied Mechanics Reviews},
  volume={71},
  number={2},
  pages={020802},
  year={2019},
  publisher={American Society of Mechanical Engineers}
}

@article{brown_roshko_1974, title={On density effects and large structure in turbulent mixing layers}, volume={64}, number={4}, journal={Journal of Fluid Mechanics}, publisher={Cambridge University Press}, author={Brown, G. L. and Roshko, A.}, year={1974}, pages={775–816}}

@article{Mollo1967,
    author = {Mollo-Christensen, E.},
    title = "{Jet Noise and Shear Flow Instability Seen From an Experimenter’s Viewpoint}",
    journal = {Journal of Applied Mechanics},
    volume = {34},
    number = {1},
    pages = {1-7},
    year = {1967},
}

@article{crow_champagne_1971, title={Orderly structure in jet turbulence}, volume={48}, number={3}, journal={Journal of Fluid Mechanics}, publisher={Cambridge University Press}, author={Crow, S. C. and Champagne, F. H.}, year={1971}, pages={547–591}}

@article{ho1981dynamics,
  title={Dynamics of an impinging jet. Part 1. The feedback phenomenon},
  author={Ho, C.-M. and Nosseir, N.S.},
  journal={Journal of Fluid Mechanics},
  volume={105},
  pages={119--142},
  year={1981},
  publisher={Cambridge University Press}
}

@article{cavalieri2013wavepackets,
  title={Wavepackets in the velocity field of turbulent jets},
  author={Cavalieri, A.V.G. and Rodr{\'\i}guez, D. and Jordan, P. and Colonius, T. and Gervais, Y.},
  journal={Journal of fluid mechanics},
  volume={730},
  pages={559--592},
  year={2013},
  publisher={Cambridge University Press}
}

@article{cavalieri2012axisymmetric,
  title={Axisymmetric superdirectivity in subsonic jets},
  author={Cavalieri, A. V. G. and Jordan, P. and Colonius, T. and Gervais, Y.},
  journal={Journal of fluid Mechanics},
  volume={704},
  pages={388--420},
  year={2012},
  publisher={Cambridge University Press}
}

@article{vanderveldenCMasher2020,
  title = {{{CMasher}}: {{Scientific}} Colormaps for Making Accessible, Informative and 'cmashing' Plots},
  shorttitle = {{{CMasher}}},
  author = {{van der Velden}, E.},
  year = {2020},
  month = feb,
  journal = {Journal of Open Source Software},
  volume = {5},
  number = {46},
  pages = {2004}
}

@article{padois2016tonal,
  title={Tonal noise of a controlled-diffusion airfoil at low angle of attack and Reynolds number},
  author={Padois, T. and Laffay, P. and Idier, A. and Moreau, S.},
  journal={The Journal of the Acoustical Society of America},
  volume={140},
  number={1},
  pages={EL113--EL118},
  year={2016},
  publisher={AIP Publishing}
}

@article{powell1961edgetone,
  title={On the edgetone},
  author={Powell, A.},
  journal={The Journal of the Acoustical Society of America},
  volume={33},
  number={4},
  pages={395--409},
  year={1961},
  publisher={Acoustical Society of America}
}

@article{richardson1931edge,
  title={Edge tones},
  author={Richardson, E. G.},
  journal={Proceedings of the Physical Society},
  volume={43},
  number={4},
  pages={394--404},
  year={1931}
}

@article{smith2024history,
  title={History, review and summary of the cavity flow phenomena},
  author={Hamilton Smith, C. O. L. and Lawson, N. and Vio, G. A.},
  journal={European Journal of Mechanics-B/Fluids},
  volume={108},
  pages={32--72},
  year={2024},
  publisher={Elsevier}
}

@techreport{rossiter1964,
    author = {Rossiter, J. E.},
    title = {Wind-tunnel experiments on the flow over rectangular cavities at subsonic and transonic speeds},
    institution = {Ministry of Aviation},
    year = {1964},
    number = {ARC/R\&M-3438}
}

@article{edgington2025vortices,
  title={Vortices, shocks and non-linear acoustic waves: the ingredients for resonance in impinging compressible jets},
  author={Edgington-Mitchell, D. M. and Weightman, J. L. and Mazharmanesh, S. and Nogueira, P.},
  journal={Journal of Fluid Mechanics},
  volume={1015},
  pages={A56},
  year={2025},
  publisher={Cambridge University Press}
}

@article{henderson2005experimental,
  title={An experimental study of the oscillatory flow structure of tone-producing supersonic impinging jets},
  author={Henderson, B. and Bridges, J. and Wernet, M.},
  journal={Journal of Fluid Mechanics},
  volume={542},
  pages={115--137},
  year={2005},
  publisher={Cambridge University Press}
}

\end{document}